\documentclass[11pt,a4paper]{article}

\usepackage[margin=1in]{geometry}

\usepackage[T1]{fontenc}
\usepackage[utf8]{inputenc}
\usepackage{mathptmx}
\usepackage{microtype}

\usepackage{amsmath,amssymb}
\usepackage[version=3]{mhchem}
\usepackage{graphicx}
\usepackage{float}
\usepackage{booktabs}
\usepackage{multirow}
\usepackage{array}
\usepackage{xcolor}

\usepackage[font=small,labelfont=bf,labelsep=period]{caption}

\usepackage[super,sort&compress,comma,numbers]{natbib}
\usepackage[colorlinks=true,
            linkcolor=blue!50!black,
            citecolor=blue!50!black,
            urlcolor=blue!50!black]{hyperref}
\usepackage{xurl}
\usepackage{titlesec}
\titleformat{\section}{\normalfont\large\bfseries}{\thesection}{1em}{}
\titleformat{\subsection}{\normalfont\normalsize\bfseries}{\thesubsection}{1em}{}
\titleformat{\subsubsection}{\normalfont\normalsize\itshape}{\thesubsubsection}{1em}{}

\titlespacing*{\section}{0pt}{12pt}{6pt}
\titlespacing*{\subsection}{0pt}{10pt}{4pt}

\usepackage{longtable}

\usepackage{fancyhdr}
\begin{document}

\begin{center}
{\Large \bfseries BatteryMat: a hierarchical machine-learning and DFT framework
for average-voltage screening of lithium-ion cathode materials\par}
\vspace{1.0em}

{\large
Jaehyung Lee$^{1}$,
Charles Rhys Campbell$^{1,3}$,
Kent Zhang$^{2}$,
and Kamal Choudhary$^{1,2,*}$\par}
\vspace{0.6em}

\begin{minipage}{0.85\textwidth}\centering\small
$^{1}$Department of Materials Science and Engineering, Johns Hopkins University, Baltimore, MD 21218, USA.\\
$^{2}$Department of Electrical and Computer Engineering, Johns Hopkins University, Baltimore, MD 21218, USA.\\
$^{3}$Department of Physics and Astronomy, West Virginia University, Morgantown, WV 26506, USA.\\
$^{*}$Corresponding author: \href{mailto:kchoudh2@jhu.edu}{kchoudh2@jhu.edu}
\end{minipage}
\end{center}

\vspace{1em}

\noindent\textbf{Abstract.}
The discovery of cathode materials for lithium-ion and other metal-ion batteries is a challenging task. We introduce \textbf{BatteryMat}, a three-tier framework that screens materials databases for cathode candidates by average voltage. A machine-learning force field first computes average cathode voltages for 7{,}193 lithium hosts among the 75{,}993 materials in JARVIS-DFT. A fast regressor trained on force-field voltages then screens candidates, and density functional theory (DFT) validates the final candidates. On held-out structures, the regressor predicts the force-field voltages with a mean absolute error of 0.17~V, and it reproduces the force-field ranking with a Spearman correlation of 0.87. The DFT pipeline agrees with experiment within 0.3~V for five commercial cathodes. For the eight candidates top-ranked at selection, DFT gave voltages of at least 3.4~V, two of them (\ce{LiCr2P2O8} and \ce{Li3FeO3}) inside the 3.0--4.5~V target window. BatteryMat is available at \href{https://github.com/atomgptlab/batterymat}{\texttt{https://github.com/}\allowbreak\texttt{atomgptlab/}\allowbreak\texttt{batterymat}} and a demo at \href{https://atomgpt.org/battery}{\texttt{https://atomgpt.org/}\allowbreak\texttt{battery}}.

\vspace{0.5em}

\vspace{1em}

\section{Introduction}

The lithium-ion battery rests on a short chain of cathode discoveries: layered \ce{TiS2} (Whittingham, 1976)\cite{whittingham1976}, \ce{LiCoO2} above 4~V (Mizushima, Goodenough and co-workers, 1980)\cite{mizushima1980lco}, olivine \ce{LiFePO4} (Padhi, Nanjundaswamy and Goodenough, 1997)\cite{padhi1997lfp} and the layered Li(Ni,Mn,Co)O\textsubscript{2} family of the modern electric-vehicle cell (Ohzuku, 2001)\cite{ohzuku2001nmc}, with Yoshino's 1985 prototype establishing the commercial cell architecture\cite{yoshino2012}. Any new cathode must still deliver a voltage inside the stability window of the electrolyte and lithium insertion that remains reversible over many cycles\cite{goodenough2013}. Computational screens turn these requirements into thresholds, for example an average voltage between $3$ and $4.5$~V, a gravimetric capacity of at least $100$~mAh/g and a bound on the energy above the convex hull\cite{li2024pipeline}, and the number of candidate oxides, phosphates, fluorides and sulfides that satisfy at least one of them exceeds the number ever synthesized by orders of magnitude.

Average voltage is the natural first screening metric: it enters the cathode energy density directly ($\mathrm{Wh\,kg^{-1}} \approx V_\mathrm{avg} \times Q_\mathrm{grav}$, with $Q_\mathrm{grav}$ the gravimetric capacity), places the cathode relative to the electrochemical window of the electrolyte\cite{goodenough2010challenges}, and, unlike ionic conductivity, cycle life or synthesizability, is computable from a static structure through a thermodynamic cycle. It is not sufficient on its own, since a candidate with the right voltage can still fail on lithium-ion mobility, on oxygen-redox stability or on synthesis, none of which the present screen evaluates, but it is the metric on which a surrogate can be benchmarked quantitatively against both DFT and experiment.

Aydinol, Ceder and co-workers showed in 1997 that the average intercalation voltage follows from density functional theory (DFT) total energies of the lithiated structure, the delithiated structure and lithium metal, without experimental input; for the oxides that could be compared with experiment the computed voltages were consistently about $0.2$~V low\cite{aydinol1997}. The GGA error in transition-metal compounds was later traced to the incomplete cancellation of electron self-interaction and corrected with GGA$+U$\cite{zhou2004ldau}, GGA and GGA$+U$ energies were made compatible within one phase diagram\cite{jain2011mphubbard}, and Hautier and co-workers applied the protocol to thousands of candidate phosphates in a single high-throughput campaign\cite{hautier2011phosphates}. DFT is now the standard quantitative reference for cathode voltages, but it is costly: a single GGA$+U$ relaxation of a 100-atom transition-metal oxide takes hundreds of CPU-core-hours, and a full delithiation curve multiplies that by the number of lithium atoms in the cell.

The Materials Project\cite{jain2013materialsproject}, AFLOW\cite{curtarolo2012aflow}, the Open Quantum Materials Database (OQMD)\cite{saal2013oqmd} and JARVIS-DFT\cite{choudhary2020jarvis,choudhary2025jarvis}, scripted through libraries such as pymatgen\cite{ong2013pymatgen}, now expose hundreds of thousands of relaxed crystal structures with formation energies and convex-hull labels. JARVIS-DFT, with about 76{,}000 structures, is a natural starting point for an intercalation-voltage screen, but the databases store relaxed structures, not the delithiation curves from which an average voltage is read, and generating those curves at database scale with DFT is not feasible.

Machine-learning surrogates close that throughput gap. Crystal graph convolutional neural networks (CGCNN)\cite{xie2018cgcnn} and MEGNet\cite{chen2019megnet} predict formation energies and band gaps from the relaxed structure at near-DFT error, and the Atomistic Line Graph Neural Network (ALIGNN)\cite{choudhary2021alignn} adds explicit three-body angular information through a coupled atom-graph and line-graph representation. Universal interatomic potentials (M3GNet\cite{chen2022m3gnet}, CHGNet\cite{deng2023chgnet}, MACE\cite{batatia2022mace}, GNoME\cite{merchant2023gnome}) extend such representations to forces and stresses; the ALIGNN-FF force field used here is built on the ALIGNN backbone and trained on 3D JARVIS-DFT energies and forces\cite{choudhary2023alignnff}. Voltage-specific surrogates trained on DFT voltages from the Materials Project, a different dataset from the JARVIS-DFT structures used here, reach mean absolute errors of about $0.43$~V (Joshi and co-workers\cite{joshi2019voltage}) and $0.32$~V over ten working ions (Louis and co-workers\cite{louis2022voltage}). These surrogates are not coupled to a tier that recomputes their top-ranked candidates, so a prediction can be checked only by another pass through the same network family.

A practical point that cuts across both surrogate and DFT campaigns is the lithium reference: voltages depend on $\mu_\mathrm{Li}$ through $V = (E_\mathrm{deLi} - E_\mathrm{Li} + n\,\mu_\mathrm{Li})/n$, so a reference computed with a different functional or different numerical settings from the cathode calculation shifts every voltage by a constant (about $1$~V if the tabulated JARVIS-DFT value, an optB88-vdW energy, is combined with PBE$+U$ cathode energies). We therefore compute the lithium-metal reference in-house for each functional (Methods).

\begin{figure}[!t]
\centering
\includegraphics[width=\textwidth]{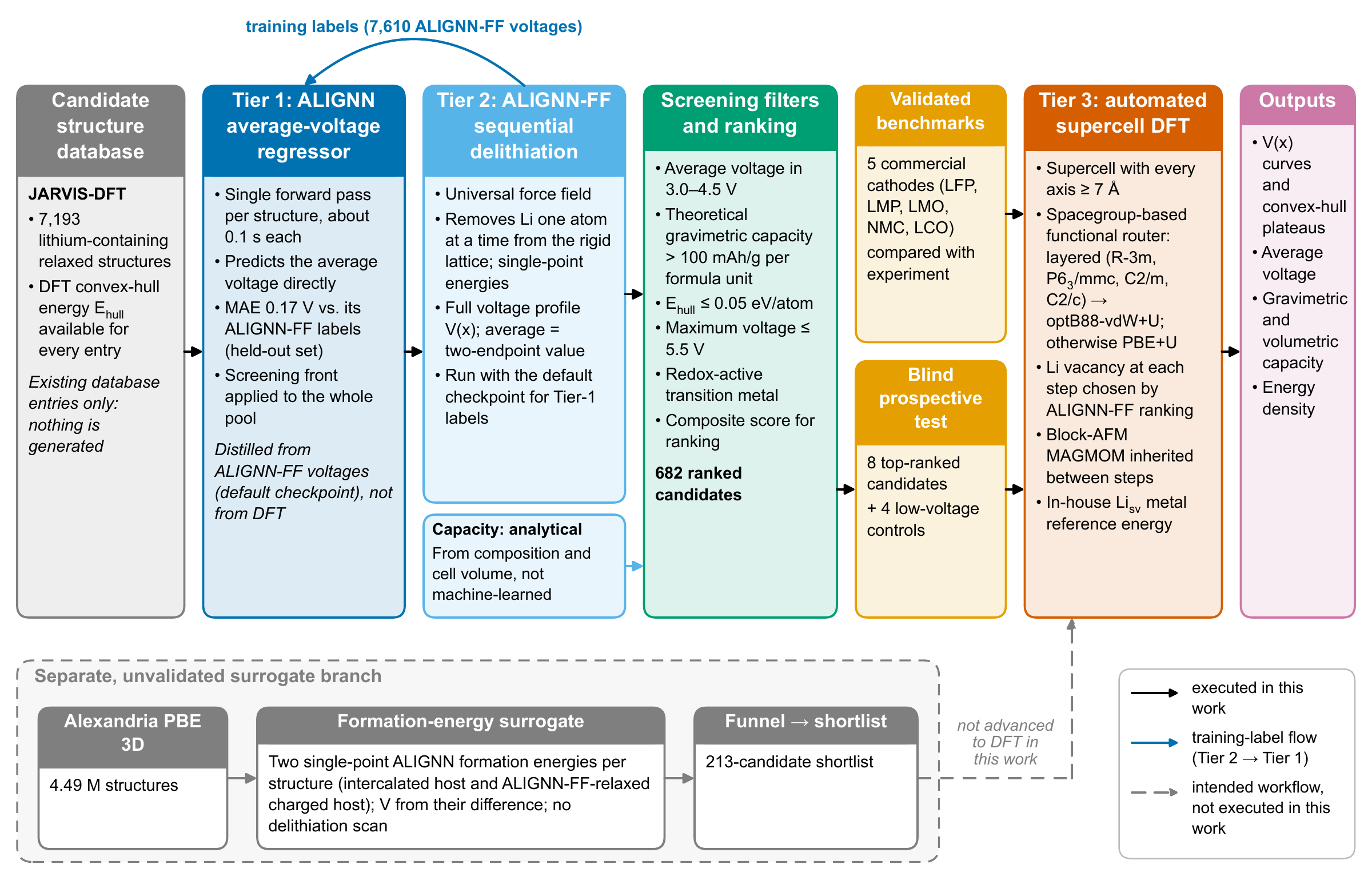}
\caption{\textbf{The BatteryMat screening pipeline.} Top row, the executed cascade: the candidate structure database (JARVIS-DFT, $7{,}193$ lithium-containing relaxed structures with convex-hull energies; nothing is generated) feeds Tier~1, an ALIGNN average-voltage regressor applied in a single forward pass to the whole pool. Tier~2, ALIGNN-FF sequential delithiation, predicts the full $V(x)$ profile; in this work it was run over the whole pool, and the Tier~1 ranking was compared with it (section ``The regressor as the screening front'') rather than used to select its inputs. Run with the package's default checkpoint, Tier~2 also supplied the $7{,}610$ voltages on which Tier~1 is trained (blue arrow; Tier~1 is a distillation of that protocol). Capacities are analytical from composition and cell volume, not machine-learned. The screening filters and composite score, applied to the Tier~2 voltages, return $682$ ranked candidates. Tier~3, automated supercell DFT with a spacegroup-based functional router, ALIGNN-FF-ranked lithium-vacancy selection, inherited block-antiferromagnetic moments and an in-house lithium reference, received two executed inputs: the five commercial benchmarks compared with experiment, and a blind prospective test of eight candidates top-ranked by an earlier version of this ranking, before the capacity correction (section ``Where the benchmarks rank, and a blind prospective DFT test''), and four low-voltage controls. Bottom row, dashed border: the separate Alexandria branch estimates voltages from two single-point ALIGNN formation energies per structure without a delithiation scan and is not validated in this work. Solid arrows mark steps executed here, with the two qualifications above; the dashed arrow marks the intended but unexecuted advance of the Alexandria shortlist to DFT.}
\label{fig:framework}
\end{figure}

BatteryMat closes that loop with a hierarchical screen whose front is a single-pass average-voltage regressor and whose back is automated DFT (Fig.~\ref{fig:framework}). (i) An ALIGNN scalar regressor trained on $7{,}610$ lithium-containing JARVIS-DFT structures predicts the average delithiation voltage from the relaxed lithiated geometry in a single forward pass, at a mean absolute error of 0.17~V relative to the ALIGNN-FF protocol it is distilled from, and reproduces the force-field ranking of the $7{,}193$-entry pool with a Spearman correlation of $0.87$ ($0.84$ outside its training partition; section ``The regressor as the screening front''). (ii) The ALIGNN-FF universal force field predicts the step-by-step delithiation profile in seconds per step; in this work it was run over the whole pool, and its ranking is the one reported and the source of the prospective DFT candidates. Run with the package's default checkpoint rather than the \texttt{alignnff\_wt01} checkpoint used for the screen and the DFT tier, it also generated the regressor's training labels. (iii) Automated PBE$+U$ or optB88-vdW$+U$ supercell DFT, with the functional selected by a spacegroup rule and the lithium reference computed with the cathode settings, validates the highest-ranked structures. On five commercial cathodes (\ce{LiFePO4}, LFP; \ce{LiMnPO4}, LMP; \ce{LiMn2O4}, LMO; a layered nickel--manganese--cobalt oxide variant, NMC; \ce{LiCoO2}, LCO), each compared over the lithium range its cell covers or over which it is cycled, the DFT tier recovers the experimental average voltage to within 0.3~V (section ``Benchmarking against experiment and error analysis''); in a blind prospective test it separates the eight candidates top-ranked at the time of selection from four low-voltage controls, although only two of the eight fall inside the $3.0$--$4.5$~V window that the screen targets (section ``Where the benchmarks rank, and a blind prospective DFT test''); and a DFT spot-check of $30$ held-out structures measures the error of the force-field labels directly (section ``DFT spot-check of the force-field labels''). Across the lithium pool the screen returns $682$ ranked candidates led by lithium-rich transition-metal oxides (section ``Top-N candidate cathodes from the screen'').

The key contributions of this work are as follows. (1) A three-tier ML+DFT pipeline in which a single ALIGNN forward pass replaces the $N_\mathrm{Li}$-step ALIGNN-FF label protocol, whose average reduces to two endpoint single points, and is exercised as the screening front over the full pool (sections ``Average voltage prediction with ALIGNN'' and ``The regressor as the screening front''). (2) An automated DFT tier with a spacegroup-based functional router, force-field-ranked vacancy selection and inherited antiferromagnetic initialization (section ``DFT validation on five commercial cathodes''; Methods). (3) A five-material benchmark of ALIGNN-FF, DFT and experiment for average voltage and theoretical volumetric capacity, with the force-field and DFT tiers compared on identical structures (section ``Benchmarking against experiment and error analysis''). (4) A blind prospective DFT test of twelve screened candidates, a DFT spot-check of the force-field labels and an exhaustive test of the greedy vacancy ordering (sections ``Where the benchmarks rank, and a blind prospective DFT test'', ``DFT spot-check of the force-field labels'' and ``Greedy versus exhaustive lithium-vacancy ordering'').

\section{Results}

\subsection{Average voltage prediction with ALIGNN}
\label{sec:alignn}

The lead capability of the pipeline is single-pass prediction of the average delithiation voltage from a relaxed lithiated structure. We trained the ALIGNN scalar regressor\cite{choudhary2021alignn} for 250 epochs on $7{,}610$ lithium-containing JARVIS-DFT entries (Methods). JARVIS-DFT contains no per-material voltage subset, and the average voltage of a host is not single-valued without a delithiation protocol, so the labels were generated in-house by sequential ALIGNN-FF delithiation of every lithium-containing structure on its rigid lattice, with the step voltages averaged into one label per material (Methods). Because the step voltages telescope, the label equals the two-endpoint average voltage between the lithiated cell and the empty rigid framework, independent of the removal order, and the regressor is a structure-to-scalar distillation of an $N_\mathrm{Li}$-step force-field protocol into one forward pass. Three caveats carry forward: the delithiated endpoint is not relaxed, the lithium reference is the force field's own, and the label pass used the default checkpoint of the ALIGNN-FF package rather than the \texttt{alignnff\_wt01} checkpoint of the screening pass and the DFT tier, so the labels sit on a different absolute scale from both (Methods). The 80/10/10 split follows the order of the input file, so the held-out test set is reproducible but the split is closer to one by database addition than to a random one (Methods). None of the five DFT-validated benchmarks is in the $761$-entry test partition on which the parity numbers below are measured: four are in the training partition and JVASP-144791 is in the validation partition.

\begin{figure}[!t]
\centering
\includegraphics[width=0.55\textwidth]{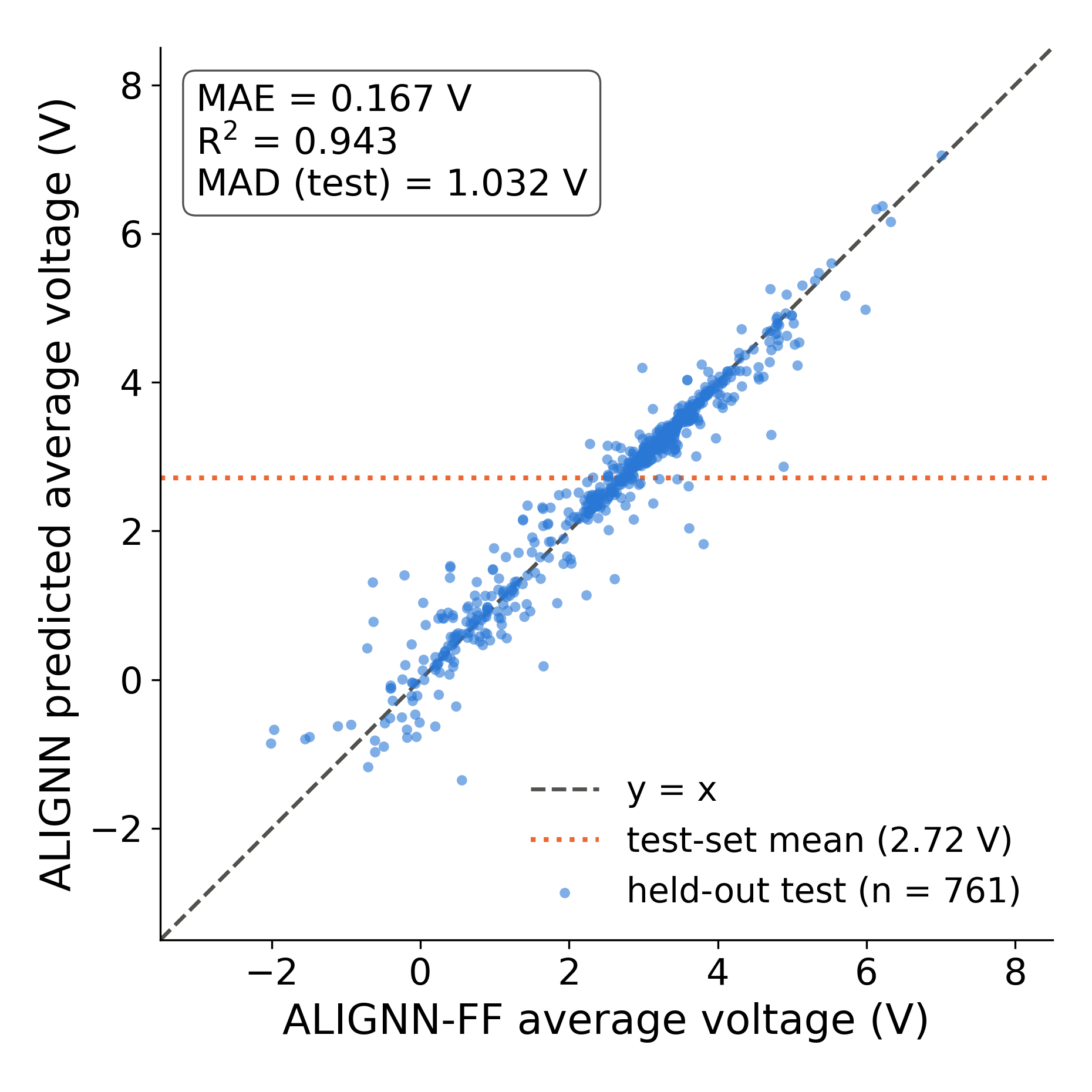}
\caption{\textbf{ALIGNN average-voltage parity plot.} Predicted versus ALIGNN-FF label average voltage on the $761$ held-out test structures after 250 training epochs. The dashed line is $y=x$; the dotted line is the test-set mean ($2.72$~V). MAE $=0.167$~V (bootstrap 95\% CI $0.148$--$0.187$~V), $R^2=0.943$, against a mean-absolute-deviation baseline of $1.03$~V. The residual distribution and the training curves are in Figures~S1 and S2.}
\label{fig:alignn}
\end{figure}

The held-out mean absolute error is 0.17~V with $R^2 = 0.94$, against a test-set-mean baseline of 1.03~V (Fig.~\ref{fig:alignn}). Predictions follow parity from $0$ to $6$~V with a mean residual of $-0.01$~V and the usual regression toward the mean at the extremes ($-0.25$~V for labels of $4.5$--$5.0$~V, $+0.23$~V below $0$~V). Because the mean error does not bound the ranking error, we also report the residual distribution: the residuals have a standard deviation of $0.32$~V, $79.5$\% lie within $0.25$~V and $90.9$\% within $0.5$~V, the central 95\% and 99\% intervals are $[-0.71, +0.66]$ and $[-1.50, +1.22]$~V, and the worst case is $2.0$~V (Figure~S1). A candidate predicted $0.5$~V above a known benchmark therefore has roughly a one-in-twenty chance that its own residual alone reverses the order ($4.9$\% of the residuals exceed $+0.5$~V), and candidates within $0.2$~V of each other should be promoted together to the ALIGNN-FF tier rather than disambiguated at the surrogate level. These are statements about fidelity to the force-field labels, not about accuracy against DFT, which is measured in the sections ``Benchmarking against experiment and error analysis'', ``Where the benchmarks rank, and a blind prospective DFT test'' and ``DFT spot-check of the force-field labels''.

\begin{table}[H]
\small
\centering
\caption{Reported mean absolute errors of voltage predictors, with the differences that prevent a like-for-like comparison. The ALIGNN row is trained and tested on a different dataset from the other two: force-field voltages of JARVIS-DFT structures, against DFT-derived voltages from the Materials Project. Pools and splits also differ between rows, so the numbers indicate the published state of the art rather than a ranking.}
\label{tab:prior}
\setlength{\tabcolsep}{4pt}
\begin{tabular}{>{\raggedright\arraybackslash}p{0.24\textwidth}>{\raggedright\arraybackslash}p{0.30\textwidth}>{\raggedright\arraybackslash}p{0.30\textwidth}c}
\toprule
Model & Label source & Pool & MAE (V) \\
\midrule
DNN, Joshi 2019\cite{joshi2019voltage} & DFT, Materials Project battery data & multi-ion hosts (Li, Mg, Ca, Zn, Y, Al) & 0.43 \\
Att-GNN, Louis 2022\cite{louis2022voltage} & DFT, Materials Project battery subset & multi-ion hosts (ten working ions) & 0.32 \\
ALIGNN (this work) & ALIGNN-FF sequential delithiation & lithium-containing JARVIS-DFT ($7{,}610$) & 0.17 \\
\bottomrule
\end{tabular}
\end{table}

Table~\ref{tab:prior} places the 0.17~V figure next to published voltage-prediction networks; the ALIGNN row differs from the other two in dataset (JARVIS-DFT rather than Materials Project), label source (ALIGNN-FF rather than DFT), pool and split, so the comparison is indicative rather than head-to-head, and the ALIGNN number measures distillation fidelity rather than accuracy against DFT.

The label protocol ran as an array of $7{,}623$ tasks, one per lithium-containing JARVIS-DFT structure; $7{,}610$ returned labels, split $6{,}088$/$761$/$761$. The screening pool of the section ``High-throughput screening across JARVIS-DFT'' is a second, independent ALIGNN-FF pass over the lithium-containing entries that JARVIS-DFT tags as intercalated-ion hosts, run with the \texttt{alignnff\_wt01} checkpoint on the $8$~\AA\ supercells and with the removal order of the label pass (Methods); $7{,}193$ entries returned a valid profile, and the offset and scatter between the two passes are quantified in the section ``The regressor as the screening front''. Table~\ref{tab:accounting} reconciles the structure counts of the two pools.

\begin{table}[!htbp]
\small
\centering
\caption{Data accounting for the structure counts of the training and screening pools. All rows start from the lithium-containing subset of JARVIS-DFT.}
\label{tab:accounting}
\begin{tabular}{>{\raggedright\arraybackslash}p{0.62\textwidth}r}
\toprule
Set & Count \\
\midrule
Lithium-containing JARVIS-DFT structures submitted to the label-generation array & $7{,}623$ \\
Structures with a completed ALIGNN-FF label (\texttt{status == 'Success'}); ALIGNN regressor training pool & $7{,}610$ \\
\quad training / validation / held-out test partitions (input-file order, 80/10/10) & $6{,}088$ / $761$ / $761$ \\
Structures with a valid rigid-lattice ALIGNN-FF profile in the screening pass & $7{,}193$ \\
Screening rows, including $281$ entries for other working ions on the same structures (not used) & $7{,}474$ \\
Survivors of the cathode filters (section ``Top-N candidate cathodes from the screen'') & $682$ \\
Commercial benchmarks advanced to DFT (four in the training partition, one in validation, none in test) & $5$ \\
Prospective candidates advanced to DFT (eight top-ranked, four controls) & $12$ \\
\bottomrule
\end{tabular}
\end{table}

\subsection{The regressor as the screening front}
\label{sec:tier1front}

The regressor is only a screening front if it is used as one. We therefore ran it over every structure of the $7{,}193$-entry lithium pool, at $0.10$--$0.14$~s per structure on a CPU (about 16 minutes for the pool), and compared the result with the ALIGNN-FF tier on three levels.

Over the whole pool the single-pass prediction and the ALIGNN-FF screening value have a Spearman rank correlation of $0.87$ (Pearson $0.88$), and $0.84$ on the $1{,}437$ entries outside the regressor's training partition; because every pool entry was in the regressor's data set (train $5{,}756$, validation $699$, test $738$), the pool-level figure is largely in-sample and the out-of-sample statement is the $1{,}437$-entry one. The mean absolute difference between the two, $0.78$~V, reflects the difference between the two ALIGNN-FF passes themselves: a mean offset of $0.63$~V (median $0.59$~V) and a structure-to-structure scatter of similar size (standard deviation $0.80$~V; $0.54$~V mean absolute difference after removing the offset). The passes share supercells and removal order and differ in the force-field checkpoint and the lithium reference (Methods); a lithium reference shifts every voltage by the same amount, so the scatter comes from the checkpoint. Against its own labels the regressor has a Spearman correlation of $0.97$ on the held-out set, and the two force-field passes themselves correlate at $0.87$, so the disagreement between the tiers is the disagreement between the two label sets, not a failure of the regressor.

Used as a binary filter for the $3.0$--$4.5$~V window, the regressor recovers the labels' decision on the held-out set with precision $0.94$ and recall $0.95$ ($366$ true positives, $24$ false positives, $19$ false negatives, $352$ true negatives; $F_1=0.945$).

Applying the filters and composite score of the section ``Top-N candidate cathodes from the screen'' with the regressor's voltage in place of the force-field voltage yields $571$ survivors, $507$ of them among the $682$ force-field survivors (Jaccard index $0.68$); over the common survivors the rank correlation is $0.85$, six of the force-field top twelve and $24$ of the top fifty are recovered, \ce{Li3FeO3} is rank~1 under both fronts, and LCO, a training-partition structure for the regressor, moves from rank $263$ to $231$ (Figure~S3). The regressor does not predict the maximum step voltage, so the $5.5$~V cap is taken from the force-field profile, where it removes $559$ of the $1{,}130$ structures that pass the other filters on the regressor's voltage (without the cap, Jaccard $0.39$ with the $682$). Most of the remaining disagreement is the offset between the label sets: of the $175$ force-field survivors that the regressor drops, $168$ are predicted below $3.0$~V although on the force-field scale they lie in the lower part of the window (median $3.45$~V), $57$ of the $64$ it admits lie above $4.5$~V on that scale, and shifting the regressor's predictions by the mean offset before applying the window raises the overlap to $640$ common survivors (Jaccard $0.83$), seven of the top twelve and $26$ of the top fifty, but pushes \ce{Li3FeO3} across the upper edge of the window. The regressor therefore reproduces the force-field ranking to within the disagreement between the two label sets when the $5.5$~V cap is taken from the force-field profile (a regressor-only front would need its own estimate of the maximum step voltage), and the voltage window should be applied on the scale of the labels it was trained on. The five commercial benchmarks are predicted at $3.25$ (LFP), $3.10$ (LMP), $3.60$ (LMO), $3.70$ (NMC) and $3.42$~V (LCO) on the label scale (compared with the other tiers in the section ``Benchmarking against experiment and error analysis'').

On $200$ randomly chosen held-out structures, Gaussian noise of $0.05$~\AA\ on every Cartesian coordinate raises the MAE from $0.161$ to $0.172$~V and shifts the individual predictions by $0.050$~V on average (at most $0.38$~V over three draws), and noise of $0.10$~\AA\ raises it to $0.197$~V (mean shift $0.089$~V, at most $0.53$~V). The regressor is thus insensitive to random noise at the scale of a force-field relaxation residual. Random displacements do not test the systematic differences between force-field- and DFT-relaxed geometries; a test on force-field-relaxed inputs was not carried out: force-converged relaxations with the \texttt{alignnff\_wt01} checkpoint were not attainable on the two cells of the enumeration test (section ``Greedy versus exhaustive lithium-vacancy ordering''), and the test with the default checkpoint, which converged most of the Alexandria relaxations (section ``Foundation-scale screening across Alexandria''), is left for future work.

The validation loss is flat from about epoch $100$ with no upturn (Figure~S2), and the checkpoint used is the final one, which is also the best on validation. The gap between the training MAE ($0.04$~V) and the validation and test MAEs ($0.25$ and $0.17$~V) shows that the network can memorize its training labels; the held-out numbers, not the training fit, are the ones quoted throughout.

Neither the production split nor a random split keeps close chemical relatives of the test structures out of training ($450$ of the $761$ test structures share a reduced formula with a training structure), so both measure interpolation. To measure transfer we retrained the regressor on grouped splits in which an entire chemistry family is absent from training and validation and forms the test set: one fold per dominant redox-active transition metal with at least $50$ members ($13$ metals, plus the $3{,}089$ structures with no redox-active metal), one fold per anion class (oxide, fluoride, chalcogenide, phosphate, other), and five random folds as the matched baseline (Methods). The random folds give test MAEs of $0.21$--$0.22$~V ($R^2=0.92$--$0.94$), slightly above the $0.17$~V of the production model, which was trained longer and with a different graph implementation. Held-out redox metals transfer well: the nine families with more than $100$ members (Co, Mn, Ni, Cr, Fe, Ti, Nb, V, Cu) give $0.09$--$0.31$~V ($R^2=0.80$--$0.96$) and the four small ones (Mo, Rh, Ru, W; $52$--$91$ members) $0.37$--$0.51$~V. Held-out anion chemistry transfers unevenly: $0.22$~V for phosphates and $0.29$~V for oxides, whose voltages span a narrow range (mean absolute deviations of $0.27$ and $0.50$~V about the class means), but $0.46$~V for chalcogenides, $0.58$~V for the heterogeneous ``other'' class and $0.91$~V with no explanatory power ($R^2\approx0$) for fluorides; the hosts with no redox-active metal give $0.61$~V. The regressor therefore extends to unseen transition-metal chemistry within the anion classes it was trained on, but its predictions for anion classes absent from training, fluorides in particular, should not be relied on, in agreement with the fluoride residuals of the DFT spot-check (section ``DFT spot-check of the force-field labels'').

\subsection{ALIGNN-FF voltage profiles}
\label{sec:alignnff}

For candidates that survive the first tier, ALIGNN-FF predicts the step-by-step delithiation profile. In the screening pass behind every pool-level number in this paper, the force field starts from the fully lithiated database structure in a supercell with all axes at least $8$~\AA, removes the lithium atoms one at a time from the rigid lattice and evaluates each intermediate as a single point, at about one second per step; force-field-ranked vacancy selection and relaxation are reserved for the DFT tier (section ``DFT validation on five commercial cathodes''), where the removal order determines the relaxed structures.

A rigid-lattice profile does not resolve the two-phase plateaus of the measured curves: the two phosphate entries (non-olivine polymorphs; section ``DFT validation on five commercial cathodes'') give sloped profiles spanning $0.6$--$0.7$~V, whereas olivine discharges on flat plateaus\cite{padhi1997lfp,delacourt2004lmp}, and spinel \ce{LiMn2O4}, which shows two plateaus near $4.0$ and $4.15$~V\cite{tomon2022lmo}, rises steadily to about $4.2$~V (web-application profiles in Figure~S7). The quantity carried forward to the screen is therefore the average of the profile, not its shape; the ALIGNN-FF profiles of all five are overlaid on the DFT profiles in Fig.~\ref{fig:curve}.

The force-field tier is also available as a web service, the AtomGPT Battery Explorer (\url{https://atomgpt.org/battery}), which takes a JARVIS ID or a POSCAR and returns the delithiation profile, the average and maximum voltages and a gravimetric capacity estimate through the same \texttt{AlignnAtomwiseCalculator} interface of the ALIGNN package (\texttt{alignn.ff.ff}) used in this work.

\subsection{High-throughput screening across JARVIS-DFT}
\label{sec:screen}

The ALIGNN-FF protocol underpins both the training labels of the ALIGNN regressor and the high-throughput distributions used here. We applied the protocol to the lithium-containing JARVIS-DFT entries tagged as intercalated-ion hosts ($7{,}193$ with a valid profile). The resulting average-voltage distribution has its main peak near $4.2$~V (Figure~S5a), within the operating window of commercial intercalation cathodes\cite{manthiram2020cathode}, and the gravimetric capacity distribution spans a broad range that reflects the diversity of host frameworks (Figure~S5b). At tens of seconds per material the screening pass over the pool takes hours of GPU time, where a DFT-only campaign over the same pool would take months.

As a transferability demonstration of the force field we extended the ALIGNN-FF screen to non-lithium working ions; the screening outputs contain pools for K, Na, Ca and Zn but none with magnesium or aluminum as the working ion. Across the K, Na, Ca and Zn screening pools, which contain all JARVIS-DFT intercalated-ion entries with the ion ($2{,}776$, $3{,}492$, $765$ and $724$ hosts, $16$--$37$\% of them oxides), the median ALIGNN-FF average voltage falls from $3.23$~V (K) to $2.70$~V (Na), $1.44$~V (Ca) and $0.50$~V (Zn); the screening outputs give the energy change per removed ion, which we divide by the ion charge for the divalent ions. The ALIGNN regressor is trained on lithium hosts only and is not applied to these ions. Of the non-lithium chemistries, one sodium and one magnesium cathode are carried to the DFT tier in the section ``Sodium and magnesium chemistries at the DFT tier''; the remaining ions are reported only as a demonstration, and no quantitative claim is made for them.

The lithium voltage distribution is bimodal, with a minor mode near $1$~V from intermetallic and alloy-type hosts and a major mode near $4.2$~V (median $3.62$~V, interquartile range $1.76$--$4.34$~V), far wider than the $\sim 0.3$~V spread of plateau voltages within any single commercial cathode family. The $3.0$--$4.5$~V band of the ranking contains $42$\% of the pool and spans the polyanion ($\sim 3.5$~V) and layered and spinel ($\sim 4.0$--$4.1$~V) operating regimes. The per-formula-unit capacity distribution is right-skewed because $Q_\mathrm{grav}$ scales as $n_\mathrm{Li}/M$ (lithium atoms per formula unit over the formula mass), and lithium-rich, light-element hosts reach compositional ceilings far above the $140$--$275$~mAh/g of the commercial chemistries.

\subsection{Foundation-scale screening across Alexandria}
\label{sec:alexandria}

To test whether a surrogate screen scales to a foundation-model-sized search space, we applied a separate and simpler protocol to the ${\sim}4.49$~million relaxed structures of the Alexandria PBE 3D dataset\cite{schmidt2023alexandria}. This branch is distinct from the three validated tiers and does not trace a delithiation curve: for each intercalated (discharged) structure the de-intercalated (charged) host is generated by removing the working ion and relaxing with ALIGNN-FF, and the average voltage is obtained in the Materials Project battery-explorer style from two single-point ALIGNN formation energies, one per endpoint, both from the same \texttt{mp\_e\_form} model so that part of the systematic error cancels; capacities are analytical from composition and cell volume. Seven working ions are admitted (Li, Na, K, Mg, Ca, Al, Zn). None of the Alexandria voltages has been checked against the force-field profile tier or against DFT, so the branch is a third, unvalidated surrogate whose only role is to prioritize candidates for the validated tiers.

\begin{figure}[!t]
\centering
\includegraphics[width=0.85\textwidth]{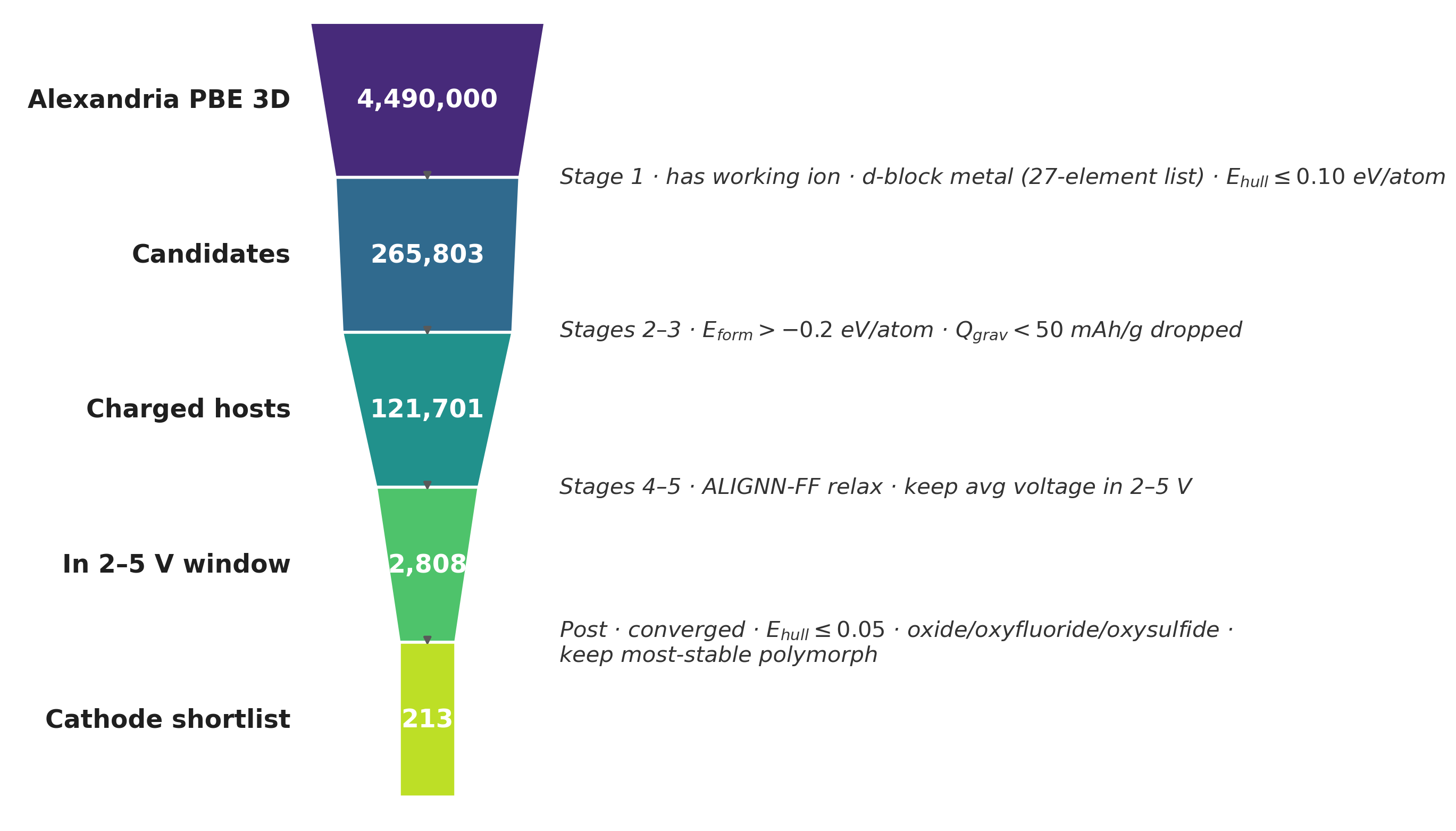}
\caption{\textbf{Foundation-scale cathode screening funnel over Alexandria PBE 3D.} The five-stage surrogate pipeline reduces ${\sim}4.49$~million relaxed structures to a $213$-candidate cathode shortlist (a ${\sim}21{,}000\times$ reduction). Each tier lists its survivor count and the criterion applied to reach the next tier. Voltages are ALIGNN-FF / \texttt{mp\_e\_form}-derived surrogate estimates, not DFT.}
\label{fig:funnel}
\end{figure}

A five-stage funnel (Fig.~\ref{fig:funnel}) reduces the $4.49$~million-structure pool to a shortlist of $213$ candidates, a reduction of roughly $2.1\times10^{4}$. The first stage keeps structures that contain a working ion, host a $d$-block metal (a $27$-element list, broader than the $13$ redox-active metals of the section ``Top-N candidate cathodes from the screen''), and lie within $0.10$~eV/atom of the convex hull ($265{,}803$ survivors). The next stages drop weakly bound hosts (formation energy $>-0.2$~eV/atom) and low-capacity hosts ($Q_\mathrm{grav}<50$~mAh/g), then relax the charged hosts with ALIGNN-FF ($121{,}701$; this branch used the default checkpoint of the ALIGNN-FF package, with which $92{,}416$ of the relaxations reached the force criterion, unlike the \texttt{alignnff\_wt01} relaxations of the section ``Greedy versus exhaustive lithium-vacancy ordering''). Ranking by surrogate average voltage retains the $2{,}808$ candidates inside the $2$--$5$~V window. A final post-filter keeps converged candidates within $0.05$~eV/atom of the hull, restricts to oxide, oxyfluoride and oxysulfide frameworks, and keeps only the most stable polymorph of each composition, yielding $213$ distinct cathode candidates ($200$ oxide-based, $23$ of them mixed-anion hosts that also contain Se, Te, Br, I or C; $9$ oxyfluorides; $4$ oxysulfides) spanning the Li, Na, K, Mg, Ca, and Al working ions.

\begin{figure}[!t]
\centering
\includegraphics[width=0.8\textwidth]{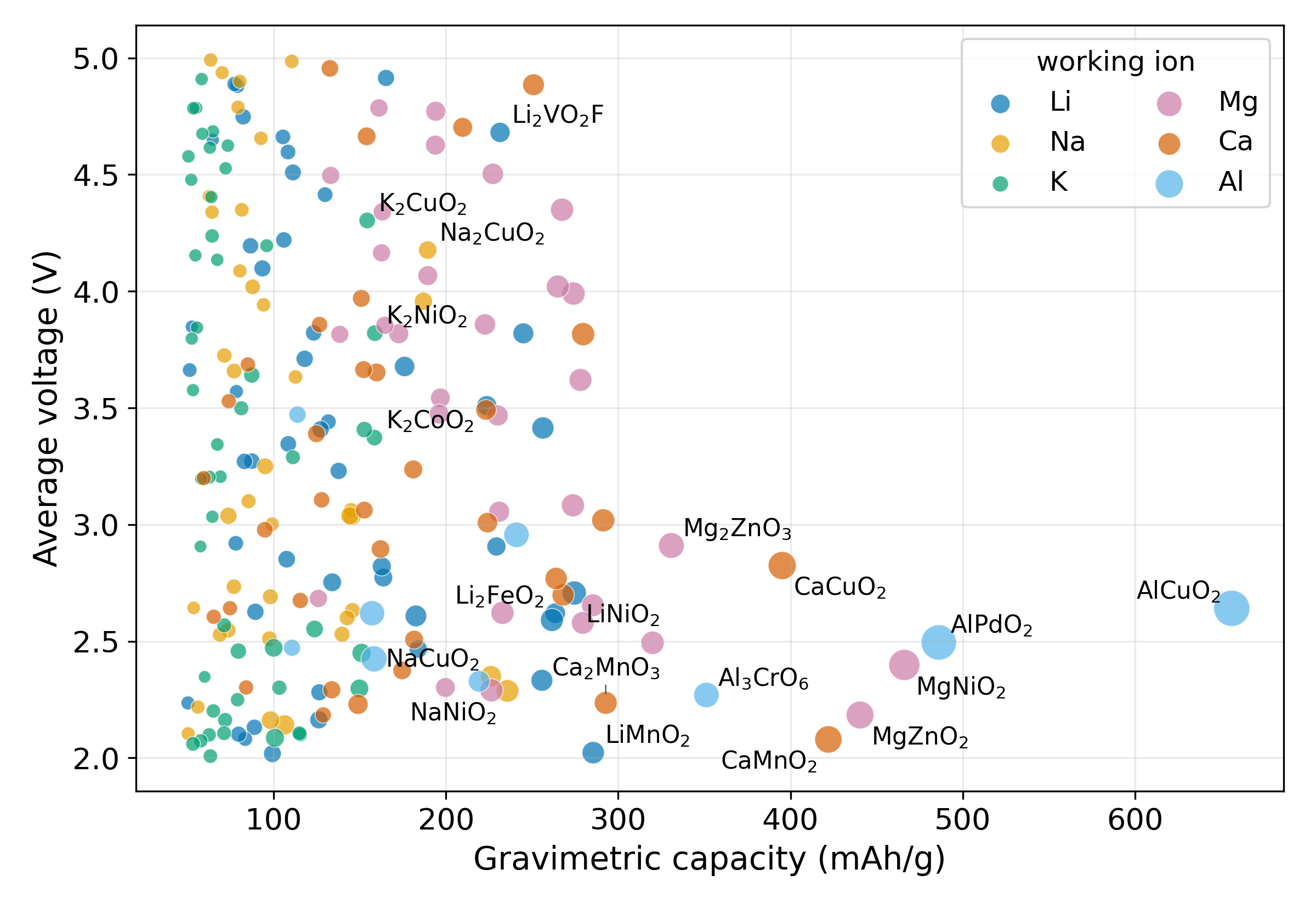}
\caption{\textbf{Voltage--capacity landscape of the 213-candidate Alexandria shortlist.} Surrogate average intercalation voltage versus gravimetric capacity for each shortlisted cathode, colored by working ion; marker area scales with volumetric capacity, which the multivalent ions (Mg, Ca, Al) maximize. All entries fall within the $2$--$5$~V ranking window, with the highest gravimetric capacities reached by the aluminum hosts.}
\label{fig:shortlist}
\end{figure}

The shortlist is anchored by chemistries that are known cathodes (Fig.~\ref{fig:shortlist}); across it, voltage and gravimetric capacity are only weakly anticorrelated (Spearman $-0.13$). Without any of them being seeded into the search, the shortlist contains \ce{LiNiO2}, \ce{NaNiO2}, \ce{LiMnO2}, and the disordered-rocksalt \ce{Li2VO2F} near the hull. The rediscovery is compositional: the surrogate places \ce{LiNiO2}, \ce{NaNiO2} and \ce{LiMnO2} at only $2.71$, $2.29$ and $2.02$~V, a further reason to treat the branch's voltages as unvalidated. The multivalent ions populate the high-volumetric-capacity region: \ce{MgNiO2}, on the hull at $466$~mAh/g, is the highest-capacity on-hull magnesium candidate in the set. We stress that every entry carries only surrogate-level voltages; the five-cathode benchmark of the section ``DFT validation on five commercial cathodes'' establishes the accuracy of the DFT tier for lithium hosts, through which a shortlisted lithium candidate would next be advanced; for the other working ions only one sodium and one magnesium host have been tested (section ``Sodium and magnesium chemistries at the DFT tier'').

\subsection{DFT validation on five commercial cathodes}
\label{sec:dft}

Five representative cathodes (\ce{LiFePO4}, \ce{LiMnPO4}, \ce{LiMn2O4}, \ce{Li4Mn3Co2Ni3O16} and \ce{LiCoO2}; JARVIS IDs 42723, 116897, 141792, 144791 and 2017; step-0 supercells in Figure~S4) were advanced to the third tier. Throughout, ``automated'' refers to the generation of every VASP input (supercell, functional and $U$ values, magnetic initialization, $k$-mesh), the force-field-ranked choice of which lithium to remove at each step, and the bookkeeping of energies and voltage curves; job submission and convergence monitoring were scripted separately, and recovery from unconverged or interrupted steps was manual (Methods). No converged step of the five reported chains was altered by hand, apart from the fully lithiated LMP endpoint, which was recomputed in the magnetic order of the rest of its chain (below). Supercells with all lattice vectors at least $7$~\AA\ give $112$-atom cells for the phosphate, spinel and NMC frameworks and a $32$-atom cell for LCO. The functional is selected by spacegroup: the 3D-bonded phosphate and spinel frameworks, which have no van-der-Waals interlayer gap, use PBE$+U$, and layered LCO (\textit{R}$\bar{3}$\textit{m}, spacegroup 166) is routed to optB88-vdW$+U$\cite{klimes2010optb88,klimes2011optb88}, which captures the interlayer dispersion that PBE misrepresents. The NMC cell (JVASP-144791) is layered but stored with triclinic \textit{P}1 symmetry, so the rule routes it to PBE$+U$; an optB88-vdW$+U$ rerun of the same cell and cell-relaxed endpoints are reported in the section ``Benchmarking against experiment and error analysis''. Lithium atoms were removed one at a time according to ALIGNN-FF-ranked vacancy energies over every remaining site, without symmetry deduplication, and each configuration was relaxed with VASP\cite{kresse1996vasp,blochl1994paw,kresse1999paw}. The step-0 magnetic moments (block antiferromagnetic for the Mn-, Fe- and Ni/Mn/Co-containing hosts) are inherited from step to step, with the removed lithium dropped, to preserve the sublattice order. The LMP chain predates this inheritance: its steps after step~0 were initialized with signs alternating atom by atom and converged to that order, whereas step~0 had converged to the block order, so we recomputed its fully lithiated endpoint in that order, $0.16$~eV per supercell below the block order, and all seventeen lithiation states share one magnetic order (Supplementary Information).

The two phosphate entries are not olivine, which we found during this revision. JVASP-42723 and JVASP-116897 have the composition and \textit{Pnma} spacegroup of olivine \ce{LiFePO4} and \ce{LiMnPO4} but place lithium on tetrahedral rather than octahedral sites (Li--O $1.96$--$2.14$~\AA), with cells of $4.92\times6.09\times9.39$~\AA\ (Fe) and $4.92\times6.36\times9.39$~\AA\ (Mn) against $4.74\times6.11\times10.44$~\AA\ for the olivine \ce{LiMnPO4} entry JVASP-91261, and neither entry has an ICSD record (Supplementary Information). We keep their delithiation chains, which exercise the automated protocol and, on identical structures, the comparison with the force-field tier (section ``Benchmarking against experiment and error analysis''), and call them the LFP and LMP polymorphs. For the comparison with experiment we computed the olivine phases from their two endpoints, outside the automated workflow but with the functional, $U$ values, plane-wave settings and $k$-mesh of the chains and the current input template (Methods), relaxing the cell at full lithiation and keeping it at $x=0$ as the chains do. Olivine \ce{LiMnPO4} was taken from JVASP-91261, and olivine \ce{LiFePO4}, which is not in the JARVIS-DFT snapshot, was built from the same cell with iron on the manganese sites; the antiferromagnetic order was set by hand (Supplementary Information). The fully lithiated olivine cells lie $29$ (Fe) and $37$ (Mn) meV/atom below the polymorphs, and their average voltages, $3.55$ and $3.92$~V, are within $0.05$~V of the polymorph chains ($3.60$ and $3.93$~V); relaxing the cell at $x=0$ as well lowers them to $3.45$ and $3.81$~V.

The stepwise total energies (Tables~S5--S8 and S11) and the olivine endpoint energies (Table~S14) yield the voltages summarized in Table~\ref{tab:dft_summary}. With the olivine values for LFP and LMP, the DFT tier reproduces the experimental average voltage to within $0.25$~V, with the smallest residuals on the NMC variant ($+0.03$~V over the lithium range its cell covers, an agreement that partly reflects cancellation; see below) and LMO ($+0.04$~V), and to within $0.3$~V with the cell also relaxed at the delithiated compositions (section ``Benchmarking against experiment and error analysis'').

\begin{table}[!htbp]
\small
\centering
\caption{DFT-validated cathodes. Voltages are the lower-convex-hull average using the in-house \ce{Li_{sv}} reference. \textsuperscript{a}Olivine phase, from its two endpoints with the cell relaxed at $x=1$ and kept at $x=0$, as in the chains ($3.45$ and $3.81$~V with the cell relaxed at both); the chains of the JARVIS-DFT entries JVASP-42723 and JVASP-116897, a \textit{Pnma} polymorph with tetrahedral lithium, give $3.60$ and $3.93$~V (section ``DFT validation on five commercial cathodes''). \textsuperscript{b}The NMC cell (JVASP-144791, a layered variant with a lithium-to-transition-metal (Li:TM) ratio of $1{:}2$) spans \ce{Li_{0.5}TMO2} to \ce{TMO2} per transition metal and is compared with the open-circuit voltage of NMC-111 over the same lithium range (to \ce{Li_{0.03}}, the end of the measured charge at $4.71$~V); over the range in which NMC-111 is normally cycled the measured average is $3.85$~V\cite{yoshizawa2003nmc}. \textsuperscript{c}LCO is compared over $x=1\to0.5$, the range over which it is cycled experimentally; the experimental value is the composition-weighted average of the measured open-circuit voltage over that range, and the DFT average over full delithiation is $4.45$~V.}
\label{tab:dft_summary}
\begin{tabular}{lccccc}
\toprule
Material & Supercell & Func. & Steps & $V_\mathrm{avg}^\mathrm{DFT}$ & $V_\mathrm{avg}^\mathrm{exp}$ \\
\midrule
LFP\textsuperscript{a} & $2{\times}2{\times}1$ & PBE$+U$ & 1 & 3.55 & 3.43\cite{kondo2018lfp} \\
LMP\textsuperscript{a} & $2{\times}2{\times}1$ & PBE$+U$ & 1 & 3.92 & 4.10\cite{aravindan2013lmp} \\
LMO & $2{\times}2{\times}2$ & PBE$+U$ & 16 & 4.08 & 4.04\cite{ohzuku1990lmo} \\
NMC & $2{\times}2{\times}1$ & PBE$+U$ & 16 & 4.40\textsuperscript{b} & 4.37\textsuperscript{b}\cite{yabuuchi2004nmc} \\
LCO & $2{\times}2{\times}2$ & optB88-vdW$+U$ & 8 & 4.21\textsuperscript{c} & 3.97\textsuperscript{c}\cite{ohzuku1994lco} \\
\bottomrule
\end{tabular}
\end{table}

For \ce{LiFePO4} (LFP polymorph, JVASP-42723, and olivine), the 2$\times$2$\times$1 supercell (16 lithium atoms, 112 atoms) was relaxed at PBE$+U$ ($U_\mathrm{eff}(\mathrm{Fe})=5.30$~eV) across all 17 lithiation states from \ce{Li16Fe16P16O64} to \ce{Fe16P16O64}. The polymorph chain gives a convex-hull average of $3.60$~V and the olivine endpoints give $3.55$~V, $+0.12$~V above the two-phase equilibrium potential of olivine, $3.43$~V, estimated from galvanostatic intermittent titration\cite{kondo2018lfp} (cycling measurements place the plateau at $3.4$--$3.5$~V\cite{padhi1997lfp,yamada2001lfp}). The hull resolves four narrow plateaus between $3.48$ and $3.67$~V, whereas olivine shows a single plateau; the extra plateaus are an artifact of the finite supercell, since a $16$-Li cell admits only a discrete set of vacancy orderings, whose intermediate compositions would not survive in the thermodynamic limit. The upward bias has the sign expected if the Materials Project $U(\mathrm{Fe})$\cite{mp_hubbard_docs}, fitted to binary oxides following Wang et al.\cite{wang2006ggau}, is larger than the phosphate requires, because the computed voltage increases with $U$ (section ``Benchmarking against experiment and error analysis'').

For \ce{LiMnPO4} (LMP polymorph, JVASP-116897, and olivine), the same supercell at PBE$+U$ ($U_\mathrm{eff}(\mathrm{Mn})=3.90$~eV) was advanced through all 17 lithiation states. The surviving earlier run of the fully delithiated step~16 (\ce{Mn16P16O64}, all manganese high-spin \ce{Mn^{3+}}) had been set up with pseudopotentials that did not match the structure ($576$ instead of $672$ valence electrons), which placed its energy about $190$~eV above the value expected from the preceding steps; rerun from the step-15 structure with the correct pseudopotentials, it converged in $46$ ionic steps with \ce{Mn^{3+}} moments of $4.0$~$\mu_\mathrm{B}$. With the fully lithiated endpoint in the alternating magnetic order of the later steps (the block-order endpoint would lower the average by $0.01$~V), the polymorph gives $3.93$~V over full delithiation, and the olivine endpoints give $3.92$~V, $0.18$~V below the experimental $\sim 4.10$~V plateau of olivine\cite{aravindan2013lmp}. The opposite sign of the residual relative to LFP under the same $U$ calibration is consistent with the oxide-fitted $U$ being too large for iron and too small for manganese in the phosphates (section ``Benchmarking against experiment and error analysis'').

For \ce{LiMn2O4} (LMO, JVASP-141792), the spinel 2$\times$2$\times$2 supercell (16 lithium atoms on tetrahedral 8a sites, 112 atoms) was advanced through all 17 lithiation states. The convex-hull average voltage is $4.08$~V, $+0.04$~V above the experimental average of $4.04$~V, which weights the measured open-circuit voltage by composition: a constant $4.11$~V where two cubic phases coexist ($0.27<x<0.60$, taken to extend to $x=0$, below which no data were taken) and a sloping one-phase region centered at $3.94$~V ($0.60<x<1$)\cite{ohzuku1990lmo}. The hull resolves two plateaus at $4.00$ and $4.17$~V that mirror this two-step profile, also seen in galvanostatic discharge curves\cite{tomon2022lmo}, although the computed step lies at $x=0.44$ rather than near $x=0.60$.

The NMC entry JVASP-144791 is a layered variant, stored in JARVIS-DFT with triclinic \textit{P}1 symmetry, of composition \ce{Li4Mn3Co2Ni3O16}, so its Li:TM ratio is $1{:}2$ rather than the $1{:}1$ of NMC-111. The 2$\times$2$\times$1 supercell (16 lithium atoms, 112 atoms) was advanced at PBE$+U$ ($U_\mathrm{eff}=3.90$, $3.32$ and $6.20$~eV for Mn, Co and Ni) across all 17 lithiation states. Per transition metal the cell runs from \ce{Li_{0.5}TMO2} to \ce{TMO2}, the upper half of the lithium range of NMC-111, so the reference is the measured open-circuit voltage of NMC-111 over that same half: $4.37$~V, averaged from \ce{Li_{0.5}} to \ce{Li_{0.03}}, where the measured charge ended at $4.71$~V\cite{yabuuchi2004nmc}. The convex-hull average is $4.40$~V, $+0.03$~V above it, but the profile rises too steeply: over the first half of the cell's range DFT gives $3.94$~V against a measured $4.19$~V and over the second half $4.86$~V against $4.58$~V, a rise of $0.92$~V against the measured $0.39$~V. Over the range in which NMC-111 is normally cycled, which starts from full lithiation and which the cell does not cover, the measured average is $3.85$~V (3.00--4.45~V at C/10)\cite{yoshizawa2003nmc}. Rerun with optB88-vdW$+U$ on the same supercell, initial magnetic moments and $U$ values, all $17$ steps converged without intervention and the average rises to $4.76$~V. Both chains relax the cell only at full lithiation, as the rule prescribes for a host outside the layered-spacegroup whitelist (Methods); relaxing it also at $x=0.5$ and $x=0$, the two compositions that fix the averages, contracts it by $5$\% and $9$--$10$\% and gives $4.18$~V with PBE$+U$ and $4.45$~V with optB88-vdW$+U$ ($-0.19$ and $+0.08$~V against the reference; section ``Benchmarking against experiment and error analysis''; Supplementary Information). Table~\ref{tab:dft_summary} keeps the $4.40$~V that the automated protocol produced.

\begin{figure}[!t]
\centering
\includegraphics[width=\textwidth]{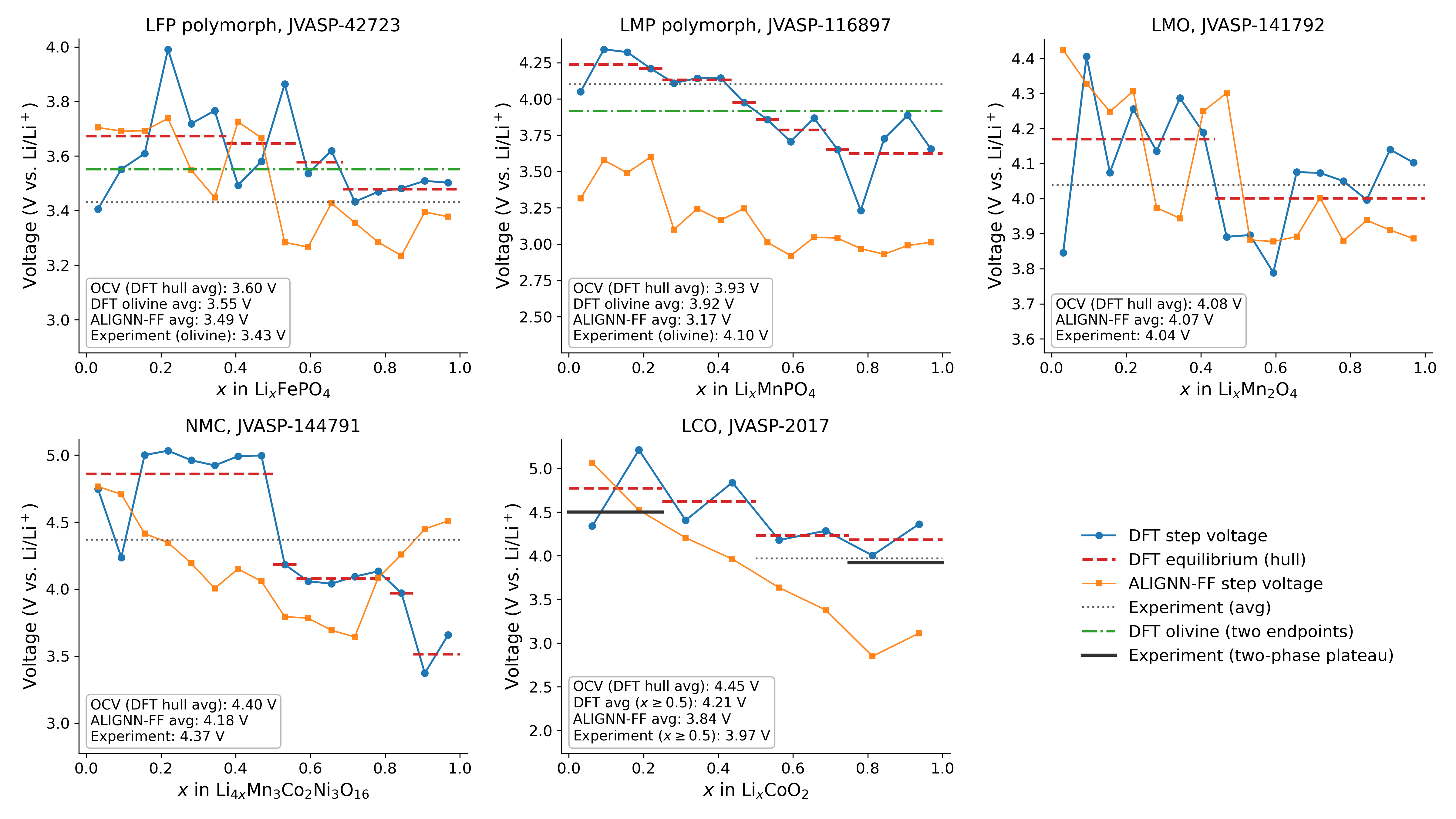}
\caption{\textbf{ALIGNN-FF versus DFT delithiation profiles for the five benchmark cathodes.} Each panel shows the DFT step voltages (blue circles) and the lower-convex-hull equilibrium plateaus (red dashed) from the supercell campaigns, the ALIGNN-FF step-voltage profile from the screening pass (orange squares; rigid-lattice protocol on the same JARVIS-DFT entry), and the experimental average voltage (dotted). The open-circuit voltage stated in each panel is the $\Delta n$-weighted hull average over the full lithium range. For NMC the experimental average is that of NMC-111 over the lithium range the cell covers (\ce{Li_{0.5}} to \ce{Li_{0.03}} per transition metal). For LCO the experimental average refers to $x=1\to0.5$, and the two measured two-phase plateaus ($3.92$ and $4.50$~V\cite{ohzuku1994lco}) are drawn as solid black lines; the optB88-vdW$+U$ hull reproduces the $0.58$~V rise between them with a uniform offset of about $+0.27$~V. The LFP and LMP panels show the chains of the JARVIS-DFT polymorph entries; their experimental averages, and the olivine DFT averages drawn as dash-dotted lines ($3.55$ and $3.92$~V), refer to olivine (section ``DFT validation on five commercial cathodes'').}
\label{fig:curve}
\end{figure}

The layered LCO framework (\ce{LiCoO2}, JVASP-2017) was routed by the spacegroup rule to optB88-vdW$+U$ ($U_\mathrm{eff}(\mathrm{Co})=3.32$~eV). The 2$\times$2$\times$2 supercell (8 lithium atoms, 32 atoms) was advanced through all 9 lithiation states from \ce{Li8Co8O16} to \ce{Co8O16} without intervention; the cell volume stays within $1$\% of the lithiated value down to $x=3/8$ and contracts to $-6.4$\% at \ce{CoO2}. Experimentally \ce{LiCoO2} is conventionally cycled only to about $x=0.5$, with a $4.2$~V charge cutoff, because deeper delithiation drives phase transitions that degrade it\cite{reimers1992lco,tan2021lco}, so, as for the sodium host of the section ``Sodium and magnesium chemistries at the DFT tier'', the benchmark compares averages over $x=1\to0.5$. Over that range the open-circuit voltage stays on the $3.92$~V two-phase plateau down to $x\approx 3/4$ and rises to $4.20$~V at $x=1/2$\cite{ohzuku1994lco}; weighted by composition it averages $3.97$~V, which we use as the reference, and the DFT average over the same range is $4.21$~V, a residual of $+0.24$~V, close to the plateau offset found below. Over the full range the DFT average is $4.45$~V, with hull plateaus at $4.18$, $4.23$, $4.62$ and $4.78$~V (Fig.~\ref{fig:curve}). The full profile can be tested against the two two-phase plateaus at which the measured voltage is flat, $3.92$~V for $3/4<x<1$ and $4.50$~V for $0<x<1/4$\cite{ohzuku1994lco}: the DFT averages over the same ranges are $4.18$ and $4.78$~V, so the calculation reproduces the $0.58$~V rise between the plateaus to within $0.02$~V, at a nearly uniform offset of $+0.26$ to $+0.28$~V. PBE$+U$ on the same supercell gives nearly equal averages over these ranges ($3.82$ and $3.81$~V) and misses the rise, its hull plateaus ($3.73$, $3.80$ and $3.98$~V) climbing by only $0.25$~V over the whole range; its averages, $3.81$~V over $x=1\to0.5$ and $3.84$~V over the full range, lie $0.39$ and $0.62$~V below the optB88-vdW$+U$ values (computed from unrounded energies). PBE$+U$ is in fact closer to the measured average over $x=1\to0.5$ ($-0.16$ against $+0.24$~V); the case for routing layered hosts to the vdW-corrected functional rests on the shape of the profile, the rise between the two-phase plateaus, which only optB88-vdW$+U$ reproduces. The run reported in the first preprint of this work (arXiv:2607.06645v1) had used the optPBE-vdW setting by mistake (Methods); it gave $4.18$~V over the full range, required a fixed-cell (ISIF $=2$) restart of its delithiated endpoint after a $28$\% volume expansion, and is retained in the Supplementary Information for comparison.

\nocite{ohzuku1994lco}%

\begin{figure}[!t]
\centering
\includegraphics[width=0.9\textwidth]{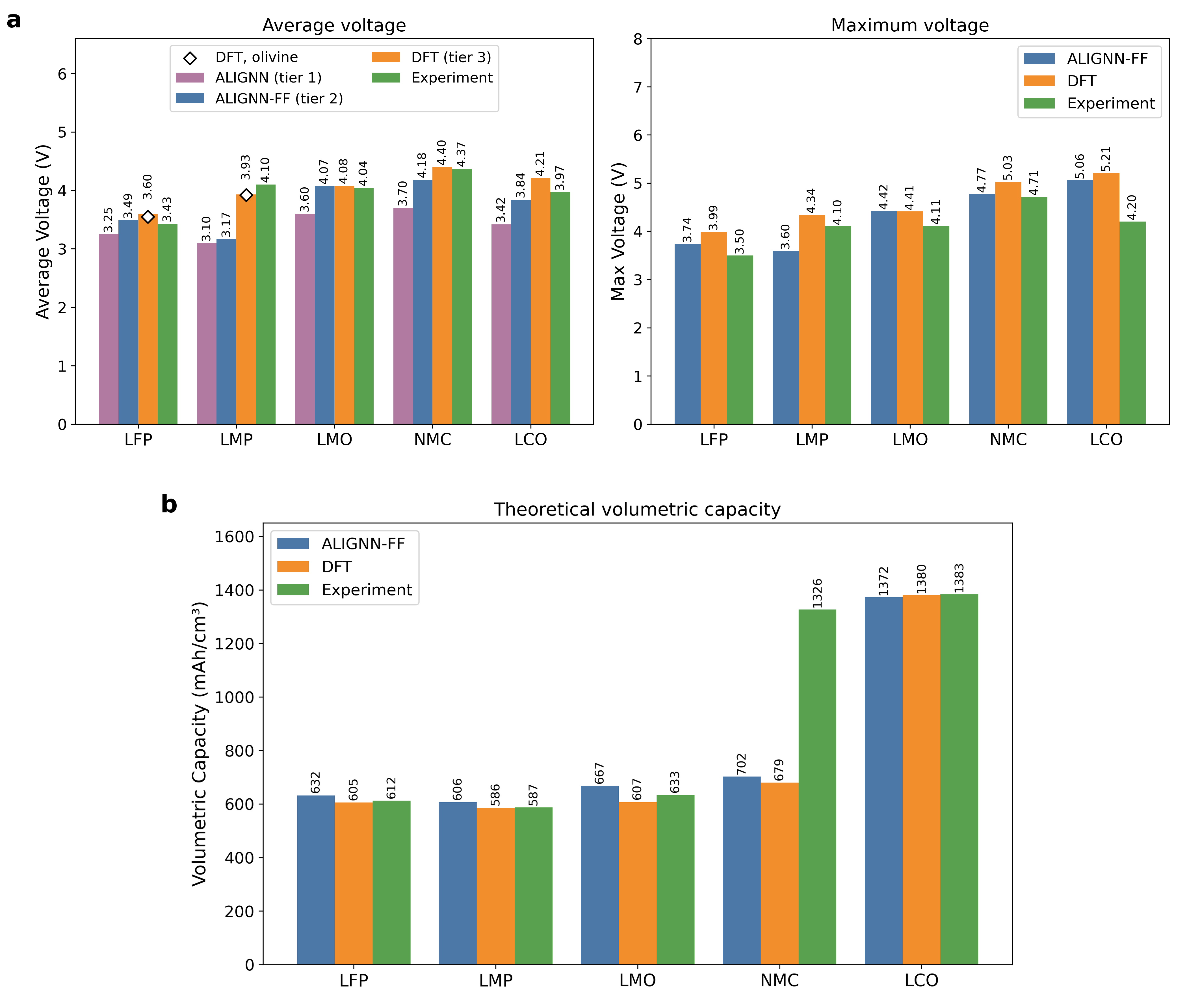}
\caption{\textbf{Voltage and volumetric-capacity benchmark.} (a) Average delithiation voltage from the ALIGNN regressor (tier~1, on the scale of its force-field labels), ALIGNN-FF (tier~2), DFT (tier~3) and experiment, and maximum voltage from ALIGNN-FF, DFT and experiment, for the five commercial cathodes. (b) Theoretical volumetric capacity from ALIGNN-FF (unrelaxed JARVIS volume), DFT (relaxed cell volume), and experiment (literature theoretical gravimetric capacity combined with measured crystal density). The NMC bars compare the JVASP-144791 cell (\ce{Li4Mn3Co2Ni3O16}, Li:TM $=1{:}2$) with experimental NMC-111 (Li:TM $=1{:}1$), so the capacity comparison for NMC is not like-for-like; its experimental voltages are taken over the lithium range the cell covers (\ce{Li_{0.5}} to \ce{Li_{0.03}} per transition metal; section ``Benchmarking against experiment and error analysis''). For LCO the DFT and experimental averages refer to $x=1\to0.5$ (full-range DFT average $4.45$~V), while the regressor and ALIGNN-FF averages and the ALIGNN-FF and DFT maximum voltages refer to the full range. Maximum voltages are the largest single-step voltages for ALIGNN-FF and DFT, which are noisy in the $32$-atom LCO cell (Table~S11). The experimental maximum is the highest plateau for LFP, LMP and LMO ($3.5$, $4.1$ and $4.11$~V) and, for the sloping NMC and LCO profiles, the highest open-circuit voltage over the compared range ($4.71$~V for NMC, $4.20$~V at $x=1/2$ for LCO). Four of the five benchmarks are in the regressor's training partition (section ``Average voltage prediction with ALIGNN''). For LFP and LMP the regressor, ALIGNN-FF and DFT bars refer to the JARVIS-DFT polymorph entries and the experimental bars to olivine; open diamonds mark the olivine DFT averages ($3.55$ and $3.92$~V; Table~\ref{tab:dft_summary}). Experimental voltage references are LFP\cite{padhi1997lfp,kondo2018lfp}, LMP\cite{aravindan2013lmp}, LMO\cite{ohzuku1990lmo}, NMC\cite{yabuuchi2004nmc}, LCO\cite{ohzuku1994lco}.}
\label{fig:benchmark}
\end{figure}

\nocite{padhi1997lfp,kondo2018lfp,aravindan2013lmp,ohzuku1990lmo,yabuuchi2004nmc,ohzuku1994lco}%

\subsection{Benchmarking against experiment and error analysis}
\label{sec:bench}

Fig.~\ref{fig:benchmark} compares ALIGNN-FF, DFT and experiment for average and maximum voltage (panel a) and theoretical volumetric capacity $Q = n_\mathrm{Li}\,F / (3.6\,V\,N_A)$ (panel b). The DFT tier agrees best with experiment for LMO ($+0.04$~V; $-0.10$~V with the cell relaxed) and, over the lithium range its cell covers, the NMC variant ($+0.03$~V, from compensating errors on the two halves of that range); the olivines show residuals of opposite sign (LFP $+0.12$~V, LMP $-0.18$~V), and LCO over its cycled range lies $0.24$~V high, close to the offset on its two-phase plateaus. The crystallographic theoretical volumetric capacity is reproduced to within $5\%$ for LFP, LMP (polymorph cells), LMO and LCO. The apparent ${\sim}50\%$ deficit for NMC reflects the stoichiometry of the JARVIS entry, a partially delithiated layered variant with Li:TM $=1{:}2$ against $1{:}1$ for NMC-111, rather than an error of the framework, so that comparison is not like-for-like. The ALIGNN-FF-to-DFT comparison (Fig.~\ref{fig:curve}) uses the same structure at both tiers, so for the phosphates it is made on the polymorphs. The largest deviation among the five is on the \ce{LiMnPO4} polymorph, where the force-field average of $3.17$~V lies $0.76$~V below DFT, more than four times the regressor's mean parity error; on the olivine \ce{LiMnPO4} entry the force-field average, $3.75$~V, lies $0.17$~V below the olivine DFT value. ALIGNN-FF spans a slightly wider range across the five chemistries than DFT ($1.01$ against $0.80$~V) but reproduces the separation of the phosphates from the oxides: both tiers place the two phosphate polymorphs below all three oxides, by $0.35$~V for ALIGNN-FF and $0.15$~V for DFT (with the LCO value over $x=1\to0.5$). Within each group the orderings differ (the \ce{LiFePO4}/\ce{LiMnPO4} and \ce{LiMn2O4}/\ce{LiCoO2} pairs swap between the tiers), so ALIGNN-FF should be read as a group-level rather than rank-level signal at the surrogate tier, which is the signal a screening cascade needs.

The olivine residuals (LFP $+0.12$~V, LMP $-0.18$~V) bracket experiment with opposite signs under the same Materials Project $U$ values\cite{mp_hubbard_docs}, which are fitted to binary-oxide formation enthalpies following Wang et al.\cite{wang2006ggau}, not to phosphates. Because the computed voltage increases with $U$, the residuals are consistent with the olivines requiring a smaller $U$ on iron and a larger one on manganese than the oxide-fitted values. With the cell relaxed at $x=0$ as well the residuals become $+0.02$ and $-0.29$~V, so part of the LFP residual comes from the fixed-cell protocol while the LMP residual does not. The GGA/GGA$+U$ mixing scheme of Jain and co-workers\cite{jain2011mphubbard} would not remove the asymmetry, because its corrections are constant per transition-metal and oxygen atom and cancel between the endpoints of an intercalation voltage. A $U$ fitted to the phosphates, or computed self-consistently by linear response, is the natural correction; Zhou et al.\ reproduced the olivine voltages to within about $0.1$~V with such values\cite{zhou2004olivine}. We report the residuals at the Materials Project values as a transparent baseline.

The optB88-vdW$+U$ functional reproduces the shape of the LCO voltage profile: between the two experimental two-phase plateaus the measured open-circuit voltage rises by $0.58$~V and the DFT averages over the same composition ranges rise by $0.59$~V, with the whole profile shifted up by about $0.27$~V. With PBE$+U$ on the same supercell the two averages differ by $0.01$~V, so the vdW correction is what produces the rise; it coincides with a different response of the cell (contraction by $6.4$\% at \ce{CoO2} with optB88-vdW$+U$, expansion by $3.5$\% with PBE$+U$), although these calculations do not isolate the mechanism. The uniform offset, rather than a growing deep-delithiation error, points to a systematic shift such as the Hubbard $U$ calibration or the lithium reference rather than a failure at low lithium content; a hybrid-functional or r$^2$SCAN\cite{furness2020r2scan} check of one endpoint pair is the natural follow-up.

For the NMC variant, against the average over the range in which NMC-111 is normally cycled ($3.85$~V\cite{yoshizawa2003nmc}), which starts from full lithiation and which the cell does not cover, the residual would be $+0.55$~V; the $+0.70$~V of the preprint was a comparison of this kind against $3.7$~V. Over the matching range the averages agree but the profiles do not (section ``DFT validation on five commercial cathodes''), and two further sets of calculations separate the protocol from the physics (Supplementary Information). First, relaxing the cell at $x=0.5$ and $x=0$, which the spacegroup rule does not do for this \textit{P}1 entry, lowers the PBE$+U$ average by $0.22$~V to $4.18$~V (residual $-0.19$~V), and over $x=1\to0.5$ to $3.73$~V, so about $0.2$~V of the fixed-cell average is the fixed-cell treatment of a layered host that the rule does not recognize. Second, replacing PBE$+U$ with optB88-vdW$+U$ raises the full-range average by $0.27$--$0.36$~V whether or not the cell is relaxed ($4.76$~V fixed, $4.45$~V relaxed; $+0.39$ and $+0.08$~V against the reference), the same direction as on LCO. In all four variants the average voltage rises by about $0.9$~V between $x>0.5$ and $x<0.5$, against $0.39$~V measured over the same lithium ranges, so the excess rise comes neither from the cell constraint nor from the dispersion treatment. Whether it is a GGA$+U$ artifact of the most oxidized states would be settled by a hybrid-functional (e.g., HSE06) or r$^2$SCAN calculation of the three compositions that fix the two half-range averages, which is outside the scope of the present work.

The same cell-relaxation test on the other PBE$+U$ benchmarks, whose cells the protocol also keeps from full lithiation, lowers the full-range averages by $0.10$ and $0.11$~V for olivine \ce{LiFePO4} and \ce{LiMnPO4} and by $0.14$~V for \ce{LiMn2O4} (and by $0.30$ and $0.44$~V for the two phosphate polymorphs; Supplementary Information). The fixed-cell protocol therefore adds $0.10$--$0.22$~V to the averages of the olivines, the spinel and NMC, and with the cell relaxed at the compositions that fix each average the five benchmarks lie within $0.3$~V of experiment (LFP $+0.02$, LMP $-0.29$, LMO $-0.10$, NMC $-0.19$ and LCO $+0.24$~V, whose chain relaxes the cell at every step), against $0.25$~V with the fixed-cell settings of the automated protocol.

\subsection{Where the benchmarks rank, and a blind prospective DFT test}
\label{sec:false-positive}

A practical question for any user of the framework is whether the surrogate ranking prioritizes materials that survive the DFT tier. We answer it in two ways: by locating the five commercial benchmarks in the ranking, and by a prospective test in which candidates were fixed before any DFT was run.

With gravimetric capacity normalized per formula unit (Methods), the five benchmarks carry their familiar theoretical capacities (LFP $170$, LMP $171$, LMO $148$, the NMC variant $144$, LCO $274$~mAh/g) and four of them pass every screening filter; the NMC variant is removed by the hull criterion ($E_\mathrm{hull}=0.079$~eV/atom). Among the $682$ survivors the benchmarks rank $91$ (LMO), $263$ (LCO), $535$ (LFP) and $562$ (LMP), the last two being the polymorph entries of the section ``DFT validation on five commercial cathodes''. The two olivine \ce{LiMnPO4} entries of JARVIS-DFT (JVASP-91261 and JVASP-121734) have force-field averages of $3.75$~V, inside the window, but their rigid-lattice profiles reach $5.80$ and $5.84$~V at the last lithium removal, so the $5.5$~V maximum-voltage filter removes them. The four ranked benchmarks sit inside the $3.0$--$4.5$~V window at the surrogate tier (ALIGNN-FF averages of $4.07$, $3.84$, $3.49$ and $3.17$~V), so their mid-to-low positions are set by the capacity and hull terms of the composite score: the lithium-rich oxides at the head of the list carry two to four times the theoretical capacity of the commercial chemistries, and LCO and LFP have hull energies of $0.023$ and $0.024$~eV/atom in JARVIS-DFT. The ranking therefore does not reproduce commercial adoption, which weighs cycle life, cost and synthesizability that the score does not encode (see Discussion), but it places all four inside the voltage window that the DFT tier subsequently confirms.

Twelve materials (eleven distinct compounds) with no prior DFT voltage data were fixed before any DFT was run: the eight top-ranked candidates of the surrogate ranking at the time of selection (which still used the per-cell capacity normalization corrected in the section ``Top-N candidate cathodes from the screen''; Supplementary Information) and four low-voltage controls with surrogate averages near $1$~V. Each was carried through the automated tier-3 protocol, and all twelve chains reached the fully delithiated composition (Table~\ref{tab:prospective}). Three endpoint relaxations stopped at the $200$-step ionic limit without converging (the fully delithiated endpoints of \ce{Li3FeO3} and \ce{LiH4SNO4} and the fully lithiated endpoint of \ce{LiMnIr2}), as did intermediate steps of the \ce{Li3FeO3} and \ce{LiH4SNO4} chains, which do not enter the full-range averages; we continued the three endpoints from their last structures, and later the \ce{Li2FeO3} step of the \ce{Li3FeO3} chain. The \ce{LiH4SNO4} endpoint converged within five further steps, leaving its average unchanged. Continuing the fully delithiated \ce{Li3FeO3} endpoint (\ce{Fe8O24} in the supercell, formally \ce{Fe^{6+}}) formed four \ce{O2} molecules (O--O $1.23$--$1.25$~\AA) and lowered the energy by more than $15$~eV, so that composition has no intercalation energy at the DFT tier; we therefore evaluate \ce{Li3FeO3} from \ce{Li3FeO3} to \ce{Li2FeO3}, the range of the \ce{Fe^{3+}}/\ce{Fe^{4+}} couple, whose endpoint converged after continuation with no O--O bond, and obtain $3.82$~V. The fully lithiated \ce{LiMnIr2} endpoint, an intermetallic to which the protocol applies $U$ on manganese, did not converge: continuing it lowered the energy by up to $5.1$~eV while the cell sheared, and $300$ further quasi-Newton steps did not converge, because the PBE$+U$ electronic state changed between ionic steps (energy jumps of up to $0.5$~eV at unchanged manganese moments). Its DFT average is $0.43$~V with the energy of the automated run and $1.72$~V with the lowest lithiated-endpoint energy found; a lower-energy lithiated state would raise it further, so neither value bounds it, and we leave it out of the mean deviation below.
Every top-ranked candidate returned a DFT average voltage of $3.40$~V or higher and every converged control $1.35$~V or lower (\ce{LiMnIr2} $1.72$~V at the lowest lithiated-endpoint energy found), with no overlap between the two groups. The separation is a weaker test than it may appear. Six of the eight top-ranked entries (five of the seven distinct materials) lie above the $4.5$~V upper edge of the window at the DFT tier, so only \ce{LiCr2P2O8} and \ce{Li3FeO3} land inside it; three of the four controls (\ce{LiCaTl2}, \ce{LiCa2Ag}, \ce{LiY2Ga}) contain no redox-active transition metal and would be removed by the redox-metal filter of the section ``Top-N candidate cathodes from the screen'', and three (\ce{LiMnIr2}, \ce{LiCa2Ag}, \ce{LiY2Ga}) lie above the $0.05$~eV/atom hull cutoff of the same screen (Table~S18); and the candidates were chosen from the force-field ranking, so the test exercises the force-field tier rather than the regressor. The two \ce{Rb2LiFeF6} entries (JVASP-141543 and JVASP-154749) are the same \textit{R}$\bar{3}$\textit{m} structure stored twice in JARVIS-DFT ($10$~meV/atom apart), which the screen does not deduplicate, and their independently run chains agree to $1$~mV, so we count the compound once, taking the mean of its two deviations. Leaving out \ce{Li3FeO3}, whose DFT value covers only the first third of its lithium, and \ce{LiMnIr2}, whose DFT value is not converged, the mean absolute deviation between the ALIGNN-FF surrogate and DFT over the remaining nine distinct materials is $0.90$~V ($0.91$ or $0.89$~V if only one of the two \ce{Rb2LiFeF6} entries is counted, $1.00$~V over the ten counted table rows), against $0.29$~V on the five commercial benchmarks ($0.34$~V if LCO is compared over full delithiation at both tiers). The surrogate underestimates six of the eight top-ranked candidates by $0.7$--$2.1$~V and overestimates the polyanionic \ce{LiCr2P2O8} by $1.0$~V, so it is a less reliable point estimate outside the oxide chemistries that dominate its training labels, and also on the \ce{Mn^{4+}} oxide \ce{Li2MgMn3O8} ($1.1$~V low). Five of the eight candidates (both \ce{Rb2LiFeF6} entries, \ce{Li2MgMn3O8}, \ce{LiCoH24C8N8O12} and \ce{LiH4SNO4}) return DFT averages above $5$~V and \ce{LiV2F7} lies at $4.92$~V, beyond the roughly $4.3$~V up to which carbonate electrolytes are stable at the cathode surface\cite{manthiram2017outlook}, while the surrogate averages ($4.0$--$4.3$~V) and maximum step voltages (at most $4.5$~V) of the five above $5$~V raise no flag; a DFT check of the upper plateau is therefore needed before a candidate is treated as compatible with a practical electrolyte. In the corrected ranking of the section ``Top-N candidate cathodes from the screen'', three of the eight keep their place among the survivors (\ce{Li3FeO3} at rank $1$, \ce{Li2MgMn3O8} at $46$, \ce{LiV2F7} at $86$), four fall below the $100$~mAh/g per-formula-unit capacity threshold (\ce{LiCr2P2O8}, both \ce{Rb2LiFeF6} entries, \ce{LiCoH24C8N8O12}) and \ce{LiH4SNO4} fails the redox-metal criterion. The force-field tier therefore separates high-voltage from low-voltage hosts among candidates with no prior DFT voltage data, but it places only two of the eight top-ranked entries inside the window, and the capacity correction changes which of them are worth carrying forward.

\begin{table}[H]
\small
\centering
\caption{Blind prospective DFT validation. Twelve materials were fixed before any DFT was run (eight top-ranked candidates and four low-voltage controls, selected on the surrogate ranking as of selection) and carried through the automated tier-3 protocol. $V^\mathrm{FF}$ is the ALIGNN-FF surrogate average voltage; $V^\mathrm{DFT}$ is the DFT average over full delithiation (except \textsuperscript{b}) with the in-house lithium reference for the stated functional. All chains reached the fully delithiated composition; three endpoint relaxations and the \ce{Li2FeO3} step were continued after reaching the ionic-step limit; the fully delithiated \ce{Li3FeO3} endpoint then released \ce{O2} and the \ce{LiMnIr2} endpoint did not converge (text). $\Delta$ is computed from unrounded values. \textsuperscript{a}These four were first run with GGA $=$ OR, which selects optPBE-vdW (Methods), giving $6.03$, $4.84$, $6.03$ and $0.41$~V (top to bottom); they were rerun with the optB88-vdW input and the reruns are reported. \textsuperscript{b}From \ce{Li3FeO3} to \ce{Li2FeO3} (\ce{Fe^{3+}}/\ce{Fe^{4+}}); the fully delithiated cell releases \ce{O2} when relaxed, so no full-range value or $\Delta$ is given. \textsuperscript{c}Lithiated endpoint not converged (text); $0.43$~V is the value of the automated run and $1.72$~V the value with the lowest lithiated-endpoint energy found, so no $\Delta$ is given.}
\label{tab:prospective}
\setlength{\tabcolsep}{5pt}
\begin{tabular}{llllccc}
\toprule
JARVIS ID & Composition & Group & Functional & $V^\mathrm{FF}$ (V) & $V^\mathrm{DFT}$ (V) & $\Delta$ (V) \\
\midrule
JVASP-117419 & \ce{LiCr2P2O8} & top & PBE$+U$ & 4.42 & 3.40 & $-1.02$ \\
JVASP-117295 & \ce{Li3FeO3} & top & PBE$+U$ & 4.39 & 3.82\textsuperscript{b} & -- \\
JVASP-154749 & \ce{Rb2LiFeF6} (B) & top & optB88-vdW$+U$\textsuperscript{a} & 4.25 & 6.13 & $+1.87$ \\
JVASP-96563 & \ce{Li2MgMn3O8} & top & PBE$+U$ & 4.23 & 5.34 & $+1.11$ \\
JVASP-116849 & \ce{LiV2F7} & top & optB88-vdW$+U$\textsuperscript{a} & 4.19 & 4.92 & $+0.73$ \\
JVASP-95533 & \ce{LiCoH24C8N8O12} & top & PBE$+U$ & 4.10 & 5.02 & $+0.91$ \\
JVASP-141543 & \ce{Rb2LiFeF6} (A) & top & optB88-vdW$+U$\textsuperscript{a} & 4.08 & 6.13 & $+2.05$ \\
JVASP-97428 & \ce{LiH4SNO4} & top & PBE & 4.01 & 5.18 & $+1.17$ \\
JVASP-81829 & \ce{LiMnIr2} & control & PBE$+U$ & 1.12 & 0.43\textsuperscript{c} & -- \\
JVASP-85076 & \ce{LiCaTl2} & control & PBE & 1.11 & 1.35 & $+0.24$ \\
JVASP-77457 & \ce{LiCa2Ag} & control & optB88-vdW\textsuperscript{a} & 1.06 & 0.45 & $-0.62$ \\
JVASP-80802 & \ce{LiY2Ga} & control & PBE & 1.04 & 0.74 & $-0.30$ \\
\bottomrule
\end{tabular}
\end{table}

\subsection{DFT spot-check of the force-field labels}
\label{sec:spotcheck}

The $0.17$~V figure measures how faithfully the regressor reproduces its ALIGNN-FF labels, not how close those labels are to DFT. To anchor the labels directly, we drew thirty structures from the regressor's held-out test partition, stratified five per $1$~V bin of label voltage over $0$--$6$~V (primitive cell of at most $20$ atoms, $E_\mathrm{hull}\le 0.10$~eV/atom, no $f$-block elements, fixed seed), and relaxed the fully lithiated and the fully delithiated primitive cell of each with the DFT tier's input template, cell included at both endpoints, the functional chosen by the same spacegroup rule (six routed to optB88-vdW and $24$ to PBE, with $+U$ on the $15$ that contain one of the eight $U$-corrected metals of the Methods). The two-endpoint DFT voltage $V=(E_\mathrm{delith}-E_\mathrm{lith})/n_\mathrm{Li}+\mu_\mathrm{Li}$ is the quantity each label represents, because the label protocol telescopes to the same two endpoints (section ``Average voltage prediction with ALIGNN''). Six delithiated endpoints did not converge within the first $200$ ionic steps and were restarted from their last structure; four then converged, and two pairs are excluded because their delithiated endpoint was still drifting after $500$ steps: both are lithium-rich oxides (\ce{Li8MnO6}, \ce{Li7FeO6}) whose fully delithiated framework is far from any stable composition.

Over the $28$ converged pairs the labels lie below DFT in $22$ cases, with a mean absolute error of $0.88$~V, a mean signed error of $+0.68$~V (DFT higher) and a mean absolute error of $0.74$~V once that offset is removed, while the ranking is largely preserved (Spearman $0.90$) and the in-window decision for the $3.0$--$4.5$~V window agrees for $21$ of the $28$. The error is concentrated where the screen does not operate. The fourteen hosts without a redox-active transition metal, which the redox-metal filter of the section ``Top-N candidate cathodes from the screen'' removes, carry a mean absolute error of $1.22$~V (up to $+2.6$~V on \ce{LiScSi2O6}), because extracting every lithium from such a host oxidizes the anion or post-transition-metal framework and the relaxed DFT endpoint departs far from the rigid framework the label assumes. On the fourteen hosts that contain a redox-active metal, the population from which all $682$ candidates are drawn, the mean absolute error is $0.54$~V with a mean offset of $+0.35$~V ($0.43$~V after removing the offset), a Spearman correlation of $0.94$ ($95$\% bootstrap interval $0.78$--$0.99$; the fourteen span $1.3$--$6.0$~V of DFT voltage, which favors a high rank correlation, and for the five inside the window it is $0.70$) and in-window agreement for $12$ of $14$; the seven oxides among them deviate from DFT by $0.13$--$0.57$~V (mean absolute error $0.34$~V), and the largest residuals are two fluorides (\ce{LiCrCdF6}, $+1.72$~V; \ce{LiCoF4}, $+1.02$~V). The regressor's own predictions for the same fourteen hosts track DFT with a Spearman correlation of $0.98$, a mean absolute error of $0.57$~V, a mean offset of $+0.53$~V and $0.37$~V after removing it.

Three differences between the tiers contribute to the offset and cannot be separated with thirty structures: the labels are single-point energies of an unrelaxed delithiated framework whereas both DFT endpoints are relaxed, the labels carry no Hubbard correction, and they are referenced to the force-field energy of lithium metal. The ranking of the section ``Top-N candidate cathodes from the screen'' and the prospective campaign use the screening pass of the force field rather than the label pass, $0.63$~V apart on average (Methods); on the same fourteen redox-metal hosts the screening-pass voltages lie $0.23$~V above DFT on average, with a mean absolute error of $0.44$~V ($0.42$~V after removing the offset), a Spearman correlation of $0.89$ ($0.62$--$1.00$) and in-window agreement for $12$ of $14$. The sign is opposite to that of the prospective test, where the screening pass lies below DFT for six of the eight top-ranked entries. Two of those six, a coordination compound and a host without a redox-active metal, lie outside the population sampled here; three are fluorides, which the spot-check does sample (on the four fluorides among its fourteen redox-metal hosts the screening pass lies between $0.14$~V below and $0.84$~V above DFT); and the three of the six that are not routed to the vdW functional keep the cell of the fully lithiated state, which on the benchmarks raises the DFT average by $0.1$--$0.4$~V, whereas the spot-check relaxes the cell at both endpoints. The difference in sign is therefore only partly explained. Against DFT the labels, and hence the regressor, are a ranking signal on redox-metal hosts, about $0.35$~V (labels) and $0.5$~V (regressor) below DFT on average with a scatter of about $0.4$~V, not a $0.17$~V-accurate voltage, while the screening pass that ranks the candidates sits about $0.2$~V above DFT on these hosts with a similar scatter, and below it on most of the prospective candidates; a candidate's position relative to the window edges should be judged with these offsets in mind (per-structure values in \texttt{analysis/spotcheck\_results.csv}).

\subsection{Greedy versus exhaustive lithium-vacancy ordering}
\label{sec:enumeration}

The DFT tier removes, at each step, the lithium whose removal gives the lowest ALIGNN-FF single-point energy, a greedy rule with no optimality guarantee. To measure what it costs, we enumerated all $2^{8}=256$ lithium-vacancy configurations of two cells with eight lithium sites, the $2{\times}2{\times}2$ \ce{LiCoO2} supercell of the benchmark campaign and the \ce{Li2MgMn3O8} cell of the prospective campaign, at the same single-point level that the rule ranks on (Supplementary Note~15 and Figure~S6). For \ce{Li2MgMn3O8} the greedy path found the exhaustive minimum at all seven intermediate compositions; for \ce{LiCoO2} it found it at two of the seven, and where it missed, the greedy configuration lay $5$--$15$~meV per formula unit above the minimum. The average voltage is the same along both paths, because it depends only on the endpoints, and the hull plateaus of the two paths differ by at most $0.12$~V. The comparison with every configuration relaxed in ALIGNN-FF could not be made: only $3$ of the $512$ relaxations with the \texttt{alignnff\_wt01} checkpoint converged, and the force field already returns residual forces of about $2$~eV/\AA\ on the DFT-relaxed parent cells (Supplementary Note~15). This is why the DFT tier ranks vacancies by single-point energy and leaves relaxation to DFT.

\subsection{Sodium and magnesium chemistries at the DFT tier}
\label{sec:nonli}

The non-lithium force-field screen (section ``High-throughput screening across JARVIS-DFT'') had not been checked against DFT for any ion other than lithium. We therefore carried one sodium and one magnesium cathode with a well-characterized experimental counterpart through the unchanged DFT tier: sequential supercell removal of the working ion with ALIGNN-FF vacancy ranking, the functional selected by the same spacegroup rule, and metal references computed in the same basis as the cathode runs (Methods). For a working ion of charge $z$ the step voltage is $V=(E_\mathrm{lo}-E_\mathrm{hi})/(z\,\Delta n)+\mu_\mathrm{M}/z$.

The sodium host is O3-\ce{NaCoO2} (JVASP-79809, \textit{R}$\bar{3}$\textit{m}, $E_\mathrm{hull}=0$), the direct analog of the LCO benchmark, in a $2{\times}2{\times}2$ supercell of $32$ atoms with eight sodium sites, routed to optB88-vdW$+U$ with the same $U_\mathrm{Co}$. Experimentally the stoichiometric O3 phase is charged through the O3, O$'$3, P$'$3 and P3 stacking sequence, with a voltage that rises from about $2.5$~V at $x=1$ to about $3.4$~V at $x=0.5$ versus \ce{Na/Na+} at C/20, an average of about $2.9$~V read from the published charge curve over that range\cite{lei2014nacoo2,delmas1981nacoo2}. Over the same range, $x=1\to0.5$, the four DFT steps fall on a single hull plateau at $3.09$~V, $0.19$~V above experiment. With the chain continued beyond the measured range, the step at $x=3/8$ converged only after about $150$ ionic steps, consistent with the layer gliding that the change from O-type to P-type stacking requires, and the step at $x=1/4$ did not converge within the $200$-step limit; both endpoints converged, so the full-range average, which depends only on them, is $3.73$~V ($4.37$~V over $x=0.5\to0$, with the cell contracting by $20$\% at \ce{CoO2}). The non-lithium force-field screen gives $3.17$~V for this entry over the same full range, $0.56$~V below DFT, the same direction as the force-field deviations on the lithium benchmarks.

The magnesium host is the tetragonal spinel \ce{MgMn2O4} (JVASP-11340, \textit{I}4$_1$/\textit{amd}, $E_\mathrm{hull}=0.017$~eV/atom), the Jahn--Teller-distorted phase obtained experimentally, in a $2{\times}2{\times}2$ supercell of $112$ atoms with sixteen magnesium sites, treated with PBE$+U$ ($U_\mathrm{Mn}=3.9$~eV) as for LMO. The relevant experiment is magnesium extraction, the \ce{Mn^{3+}/Mn^{4+}} couple, at about $3.4$~V versus \ce{Mg/Mg^{2+}} in an ionic-liquid electrolyte at $150\,^\circ$C\cite{okamoto2015mgspinel}, with about $40$\% of the magnesium extractable through a two-phase reaction between tetragonal and cubic spinel\cite{hatakeyama2019mgmn2o4}; the lower potentials often quoted for this compound belong to magnesium insertion (\ce{Mn^{3+}/Mn^{2+}}), a different reaction. Over sixteen to ten magnesium per cell the DFT two-endpoint average is $3.10$~V, $0.30$~V below experiment; the range lies within the first hull plateau ($3.08$~V, $x=1\to0.56$). Continued past the measured range, the chain rises through hull plateaus at $3.11$, $3.32$, $3.43$ and $3.66$~V, with an average of $3.22$~V from sixteen to one magnesium; the last step, to \ce{Mn2O4}, did not converge within the $200$-step limit or a further $300$-step restart and is not used. The force-field screen contains no magnesium working-ion entries, so only the DFT-to-experiment comparison is available; the reference is measured at $150\,^\circ$C, and because multivalent intercalation shows large voltage hysteresis it is an upper bound on the equilibrium value.

Both non-lithium hosts therefore lie within $0.3$~V of experiment with the DFT tier unchanged ($+0.19$~V for sodium, $-0.30$~V for magnesium against a reference that is itself an upper bound); two hosts do not establish general transfer, and the one host with both tiers ($0.56$~V apart) does not establish the accuracy of the non-lithium force-field screen, which remains a transferability demonstration (see Discussion).

\subsection{Top-N candidate cathodes from the screen}
\label{sec:topn}

The screen returns $682$ lithium-intercalation candidates from the $7{,}193$-entry pool after filtering on average voltage between $3.0$ and $4.5$~V, theoretical gravimetric capacity above $100$~mAh/g per formula unit, hull energy at or below $0.05$~eV/atom, maximum step voltage at or below $5.5$~V and the presence of a redox-active transition metal (thresholds justified in the Methods). Survivors are scored by $\mathrm{score}=\frac{1}{3}\,\widetilde{V}_\mathrm{avg}+\frac{1}{3}\,\widetilde{Q}_\mathrm{grav}-\frac{1}{3}\,\widetilde{E}_\mathrm{hull}$, each tilde denoting min--max normalization across the survivors, with $Q_\mathrm{grav}=n_\mathrm{Li}F/(3.6\,M)$ the theoretical capacity for extraction of all $n_\mathrm{Li}$ lithium atoms from a formula unit of mass $M$, which does not depend on how many formula units the cell holds (\ce{LiFePO4} $170$~mAh/g, \ce{LiCoO2} $274$~mAh/g). Table~\ref{tab:topn} lists the top twelve from the unaltered output of the screen (top fifty in the Supplementary Information). The list is led by lithium-rich transition-metal oxides: \ce{Li3FeO3} at rank~1, which the prospective test places at $3.82$~V over its \ce{Fe^{3+}}/\ce{Fe^{4+}} range; \ce{Li2MnO3} at rank~2 (ICSD 202639), the lithium-rich layered end member and a component of lithium-rich NMC cathodes; and, further down, \ce{Li3CrClO4}, \ce{Li3RuO4}, \ce{Li3CrO4}, three \ce{Li5Mn3O8} polymorphs, \ce{Li2NiO2} and \ce{Li2RhO3}. Ten of the twelve are Mn, Fe, Cr, Ni, Ru or Rh oxides, one is an oxychloride and one is polyanionic (\ce{Li2VBPO7}). This dominance follows from the capacity term, which counts every lithium whether or not the transition metal can supply the charge: \ce{Li3FeO3} would require \ce{Fe^{6+}} (the \ce{Fe^{3+}}/\ce{Fe^{4+}} couple alone gives $215$ of its $645$~mAh/g), \ce{Li3CrClO4} and \ce{Li2CrO4} (rank $32$) already contain \ce{Cr^{6+}}, and the $459$~mAh/g of \ce{Li2MnO3} (\ce{Mn^{4+}}) rests on oxygen redox and is well above the $\sim 250$~mAh/g realized in lithium-rich layered oxides once oxygen loss and voltage fade set in. For \ce{Li3FeO3} the DFT tier bears this out: \ce{Li2FeO3} relaxes without O--O bonds whereas the fully delithiated cell releases \ce{O2} (section ``Where the benchmarks rank, and a blind prospective DFT test''), so its DFT voltage refers to the first $215$~mAh/g. The composite score does not encode oxygen-redox stability, synthesizability or kinetics (see Discussion), so we treat the top-12 as a list to be re-filtered by accessible oxidation states before it is taken to the DFT tier, not as a definitive ranking. Of the $71$ candidates returned by the earlier per-cell normalization, $40$ remain among the $682$, and of the four benchmarks that the per-cell normalization had removed, LFP, LMP and LMO are recovered while the NMC variant is now excluded by the hull criterion.

\begin{table}[!htbp]
\small
\centering
\caption{Top-12 cathode candidates from the BatteryMat screen of the lithium-containing JARVIS-DFT pool, ranked by composite score. $V_\mathrm{avg}$ is the ALIGNN-FF prediction; $Q_\mathrm{grav}$ is the theoretical capacity per formula unit for full lithium extraction; $E_\mathrm{hull}$ and the relaxed structure are JARVIS-DFT labels. The full top-50 with the surrogate maximum voltage, and the ICSD codes and Materials Project source entries recorded in JARVIS-DFT for the benchmark, top-12 and prospective materials, are in the Supplementary Information.}
\label{tab:topn}
\setlength{\tabcolsep}{3pt}
\footnotesize
\begin{tabular}{rllccccl}
\toprule
Rank & JARVIS ID & Composition & $V_\mathrm{avg}$ & $Q_\mathrm{grav}$ & $E_\mathrm{hull}$ & Score & Notes \\
 & & & (V) & (mAh/g) & (eV/atom) & & \\
\midrule
1 & JVASP-117295 & \ce{Li3FeO3} & 4.39 & 645 & 0.001 & 0.639 & Fe oxide; DFT $3.82$~V to \ce{Li2FeO3} (Table~\ref{tab:prospective}) \\
2 & JVASP-11735 & \ce{Li2MnO3} & 4.45 & 459 & 0.000 & 0.542 & Li-rich layered end member, \textit{C}2/\textit{m} \\
3 & JVASP-122347 & \ce{KLi4FeO5} & 4.35 & 529 & 0.011 & 0.492 & mixed-alkali Fe oxide \\
4 & JVASP-122531 & \ce{Li3CrClO4} & 4.08 & 467 & 0.000 & 0.465 & Cr oxychloride \\
5 & JVASP-31181 & \ce{Li3RuO4} & 4.06 & 433 & 0.000 & 0.439 & Li-rich Ru oxide \\
6 & JVASP-42740 & \ce{Li3CrO4} & 3.61 & 588 & 0.000 & 0.433 & Li-rich Cr oxide \\
7 & JVASP-148139 & \ce{Li5Mn3O8} & 4.09 & 409 & 0.002 & 0.419 & Li-rich Mn oxide, \textit{P}2/\textit{m} \\
8 & JVASP-112917 & \ce{Li5Mn3O8} & 4.03 & 409 & 0.000 & 0.418 & polymorph, \textit{P}$\bar{1}$ \\
9 & JVASP-118606 & \ce{Li2NiO2} & 3.73 & 513 & 0.000 & 0.414 & Li-rich Ni oxide, \textit{Immm} \\
10 & JVASP-49957 & \ce{Li2RhO3} & 4.24 & 325 & 0.000 & 0.412 & Li-rich Rh oxide \\
11 & JVASP-147246 & \ce{Li5Mn3O8} & 4.06 & 409 & 0.004 & 0.400 & polymorph, \textit{C}2/\textit{m} \\
12 & JVASP-44367 & \ce{Li2VBPO7} & 4.38 & 245 & 0.000 & 0.396 & V borophosphate \\
\bottomrule
\end{tabular}
\end{table}

\section{Discussion}
\label{sec:discussion}

The Materials Project Battery Explorer\cite{jain2013materialsproject,ong2013pymatgen} provides DFT-computed average voltages, capacities and stability indicators for thousands of Li, Na, Mg, Ca and Zn intercalation pairs; it is broader than BatteryMat in coverage and narrower in protocol flexibility, and BatteryMat is complementary, an open framework that can be rerun on a user's own structures with an auto-selected functional and an in-house lithium reference. We do not compare the five benchmarks numerically with Battery Explorer values, because the two frameworks differ in four respects that can each move a voltage by as much as the residuals under discussion: supercell construction (database primitive cells between two tabulated endpoints there, $\geq 7$~\AA\ supercells along a sequential path here), functional choice (uniform PBE$+U$ against the spacegroup-routed PBE$+U$/optB88-vdW$+U$), lithium-reference convention (GGA/GGA$+U$ mixing with fitted elemental corrections\cite{jain2011mphubbard} against an unmixed in-house \ce{Li_{sv}} reference) and endpoint definition (whichever charged and discharged entries exist in the database, against the fully lithiated and fully delithiated compositions used here); a like-for-like cross-validation would require recomputing one side under the other's conventions and is left as future work. The phosphate campaign of Hautier and co-workers\cite{hautier2011phosphates} established the high-throughput-DFT template for cathode discovery before the relevant surrogate models had matured; in BatteryMat the two ML tiers cost about $10^{-1}$~s per structure and $1$~s per delithiation step, against $30$--$1300$ core-hours for a step of a $112$-atom PBE$+U$ chain (Supplementary Note~18), which frees the DFT tier to validate a handful of survivors rather than evaluate the whole pool. Surrogate-only voltage networks\cite{joshi2019voltage,louis2022voltage} (Table~\ref{tab:prior}) lack that validation tier, which is what supplies a voltage curve comparable with experiment and a measured bound on the surrogate error.

The second tier is not hard-wired to ALIGNN-FF. It needs only the energy of a partially delithiated supercell and a ranking of lithium vacancies by relative energy, relaxation being left to DFT (section ``Greedy versus exhaustive lithium-vacancy ordering''), so M3GNet\cite{chen2022m3gnet}, CHGNet\cite{deng2023chgnet} or MACE\cite{batatia2022mace} could replace it through their standard calculator interfaces; we have not benchmarked these alternatives on the delithiation task and make no claim about their relative accuracy (Supplementary Note~19). Two caveats about ALIGNN-FF itself belong here. Its training set, the JARVIS-DFT 3D dataset of energies and forces, is dominated by stoichiometric ground-state structures, so the vacancy-rich intermediates of a delithiation path are an extrapolation from the training distribution; the direct evidence on that extrapolation is the DFT comparison on delithiated cells in the sections ``Benchmarking against experiment and error analysis'', ``Where the benchmarks rank, and a blind prospective DFT test'' and ``DFT spot-check of the force-field labels''. Its training labels are optB88-vdW energies, whereas the DFT tier uses PBE$+U$ for three-dimensionally bonded frameworks, so the surrogate and validation tiers are not on a single energy scale; on \ce{LiCoO2} the two functionals differ by $0.39$~V over $x=1\to0.5$ and by $0.62$~V over the full range (section ``DFT validation on five commercial cathodes''). The framework is released as a Python package (Code availability) whose \texttt{dft\_prep} module generates the inputs of each DFT step, ranks the lithium vacancies and computes the hull voltage from the command line (Supplementary Note~5; its use on structures outside JARVIS-DFT is described in Supplementary Note~19). The AtomGPT Battery Explorer web service mirrors the second tier, and the AtomGPT chat interface\cite{choudhary2024atomgpt} implements the first tier as a tool within the open-access AtomGPT.org agentic platform\cite{lee2026agapi}, which also exposes related atomistic tools such as the SlaKoNet tight-binding framework\cite{choudhary2025slakonet}.

Three directions follow directly. A hybrid-functional or r$^2$SCAN calculation of the compositions that fix each layered-oxide average would show where the remaining residuals come from: PBE$+U$ and optB88-vdW$+U$ bracket the measured LCO average over $x=1\to0.5$, and both make the NMC profile rise between the two halves of its range by more than twice the measured amount (section ``Benchmarking against experiment and error analysis''). A DFT campaign over the head of the sodium candidate pool would establish whether the surrogate ranking can be trusted for that ion, since its transfer to DFT has been measured on only one non-lithium host (section ``Sodium and magnesium chemistries at the DFT tier''). Universal force fields could replace or ensemble with the second tier, and one trained on meta-GGA data, such as a MACE model trained on the r$^2$SCAN MatPES set\cite{kaplan2025matpes}, would put all three tiers on one energy scale and remove the need for the spacegroup-based functional router, which we regard as the preferred long-term design. Longer term, replacing the force-field labels with DFT voltages, generated by the present framework on a curated subset or taken from a public DFT-voltage corpus, would turn the regressor into a direct DFT distillation, possibly at a higher MAE because DFT labels carry more variance than force-field labels, that approximates what the user will see at the validation tier.

Ten limitations frame the scope of the present claims. First, the 0.17~V parity error of the regressor (section ``Average voltage prediction with ALIGNN'') is an ML-against-ML metric: it measures how faithfully a single forward pass reproduces the ALIGNN-FF protocol that generated the labels, not how faithfully ALIGNN reproduces DFT. Against DFT the force-field tier deviates by $0.29$~V on average on the five benchmarks ($0.34$~V if LCO is compared over full delithiation at both tiers; $0.76$~V on the \ce{LiMnPO4} polymorph), by $0.90$~V on the nine distinct prospective materials with converged full-range values, most of them outside the oxide core of the training labels, and, on the redox-metal hosts of the spot-check, the labels sit $0.35$~V below DFT with a scatter of $0.43$~V while preserving the DFT ranking (Spearman $0.94$); the DFT-to-experiment residuals of Table~\ref{tab:dft_summary} are a separate matter.

Second, no tier carries an uncertainty estimate: the DFT calculations are single-shot relaxations, the force-field profiles single-model predictions and the regressor reports a point value. The uncertainty statements in the paper are empirical (the held-out residual distribution and its bootstrap interval, the spread across the grouped folds, the spot-check scatter and the spread across the prospective materials); an ensemble over universal force fields, with the spread propagated into a screening confidence interval, is future work, and the chemistry-grouping claim of the section ``Benchmarking against experiment and error analysis'' rests on five materials.

Third, the DFT tier has been tested on lithium hosts and on one sodium and one magnesium host (section ``Sodium and magnesium chemistries at the DFT tier''); the non-lithium force-field screen is a transferability demonstration, and the regressor is lithium-only.

Fourth, the finite supercells do not capture two-phase coexistence: a $2{\times}2{\times}1$ phosphate cell admits only a discrete set of vacancy orderings, so the single plateau of olivine LFP becomes a staircase of computed plateaus spanning ${\sim}0.2$~V on the LFP polymorph chain (section ``DFT validation on five commercial cathodes'').

Fifth, the functional router is a spacegroup whitelist (\textit{R}$\bar{3}$\textit{m}, \textit{P}6$_3$/\textit{mmc}, \textit{C}2/\textit{m}, \textit{C}2/\textit{c}), not a bonding analysis: a layered host in another spacegroup is routed to PBE$+U$, as happens for the NMC benchmark, which JARVIS-DFT stores as \textit{P}1, and a three-dimensionally bonded host in a whitelisted spacegroup receives a vdW correction it does not need. Candidates on the two routes sit on different absolute voltage scales ($0.25$--$0.62$~V apart on the two layered oxides tested with both functionals), and hosts outside the whitelist keep the cell of the fully lithiated state, which adds $0.10$--$0.22$~V to the averages of the olivines, the spinel and NMC and $0.30$ and $0.44$~V to those of the two phosphate polymorphs (section ``Benchmarking against experiment and error analysis''). The router is a single function in the code and can be replaced by a bonding-based detector (interlayer gap or connectivity), and a false negative reproduces the uniform PBE$+U$ treatment of earlier campaigns.

Sixth, the DFT tier and the label generator assume that the database entry is the fully lithiated end member, which the JVASP-144791 cell (Li:TM $=1{:}2$) is not, and the framework does not lithiate a host beyond its database composition. As an indication of how often the assumption fails, $2{,}026$ of the $7{,}193$ pool entries ($28$\%) have another JARVIS-DFT entry with the same host stoichiometry and a higher lithium content; this over-counts, because the more lithiated composition is often an unstable distinct compound or lies outside the practical cycling window (\ce{LiCoO2} is flagged by \ce{Li2CoO2} entries $0.07$--$0.23$~eV/atom above the hull, \ce{LiMn2O4} by its over-lithiated tetragonal spinel), and it misses hosts with no more lithiated entry, JVASP-144791 among them; a framework-aware test would require structure matching and is left as future work. Nor does the screen check the phase of an entry: the LFP and LMP benchmark entries are a polymorph with tetrahedral lithium rather than olivine (section ``DFT validation on five commercial cathodes''), and the olivine \ce{LiMnPO4} entries are removed by the maximum-voltage filter (section ``Where the benchmarks rank, and a blind prospective DFT test'').

Seventh, the sequential path uses a greedy vacancy choice whose error is bounded by exhaustive enumeration on two small cells only (at most $15$~meV per formula unit and $0.12$~V in any plateau; section ``Greedy versus exhaustive lithium-vacancy ordering''), not in general.

Eighth, the composite score and the filters encode voltage, theoretical capacity and thermodynamic stability only; the theoretical capacity counts every lithium whether or not the redox-active metal can be oxidized that far (section ``Top-N candidate cathodes from the screen''), and lithium-ion mobility, oxygen-redox stability at high state of charge, synthesizability and cost, which separate the lithium-rich oxides at the head of Table~\ref{tab:topn} from a practical cathode, are not encoded.

Ninth, the regressor transfers across transition metals but not reliably across anion chemistry (withheld chalcogenides $0.46$~V, hosts without a redox-active metal $0.61$~V, fluorides $0.91$~V with no explanatory power; section ``The regressor as the screening front''), so candidates from anion classes that are sparse in the training labels should be taken to the force-field or DFT tier before being ranked.

Tenth, the automation covers input generation and bookkeeping but not recovery from failed or anomalous relaxations: the LCO run of the preprint needed one manual intervention (an ISIF $=2$ rerun of its delithiated endpoint; the optB88-vdW$+U$ rerun reported here needed none), the fully delithiated LMP step was rerun after a setup error in its pseudopotentials and the fully lithiated LMP endpoint was recomputed in the magnetic order of the rest of its chain, three endpoints and one intermediate step of the prospective campaign were continued by hand (after which fully delithiated \ce{Li3FeO3} released oxygen and fully lithiated \ce{LiMnIr2} did not converge), and the code only warns when a relaxation changes the cell volume by more than $10$\% (Methods).

In summary, BatteryMat ranks lithium-ion cathode candidates with a single-pass ALIGNN average-voltage regressor, refines them with a force-field delithiation profile and validates them with automated supercell DFT. The regressor reaches a $0.17$~V mean absolute error ($R^2=0.94$) against its force-field labels on a held-out test and, applied as the front of the screen, reproduces the force-field ranking of the full pool with a Spearman correlation of $0.87$ ($0.84$ on the entries outside its training partition). The DFT tier reproduces the experimental average voltage of five commercial cathodes (the phosphates computed as olivine) to within $0.3$~V whether or not the cell is relaxed at the delithiated compositions, each over the lithium range its cell covers or over which it is cycled, and the crystallographic theoretical volumetric capacity to within $5\%$ on four; on the NMC variant, whose cell holds half the lithium of NMC-111, the average agrees only because a low first half and a high second half compensate, which we flag rather than absorb into the headline. Reproducing these results depends on three implementation details: the spacegroup-based functional rule, a lithium reference computed with the functional and plane-wave settings of the cathode runs, and magnetic moments inherited along the delithiation path (in every chain except LMP and the superseded optPBE-vdW LCO run). In a blind prospective test the eight candidates top-ranked at selection and four low-voltage controls, fixed before any DFT was run, separate by voltage at the DFT tier, although only two of the eight land inside the target window, and a DFT spot-check of $30$ held-out structures shows that, on the redox-metal hosts the screen admits, the force-field labels preserve the DFT ranking (Spearman $0.94$) but sit $0.35$~V below DFT with $0.43$~V scatter, so the $0.17$~V figure is a distillation fidelity rather than an accuracy against DFT. The screen returns $682$ ranked candidates from the $7{,}193$-entry pool, led by lithium-rich Fe, Mn and Cr oxides; like the $213$-candidate Alexandria shortlist, these are existing database structures prioritized for validation rather than newly generated cathodes, and apart from four commercial benchmarks, three prospective candidates and three spot-check hosts (the last with endpoint DFT only), they have not been carried through the DFT tier. We read these results as showing that an average-voltage regressor can serve as the front of a hierarchical screen rather than as a benchmark-only ML target, provided that the ranking it feeds is tested at a validation tier (here prospectively for the force-field ranking it reproduces, and on the spot-check hosts for the regressor itself) at which quantitative agreement with experiment can be recovered.

\section{Methods}
\label{sec:methods}

\subsection{ALIGNN architecture, training and inference}
ALIGNN\cite{choudhary2021alignn} represents a crystal as an atom graph, with edges between atoms within $8$~\AA\ (at most $12$ neighbors per atom), and its line graph, whose edges carry bond angles. Four ALIGNN layers, which update atom, bond and triplet features on both graphs, are followed by four edge-gated GCN layers on the atom graph, with $256$ hidden features and ${\sim}4$~M parameters; mean-pooled atom features are projected by a final linear layer to a single scalar, the predicted average voltage (feature definitions in Supplementary Note~11). The regressor was trained with the \texttt{alignn\_atomwise} implementation for $250$ epochs on $7{,}610$ lithium-containing JARVIS-DFT structures with an L1 (mean-absolute-error) loss, the AdamW optimizer at a constant learning rate of $10^{-3}$ (weight decay $10^{-5}$) and batch size $32$; the one-cycle schedule named in the configuration file has no effect for this model class (Supplementary Note~11). Training labels are average delithiation voltages produced in-house with ALIGNN-FF, using the default checkpoint of the ALIGNN-FF package (\texttt{default\_path} in \texttt{alignn.ff.ff}) rather than \texttt{alignnff\_wt01}: for each lithium-containing JARVIS-DFT structure a supercell with all axes at least $8$~\AA\ is built, the lithium atoms are removed one at a time in site order from the rigid lattice, each intermediate is evaluated as a single point, and the step voltages, referenced to the force-field energy of the JARVIS-DFT elemental lithium entry (JVASP-25117), are averaged. A second, independent ALIGNN-FF pass with the \texttt{alignnff\_wt01} checkpoint, used for the screening pool and for every pool-level force-field voltage in this paper, builds the same supercells, removes the lithium atoms in the same order and references the voltages to an effective lithium energy of $-0.957$~eV/atom. We infer this value from the screening outputs ($0.03$~eV/atom below the tabulated JARVIS-DFT energy of $-0.925$~eV/atom) because the script that produced them is not part of the code release (Code availability). The two passes differ by $0.63$~V on average, with a structure-to-structure standard deviation of $0.80$~V (section ``The regressor as the screening front''), so voltages on the label scale and on the screening scale are not interchangeable. The 80/10/10 train/validation/test split follows the order of the input file: with the \texttt{keep\_data\_order} option of the ALIGNN training script the entries are not shuffled, so the random seed in the configuration ($123$) has no effect, and the validation and test partitions are the last $20$\% of the entries, all with JARVIS identifiers above $100{,}000$. Four of the five DFT-validated benchmark cathodes (JVASP-2017, JVASP-42723, JVASP-116897, JVASP-141792) fall in the training partition and JVASP-144791 in the validation partition; none is in the held-out test partition. The identifier lists of the three partitions and the training configuration are available on request (Code availability). The grouped-split study of the section ``The regressor as the screening front'' retrains the same architecture from scratch for each fold ($100$ epochs, otherwise the production hyperparameters), with families assigned by the most frequent redox-active transition metal in the formula (families with fewer than $50$ members are not held out) or by anion class, with $10$\% of the remaining structures used for validation; these runs used the pure-PyTorch implementation \texttt{alignn\_atomwise\_pure} of the ALIGNN code base (Supplementary Note~11).

\subsection{Screening filters and their justification}
The screening file holds $7{,}474$ rows, of which $281$ are entries for other working ions on the same structures and are excluded, leaving the $7{,}193$ lithium hosts of the pool. The candidate filters of the section ``Top-N candidate cathodes from the screen'' follow the thresholds of a recent computational cathode screen\cite{li2024pipeline} and give numerical form to the requirements set out by Goodenough and Park\cite{goodenough2013}. The $3.0$--$4.5$~V average-voltage window, the window used in that screen, spans the polyanion ($\sim 3.5$~V) and layered/spinel ($\sim 4$~V) operating regimes; its upper edge lies near the oxidative limit of carbonate electrolytes, which are not stable in contact with the cathode surface above about $4.3$~V versus \ce{Li/Li+}\cite{manthiram2017outlook}. The $5.5$~V cap on the maximum step voltage is set $1$~V above the window so that a single high surrogate plateau does not remove an otherwise viable host. The theoretical gravimetric capacity threshold of $100$~mAh/g is the minimum used in the same screen\cite{li2024pipeline}, evaluated here per formula unit for full lithium extraction, $Q_\mathrm{grav}=n_\mathrm{Li}F/(3.6\,M)$; the five commercial benchmarks lie between $144$ and $274$~mAh/g on this scale. The hull-energy cutoff of $0.05$~eV/atom is a conventional metastability window: Sun et al.\cite{sun2016metastability} found that about half of $29{,}902$ observed inorganic crystalline phases are metastable, with a median metastability of $15\pm0.5$~meV/atom and a 90th percentile of $67\pm2$~meV/atom, so $50$~meV/atom admits most experimentally realized metastable phases while excluding the long tail. The redox-metal criterion (Ti, V, Cr, Mn, Fe, Co, Ni, Cu, Nb, Mo, Ru, Rh, W) requires a charge-compensating transition metal (the first stage of the Alexandria funnel uses a broader list of $27$ $d$-block metals); it removes lithium-rich electrolyte-like hosts such as \ce{Li14P2N6O3} and \ce{Li2CN2}, for which a force-field delithiation voltage is not a cathode voltage. Without it the same filters return $740$ candidates. The composite score weights the three normalized signals equally; it is a prioritization heuristic, not a figure of merit, and encodes neither synthesizability nor kinetics.

\subsection{DFT calculations}
First-principles calculations use VASP\cite{kresse1996vasp} with PAW pseudopotentials\cite{blochl1994paw,kresse1999paw}, a plane-wave cutoff $E_\mathrm{cut} = 520$~eV, Gaussian smearing ($\sigma = 0.05$~eV) and spin polarization (ISPIN $= 2$). Two input templates were used. The earlier template uses conjugate-gradient ionic relaxation (IBRION $= 2$), \texttt{EDIFF} $= 10^{-4}$~eV with the default energy-based ionic criterion, a cap of $500$ self-consistency iterations (NELM $= 500$), cold starts and the default charge mixing; it was used for most LMP and LMO steps and for the first steps of the LFP and NMC chains. The current template keeps \texttt{EDIFF} and the ionic criterion but uses quasi-Newton relaxation (IBRION $= 1$), NELM $= 100$ so that a non-converging step fails quickly, a restart from the previous step's wave function and charge density and a conservative DFT$+U$ mixing schedule; it was used for the remaining benchmark steps, for every step of the reported LCO chain, and for all other cathode calculations in this paper except the first seven steps of the superseded optPBE-vdW LCO run (exceptions, per-chain counts and performance tags in Supplementary Notes~2 and 17). $\Gamma$-centered $k$-meshes are scaled inversely with supercell size from a $3\times 3\times 3$ base (for the primitive cells of the spot-check, $n_i=\lceil 21\,\text{\AA}/|\mathbf{a}_i|\rceil$). The PBE\cite{perdew1996pbe} functional with a Hubbard $U$ correction (PBE$+U$) is used for the phosphate (LFP, LMP) and spinel (LMO) frameworks and for NMC, whose JARVIS-DFT entry has \textit{P}1 symmetry; optB88-vdW$+U$ (GGA $=$ BO, PARAM1 $=0.1833333333$, PARAM2 $=0.22$, LUSE\_VDW $=$ .TRUE., AGGAC $=0.0$)\cite{klimes2010optb88,klimes2011optb88} is used for layered LCO, which is identified automatically by its \textit{R}$\bar{3}$\textit{m} spacegroup. The LCO campaign of the preprint, and the first pass of the four prospective materials that the spacegroup rule routed to the vdW functional (Table~\ref{tab:prospective}), were run with GGA $=$ OR, LUSE\_VDW $=$ .TRUE. and AGGAC $=0.0$, which selects the optPBE-vdW functional rather than optB88-vdW; the input generator has been corrected, the LCO benchmark and the four prospective materials reported here are reruns with the optB88-vdW tags, and the earlier optPBE-vdW LCO run is kept in the Supplementary Information as a functional-sensitivity comparison. A Dudarev DFT$+U$ correction\cite{dudarev1998ldau} (LDAUTYPE $= 2$, $U_\mathrm{eff} = U - J$, $J = 0$) is applied to $d$ orbitals with the Materials Project $U_\mathrm{eff}$ values\cite{mp_hubbard_docs}, fitted to binary-oxide formation enthalpies following Wang et al.\cite{wang2006ggau}: Mn $3.90$, Fe $5.30$, Co $3.32$, Ni $6.20$, V $3.25$, Cr $3.70$, Mo $4.38$, W $6.20$~eV. The correction is applied whenever one of these metals is present, so it also enters hosts outside the oxide and fluoride chemistries for which the Materials Project uses these values (fitted on oxides): the telluride \ce{LiInTe6Mo6}, the nitride \ce{Sr2LiFe2N3} and the intermetallic \ce{Li2MnRh} of the spot-check, and the cobalt coordination compound \ce{LiCoH24C8N8O12} and the intermetallic control \ce{LiMnIr2} of the prospective campaign; for these the $U$ values are untested, and for \ce{LiMnIr2} the PBE$+U$ lithiated endpoint did not converge (section ``Where the benchmarks rank, and a blind prospective DFT test'').

\subsection{Supercell construction and delithiation}
Supercells are diagonal expansions sized so that every lattice vector is at least $7$~\AA\ (per-axis multiplier $\lceil 7\,\text{\AA} / |\mathbf{a}_i| \rceil$, default cap $300$ atoms; Supplementary Note~3). Because the rule tests the lengths of the lattice vectors rather than the shortest distance between periodic images, non-orthogonal cells can fall below $7$~\AA\ by the second measure: the shortest lattice translation is $5.69$~\AA\ in the rhombohedral LCO supercell, $5.84$~\AA\ in the \ce{NaCoO2} supercell and $3.3$--$5.8$~\AA\ in seven of the twelve prospective cells, while the phosphate, spinel and NMC benchmark cells meet $7$~\AA\ by both measures. Initial magnetic moments follow a block-structured antiferromagnetic pattern ($\pm 5\,\mu_B$ Fe, $\pm 4\,\mu_B$ Mn, $\pm 3\,\mu_B$ V, Cr and Mo, $\pm 2\,\mu_B$ Ni and W, $\pm 0.6\,\mu_B$ Co, $0$ otherwise; in LCO, whose primitive cell holds a single cobalt atom, every Co starts at $+0.6\,\mu_B$), mapped from the primitive cell to the supercell, and are inherited from the previous step's INCAR with the removed lithium's entry dropped (in every chain except LMP and the superseded optPBE-vdW LCO run, whose steps after step~0 were initialized with alternating signs); this preserves the antiferromagnetic sublattice across the curve and avoids the per-atom alternating-sign pattern that caused DFT$+U$ self-consistent-field divergence in early test runs (Supplementary Note~4). Step 0 is relaxed with full cell and ion relaxation (ISIF $= 3$); later steps use ions-only relaxation (ISIF $= 2$) for hosts outside the layered whitelist (the NMC entry among them), which keeps every composition on the cell of the fully lithiated state and avoids Pulay-stress artifacts of repeated cell relaxations (the section ``Benchmarking against experiment and error analysis'' quantifies what this choice costs), and full ISIF $= 3$ for whitelisted layered hosts such as LCO, whose interlayer spacing changes on delithiation. At each step every remaining lithium site is ranked by ALIGNN-FF without symmetry deduplication (the symmetry is broken after the first vacancy), the lowest-energy vacancy is selected and the new configuration is relaxed in VASP; this greedy choice has no optimality guarantee, and its cost is bounded on two cells in the section ``Greedy versus exhaustive lithium-vacancy ordering''.

\subsection{Endpoints, convergence and fallbacks}
The fully lithiated endpoint is the JARVIS-DFT entry as stored (the Discussion notes the one case where that entry is itself partially delithiated); the fully delithiated endpoint is defined operationally as the same framework with every lithium site vacated, and the average voltage is evaluated over the converged range or, where the experiment covers less, over the experimental range, with the range stated (all $16$ lithium atoms for LFP, LMP, LMO and NMC; $x=1\to0.5$ for LCO and for the sodium chain of the section ``Sodium and magnesium chemistries at the DFT tier'', with the full-range values also given; for the magnesium chain, from $16$ to $10$ of its $16$ magnesium atoms, with the final \ce{Mn2O4} step unconverged). Across the five benchmark campaigns all $77$ supercell relaxations reached electronic and ionic convergence, the fully delithiated LMP endpoint only when rerun in October 2026 after a setup error in its pseudopotentials (section ``DFT validation on five commercial cathodes''). In the LCO run of the preprint one converged step, the delithiated endpoint, was rerun under ISIF $=2$ after a 28\% volume expansion; the optB88-vdW$+U$ LCO rerun that replaces it converged at every step without intervention. Apart from the fully delithiated LMP endpoint, ten benchmark relaxations ran in more than one job: one recovery with reduced charge mixing (LFP step~1, whose first run reached the self-consistency cap) and nine continuations or restarts after their jobs ended before convergence; Supplementary Note~17 lists them and gives the number of steps run with each input template and the convergence statistics of the five benchmark chains. No automated fallback exists for an unconverged step. The chain scripts used for the prospective campaign run each relaxation once: a VASP crash stops the chain, whereas a relaxation that reaches the ionic-step limit is recorded, flagged in the job log and passed to the next step, and recovery (restart, mixing change, ISIF change) is a human decision. In the prospective campaign three endpoints and several intermediate steps reached the limit; the three endpoints and the \ce{Li2FeO3} step that ends the reported \ce{Li3FeO3} range were continued from their last structures and wave functions with the conjugate-gradient optimizer, because the quasi-Newton runs had oscillated, and the \ce{LiMnIr2} endpoint was then rerun with the quasi-Newton optimizer from a cold start (section ``Where the benchmarks rank, and a blind prospective DFT test''). The olivine and cell-relaxed endpoints of the sections ``DFT validation on five commercial cathodes'' and ``Benchmarking against experiment and error analysis'' were run separately, with a cold start, and all converged, two of the cell-relaxed ones only after a rerun with the current template (Supplementary Note~17). Since the LCO case the \texttt{next} command warns when a relaxation changes the cell volume by more than $10$\%, so that the condition is flagged before the chain continues; it does not rerun the step.

\subsection{Lithium reference and voltage curve}
The lithium chemical potential is computed in-house with the \ce{Li_{sv}} PAW and $E_\mathrm{cut} = 520$~eV of the cathode runs, Methfessel--Paxton smearing (ISMEAR $= 1$, $\sigma = 0.1$~eV) and a $17^3$ $k$-mesh on BCC lithium, giving $\mu_\mathrm{Li}^\mathrm{PBE} = -1.9031$~eV/atom and $\mu_\mathrm{Li}^\mathrm{optB88\text{-}vdW} = -0.9778$~eV/atom ($-0.9646$~eV/atom with the optPBE-vdW tags of the original LCO run and of the first pass of four prospective runs); each is used with the matching cathode functional. The sodium and magnesium references of the section ``Sodium and magnesium chemistries at the DFT tier'' are computed the same way on BCC sodium (\ce{Na_{pv}}, $17^3$ $k$-mesh) and HCP magnesium (\ce{Mg_{pv}}, $17\times17\times11$), giving $\mu_\mathrm{Na}=-1.3105$ (PBE) and $+0.9196$~eV/atom (optB88-vdW) and $\mu_\mathrm{Mg}=-1.5058$ (PBE) and $+1.1051$~eV/atom (optB88-vdW); the positive vdW-DF values reflect the different absolute energy zero of the nonlocal functional, which cancels in any voltage evaluated within that functional. The reference is recomputed within each functional because only the energy difference between the cathode endpoints and the metal, all evaluated with the same functional and PAW set, enters a voltage; the two absolute values differ by $0.93$~eV/atom for the same BCC cell and are not in themselves inconsistent, although candidates treated with different functionals do sit on different absolute scales (see Discussion). Table~\ref{tab:liref} compares the in-house values with the tabulated JARVIS-DFT value of $-0.925$~eV/atom, which was obtained with the optB88-vdW functional: used with the PBE cathode runs, the tabulated value would shift every PBE voltage in this paper by $+0.98$~V (the LFP polymorph chain from $3.60$ to $4.57$~V), whereas it differs from the in-house optB88-vdW value by only $0.05$~eV/atom, the residual effect of different numerical settings.

\begin{table}[H]
\small
\centering
\caption{Lithium-metal reference energies. In-house values use the \ce{Li_{sv}} PAW, $E_\mathrm{cut}=520$~eV, ISMEAR $=1$, $\sigma=0.1$~eV and a $17^3$ $k$-mesh on BCC lithium, with the same PAW family and cutoff as the cathode runs; the tabulated value is the JARVIS-DFT optB88-vdW energy per atom of its elemental lithium reference entry (JVASP-25117), computed with JARVIS's own numerical settings. The last column is the shift that using the tabulated value with the PBE$+U$ cathode energies would produce in every PBE voltage.}
\label{tab:liref}
\setlength{\tabcolsep}{4pt}
\begin{tabular}{llcc}
\toprule
Reference & Functional & $\mu_\mathrm{Li}$ (eV/atom) & Shift of PBE voltages (V) \\
\midrule
In-house & PBE & $-1.9031$ & 0 \\
In-house & optB88-vdW & $-0.9778$ & n/a \\
In-house & optPBE-vdW & $-0.9646$ & n/a (superseded runs) \\
Tabulated (JARVIS-DFT) & optB88-vdW & $-0.925$ & $+0.98$ \\
\bottomrule
\end{tabular}
\end{table} Plateau voltages follow $V = (E_\mathrm{lo} - E_\mathrm{hi})/\Delta n + \mu_\mathrm{Li}$, where $E_\mathrm{hi}$ and $E_\mathrm{lo}$ are the more- and less-lithiated step energies. Equilibrium voltages are extracted from the lower convex hull of the formation energies $\Delta E(x) = E(x) - x\,E(1) - (1-x)\,E(0)$, and the reported $V_\mathrm{avg}$ is the $\Delta n$-weighted mean over hull plateaus, which equals the two-endpoint average $[E(x_\mathrm{lo})-E(x_\mathrm{hi})]/\Delta n+\mu_\mathrm{Li}$ over the same range because the plateau energies telescope.

\section*{Data availability}

The BatteryMat source code, including the cathode screen (filters and ranking) and DFT input generation, is openly available at \href{https://github.com/atomgptlab/batterymat}{\texttt{https://github.com/}\allowbreak\texttt{atomgptlab/}\allowbreak\texttt{batterymat}}. The DFT input files (POSCAR, INCAR, KPOINTS) and recorded energies (energies.json) for the five validated cathodes and for the comparison, prospective, spot-check, sodium and magnesium calculations, including the twelve prospective materials, are in the same repository. Each INCAR in the repository is the one used for the run that produced the recorded energy, as echoed in the header of its OUTCAR (for the fully lithiated LMP endpoint, whose recorded energy is the recompute in the magnetic order of the rest of its chain, that run's inputs are in the directory named in its \texttt{energies.json} entry). The ALIGNN-FF training labels for the lithium-containing JARVIS-DFT subset, the trained ALIGNN model weights and the OUTCAR files are available from the corresponding author on reasonable request. The underlying JARVIS-DFT structures are publicly available at \url{https://jarvis.nist.gov}.

\section*{Code availability}

The cathode-screening, DFT-tier and analysis code is available at \href{https://github.com/atomgptlab/batterymat}{\texttt{https://github.com/}\allowbreak\texttt{atomgptlab/}\allowbreak\texttt{batterymat}} under the MIT license; file paths given in the text and the Supplementary Information are relative to \texttt{batterymat/\allowbreak screening\_cathode/} in that repository unless stated otherwise. The screening-pass voltages used in this paper are distributed there as the files \texttt{batterymat/\allowbreak average\_voltage/\allowbreak[Ion]\_min.csv}; the script \texttt{batterymat/\allowbreak screening\_anode/\allowbreak screen.py} in the same repository is an earlier database-cell variant that enters the metal reference energy with the opposite sign to the voltage equation of the Methods and was not used to produce them. The label-generation script, the regressor configuration and partition lists, and the Alexandria-branch scripts are available from the corresponding author on reasonable request. A web demonstration is available at \url{https://atomgpt.org/battery}.

\section*{Author contributions}

J.L. designed and implemented the BatteryMat pipeline, ran the ALIGNN-FF and DFT campaigns, and wrote the manuscript. C.R.C.\ contributed to the DFT validation and analysis. K.Z.\ enabled the battery discovery tool in the AtomGPT user chat interface. K.C.\ supervised the project, integrated the AtomGPT Battery Explorer interface, and revised the manuscript. All authors discussed the results and approved the final version.

\section*{Competing interests}

Kamal Choudhary is an Associate Editor of \textit{npj Computational Materials}. Kamal Choudhary was not involved in the journal's review of, or decisions related to, this manuscript. The other authors do not have a competing interest.

\section*{Acknowledgements}

The authors acknowledge the Skipjack high-performance computing cluster at Johns Hopkins University for computational resources.

\section*{Funding}

The authors received no specific funding for this work.

\bibliography{ref}

\begin{thebibliography}{10}
\expandafter\ifx\csname url\endcsname\relax
  \def\url#1{\texttt{#1}}\fi
\expandafter\ifx\csname urlprefix\endcsname\relax\def\urlprefix{URL }\fi
\providecommand{\bibinfo}[2]{#2}
\providecommand{\eprint}[2][]{\url{#2}}

\bibitem{whittingham1976}
\bibinfo{author}{Whittingham, M.~S.}
\newblock \bibinfo{title}{Electrical energy storage and intercalation
  chemistry}.
\newblock \emph{\bibinfo{journal}{Science}} \textbf{\bibinfo{volume}{192}},
  \bibinfo{pages}{1126--1127} (\bibinfo{year}{1976}).

\bibitem{mizushima1980lco}
\bibinfo{author}{Mizushima, K.}, \bibinfo{author}{Jones, P.~C.},
  \bibinfo{author}{Wiseman, P.~J.} \& \bibinfo{author}{Goodenough, J.~B.}
\newblock \bibinfo{title}{{Li$_x$CoO$_2$} ($0<x\leq 1$): A new cathode material
  for batteries of high energy density}.
\newblock \emph{\bibinfo{journal}{Mater. Res. Bull.}}
  \textbf{\bibinfo{volume}{15}}, \bibinfo{pages}{783--789}
  (\bibinfo{year}{1980}).

\bibitem{padhi1997lfp}
\bibinfo{author}{Padhi, A.~K.}, \bibinfo{author}{Nanjundaswamy, K.~S.} \&
  \bibinfo{author}{Goodenough, J.~B.}
\newblock \bibinfo{title}{Phospho-olivines as positive-electrode materials for
  rechargeable lithium batteries}.
\newblock \emph{\bibinfo{journal}{J. Electrochem. Soc.}}
  \textbf{\bibinfo{volume}{144}}, \bibinfo{pages}{1188--1194}
  (\bibinfo{year}{1997}).

\bibitem{ohzuku2001nmc}
\bibinfo{author}{Ohzuku, T.} \& \bibinfo{author}{Makimura, Y.}
\newblock \bibinfo{title}{Layered lithium insertion material of
  {LiCo$_{1/3}$Ni$_{1/3}$Mn$_{1/3}$O$_2$} for lithium-ion batteries}.
\newblock \emph{\bibinfo{journal}{Chem. Lett.}} \textbf{\bibinfo{volume}{30}},
  \bibinfo{pages}{642--643} (\bibinfo{year}{2001}).

\bibitem{yoshino2012}
\bibinfo{author}{Yoshino, A.}
\newblock \bibinfo{title}{The birth of the lithium-ion battery}.
\newblock \emph{\bibinfo{journal}{Angew. Chem. Int. Ed.}}
  \textbf{\bibinfo{volume}{51}}, \bibinfo{pages}{5798--5800}
  (\bibinfo{year}{2012}).

\bibitem{goodenough2013}
\bibinfo{author}{Goodenough, J.~B.} \& \bibinfo{author}{Park, K.-S.}
\newblock \bibinfo{title}{The {L}i-ion rechargeable battery: A perspective}.
\newblock \emph{\bibinfo{journal}{J. Am. Chem. Soc.}}
  \textbf{\bibinfo{volume}{135}}, \bibinfo{pages}{1167--1176}
  (\bibinfo{year}{2013}).

\bibitem{li2024pipeline}
\bibinfo{author}{Li, H.~H.}, \bibinfo{author}{Shen, J.-X.} \&
  \bibinfo{author}{Persson, K.~A.}
\newblock \bibinfo{title}{A rapid lithium-ion cathode discovery pipeline and
  its exemplary application}.
\newblock \emph{\bibinfo{journal}{Energy Adv.}} \textbf{\bibinfo{volume}{3}},
  \bibinfo{pages}{255--262} (\bibinfo{year}{2024}).

\bibitem{goodenough2010challenges}
\bibinfo{author}{Goodenough, J.~B.} \& \bibinfo{author}{Kim, Y.}
\newblock \bibinfo{title}{Challenges for rechargeable {Li} batteries}.
\newblock \emph{\bibinfo{journal}{Chem. Mater.}} \textbf{\bibinfo{volume}{22}},
  \bibinfo{pages}{587--603} (\bibinfo{year}{2010}).

\bibitem{aydinol1997}
\bibinfo{author}{Aydinol, M.~K.}, \bibinfo{author}{Kohan, A.~F.},
  \bibinfo{author}{Ceder, G.}, \bibinfo{author}{Cho, K.} \&
  \bibinfo{author}{Joannopoulos, J.}
\newblock \bibinfo{title}{\textit{Ab initio} study of lithium intercalation in
  metal oxides and metal dichalcogenides}.
\newblock \emph{\bibinfo{journal}{Phys. Rev. B}} \textbf{\bibinfo{volume}{56}},
  \bibinfo{pages}{1354--1365} (\bibinfo{year}{1997}).

\bibitem{zhou2004ldau}
\bibinfo{author}{Zhou, F.}, \bibinfo{author}{Cococcioni, M.},
  \bibinfo{author}{Marianetti, C.~A.}, \bibinfo{author}{Morgan, D.} \&
  \bibinfo{author}{Ceder, G.}
\newblock \bibinfo{title}{First-principles prediction of redox potentials in
  transition-metal compounds with {LDA}+{$U$}}.
\newblock \emph{\bibinfo{journal}{Phys. Rev. B}} \textbf{\bibinfo{volume}{70}},
  \bibinfo{pages}{235121} (\bibinfo{year}{2004}).

\bibitem{jain2011mphubbard}
\bibinfo{author}{Jain, A.} \emph{et~al.}
\newblock \bibinfo{title}{Formation enthalpies by mixing {GGA} and {GGA}+{$U$}
  calculations}.
\newblock \emph{\bibinfo{journal}{Phys. Rev. B}} \textbf{\bibinfo{volume}{84}},
  \bibinfo{pages}{045115} (\bibinfo{year}{2011}).

\bibitem{hautier2011phosphates}
\bibinfo{author}{Hautier, G.} \emph{et~al.}
\newblock \bibinfo{title}{Phosphates as lithium-ion battery cathodes: An
  evaluation based on high-throughput {\textit{ab initio}} calculations}.
\newblock \emph{\bibinfo{journal}{Chem. Mater.}} \textbf{\bibinfo{volume}{23}},
  \bibinfo{pages}{3495--3508} (\bibinfo{year}{2011}).

\bibitem{jain2013materialsproject}
\bibinfo{author}{Jain, A.} \emph{et~al.}
\newblock \bibinfo{title}{Commentary: The {M}aterials {P}roject: A materials
  genome approach to accelerating materials innovation}.
\newblock \emph{\bibinfo{journal}{APL Mater.}} \textbf{\bibinfo{volume}{1}},
  \bibinfo{pages}{011002} (\bibinfo{year}{2013}).

\bibitem{curtarolo2012aflow}
\bibinfo{author}{Curtarolo, S.} \emph{et~al.}
\newblock \bibinfo{title}{{AFLOW}: An automatic framework for high-throughput
  materials discovery}.
\newblock \emph{\bibinfo{journal}{Comput. Mater. Sci.}}
  \textbf{\bibinfo{volume}{58}}, \bibinfo{pages}{218--226}
  (\bibinfo{year}{2012}).

\bibitem{saal2013oqmd}
\bibinfo{author}{Saal, J.~E.}, \bibinfo{author}{Kirklin, S.},
  \bibinfo{author}{Aykol, M.}, \bibinfo{author}{Meredig, B.} \&
  \bibinfo{author}{Wolverton, C.}
\newblock \bibinfo{title}{Materials design and discovery with high-throughput
  density functional theory: The {O}pen {Q}uantum {M}aterials {D}atabase
  ({OQMD})}.
\newblock \emph{\bibinfo{journal}{JOM}} \textbf{\bibinfo{volume}{65}},
  \bibinfo{pages}{1501--1509} (\bibinfo{year}{2013}).

\bibitem{choudhary2020jarvis}
\bibinfo{author}{Choudhary, K.} \emph{et~al.}
\newblock \bibinfo{title}{The joint automated repository for various integrated
  simulations ({JARVIS}) for data-driven materials design}.
\newblock \emph{\bibinfo{journal}{npj Comput. Mater.}}
  \textbf{\bibinfo{volume}{6}}, \bibinfo{pages}{173} (\bibinfo{year}{2020}).

\bibitem{choudhary2025jarvis}
\bibinfo{author}{Choudhary, K.}
\newblock \bibinfo{title}{The {JARVIS} infrastructure is all you need for
  materials design}.
\newblock \emph{\bibinfo{journal}{Comput. Mater. Sci.}}
  \textbf{\bibinfo{volume}{259}}, \bibinfo{pages}{114063}
  (\bibinfo{year}{2025}).

\bibitem{ong2013pymatgen}
\bibinfo{author}{Ong, S.~P.} \emph{et~al.}
\newblock \bibinfo{title}{Python materials genomics (pymatgen): A robust,
  open-source python library for materials analysis}.
\newblock \emph{\bibinfo{journal}{Comput. Mater. Sci.}}
  \textbf{\bibinfo{volume}{68}}, \bibinfo{pages}{314--319}
  (\bibinfo{year}{2013}).

\bibitem{xie2018cgcnn}
\bibinfo{author}{Xie, T.} \& \bibinfo{author}{Grossman, J.~C.}
\newblock \bibinfo{title}{Crystal graph convolutional neural networks for an
  accurate and interpretable prediction of material properties}.
\newblock \emph{\bibinfo{journal}{Phys. Rev. Lett.}}
  \textbf{\bibinfo{volume}{120}}, \bibinfo{pages}{145301}
  (\bibinfo{year}{2018}).

\bibitem{chen2019megnet}
\bibinfo{author}{Chen, C.}, \bibinfo{author}{Ye, W.}, \bibinfo{author}{Zuo,
  Y.}, \bibinfo{author}{Zheng, C.} \& \bibinfo{author}{Ong, S.~P.}
\newblock \bibinfo{title}{Graph networks as a universal machine learning
  framework for molecules and crystals}.
\newblock \emph{\bibinfo{journal}{Chem. Mater.}} \textbf{\bibinfo{volume}{31}},
  \bibinfo{pages}{3564--3572} (\bibinfo{year}{2019}).

\bibitem{choudhary2021alignn}
\bibinfo{author}{Choudhary, K.} \& \bibinfo{author}{DeCost, B.}
\newblock \bibinfo{title}{Atomistic line graph neural network for improved
  materials property predictions}.
\newblock \emph{\bibinfo{journal}{npj Comput. Mater.}}
  \textbf{\bibinfo{volume}{7}}, \bibinfo{pages}{185} (\bibinfo{year}{2021}).

\bibitem{chen2022m3gnet}
\bibinfo{author}{Chen, C.} \& \bibinfo{author}{Ong, S.~P.}
\newblock \bibinfo{title}{A universal graph deep learning interatomic potential
  for the periodic table}.
\newblock \emph{\bibinfo{journal}{Nat. Comput. Sci.}}
  \textbf{\bibinfo{volume}{2}}, \bibinfo{pages}{718--728}
  (\bibinfo{year}{2022}).

\bibitem{deng2023chgnet}
\bibinfo{author}{Deng, B.} \emph{et~al.}
\newblock \bibinfo{title}{{CHG}{N}et as a pretrained universal neural network
  potential for charge-informed atomistic modelling}.
\newblock \emph{\bibinfo{journal}{Nat. Mach. Intell.}}
  \textbf{\bibinfo{volume}{5}}, \bibinfo{pages}{1031--1041}
  (\bibinfo{year}{2023}).

\bibitem{batatia2022mace}
\bibinfo{author}{Batatia, I.}, \bibinfo{author}{Kov{\'a}cs, D.~P.},
  \bibinfo{author}{Simm, G. N.~C.}, \bibinfo{author}{Ortner, C.} \&
  \bibinfo{author}{Cs{\'a}nyi, G.}
\newblock \bibinfo{title}{{MACE}: Higher order equivariant message passing
  neural networks for fast and accurate force fields}.
\newblock In \emph{\bibinfo{booktitle}{Advances in Neural Information
  Processing Systems}}, vol.~\bibinfo{volume}{35},
  \bibinfo{pages}{11423--11436} (\bibinfo{year}{2022}).

\bibitem{merchant2023gnome}
\bibinfo{author}{Merchant, A.} \emph{et~al.}
\newblock \bibinfo{title}{Scaling deep learning for materials discovery}.
\newblock \emph{\bibinfo{journal}{Nature}} \textbf{\bibinfo{volume}{624}},
  \bibinfo{pages}{80--85} (\bibinfo{year}{2023}).

\bibitem{choudhary2023alignnff}
\bibinfo{author}{Choudhary, K.} \emph{et~al.}
\newblock \bibinfo{title}{Unified graph neural network force-field for the
  periodic table: solid state applications}.
\newblock \emph{\bibinfo{journal}{Digit. Discov.}}
  \textbf{\bibinfo{volume}{2}}, \bibinfo{pages}{346--355}
  (\bibinfo{year}{2023}).

\bibitem{joshi2019voltage}
\bibinfo{author}{Joshi, R.~P.} \emph{et~al.}
\newblock \bibinfo{title}{Machine learning the voltage of electrode materials
  in metal-ion batteries}.
\newblock \emph{\bibinfo{journal}{ACS Appl. Mater. Interfaces}}
  \textbf{\bibinfo{volume}{11}}, \bibinfo{pages}{18494--18503}
  (\bibinfo{year}{2019}).

\bibitem{louis2022voltage}
\bibinfo{author}{Louis, S.-Y.} \emph{et~al.}
\newblock \bibinfo{title}{Accurate prediction of voltage of battery electrode
  materials using attention-based graph neural networks}.
\newblock \emph{\bibinfo{journal}{ACS Appl. Mater. Interfaces}}
  \textbf{\bibinfo{volume}{14}}, \bibinfo{pages}{26587--26594}
  (\bibinfo{year}{2022}).

\bibitem{delacourt2004lmp}
\bibinfo{author}{Delacourt, C.}, \bibinfo{author}{Poizot, P.},
  \bibinfo{author}{Morcrette, M.}, \bibinfo{author}{Tarascon, J.-M.} \&
  \bibinfo{author}{Masquelier, C.}
\newblock \bibinfo{title}{One-step low-temperature route for the preparation of
  electrochemically active {LiMnPO$_4$} powders}.
\newblock \emph{\bibinfo{journal}{Chem. Mater.}} \textbf{\bibinfo{volume}{16}},
  \bibinfo{pages}{93--99} (\bibinfo{year}{2004}).

\bibitem{tomon2022lmo}
\bibinfo{author}{Tomon, C.} \emph{et~al.}
\newblock \bibinfo{title}{Core-shell structure of {LiMn$_2$O$_4$} cathode
  material reduces phase transition and {Mn} dissolution in {Li}-ion
  batteries}.
\newblock \emph{\bibinfo{journal}{Commun. Chem.}} \textbf{\bibinfo{volume}{5}},
  \bibinfo{pages}{54} (\bibinfo{year}{2022}).

\bibitem{manthiram2020cathode}
\bibinfo{author}{Manthiram, A.}
\newblock \bibinfo{title}{A reflection on lithium-ion battery cathode
  chemistry}.
\newblock \emph{\bibinfo{journal}{Nat. Commun.}} \textbf{\bibinfo{volume}{11}},
  \bibinfo{pages}{1550} (\bibinfo{year}{2020}).

\bibitem{schmidt2023alexandria}
\bibinfo{author}{Schmidt, J.} \emph{et~al.}
\newblock \bibinfo{title}{Machine-learning-assisted determination of the global
  zero-temperature phase diagram of materials}.
\newblock \emph{\bibinfo{journal}{Adv. Mater.}} \textbf{\bibinfo{volume}{35}},
  \bibinfo{pages}{2210788} (\bibinfo{year}{2023}).

\bibitem{klimes2010optb88}
\bibinfo{author}{Klime{\v{s}}, J.}, \bibinfo{author}{Bowler, D.~R.} \&
  \bibinfo{author}{Michaelides, A.}
\newblock \bibinfo{title}{Chemical accuracy for the van der {W}aals density
  functional}.
\newblock \emph{\bibinfo{journal}{J. Phys.: Condens. Matter}}
  \textbf{\bibinfo{volume}{22}}, \bibinfo{pages}{022201}
  (\bibinfo{year}{2010}).

\bibitem{klimes2011optb88}
\bibinfo{author}{Klime{\v{s}}, J.}, \bibinfo{author}{Bowler, D.~R.} \&
  \bibinfo{author}{Michaelides, A.}
\newblock \bibinfo{title}{Van der {W}aals density functionals applied to
  solids}.
\newblock \emph{\bibinfo{journal}{Phys. Rev. B}} \textbf{\bibinfo{volume}{83}},
  \bibinfo{pages}{195131} (\bibinfo{year}{2011}).

\bibitem{kresse1996vasp}
\bibinfo{author}{Kresse, G.} \& \bibinfo{author}{Furthm{\"u}ller, J.}
\newblock \bibinfo{title}{Efficient iterative schemes for \textit{ab initio}
  total-energy calculations using a plane-wave basis set}.
\newblock \emph{\bibinfo{journal}{Phys. Rev. B}} \textbf{\bibinfo{volume}{54}},
  \bibinfo{pages}{11169--11186} (\bibinfo{year}{1996}).

\bibitem{blochl1994paw}
\bibinfo{author}{Bl{\"o}chl, P.~E.}
\newblock \bibinfo{title}{Projector augmented-wave method}.
\newblock \emph{\bibinfo{journal}{Phys. Rev. B}} \textbf{\bibinfo{volume}{50}},
  \bibinfo{pages}{17953--17979} (\bibinfo{year}{1994}).

\bibitem{kresse1999paw}
\bibinfo{author}{Kresse, G.} \& \bibinfo{author}{Joubert, D.}
\newblock \bibinfo{title}{From ultrasoft pseudopotentials to the projector
  augmented-wave method}.
\newblock \emph{\bibinfo{journal}{Phys. Rev. B}} \textbf{\bibinfo{volume}{59}},
  \bibinfo{pages}{1758--1775} (\bibinfo{year}{1999}).

\bibitem{yoshizawa2003nmc}
\bibinfo{author}{Yoshizawa, H.} \& \bibinfo{author}{Ohzuku, T.}
\newblock \bibinfo{title}{Prismatic lithium-ion batteries of
  {LiCo$_{1/3}$Ni$_{1/3}$Mn$_{1/3}$O$_2$} with graphite}.
\newblock \emph{\bibinfo{journal}{Electrochemistry}}
  \textbf{\bibinfo{volume}{71}}, \bibinfo{pages}{1177--1181}
  (\bibinfo{year}{2003}).

\bibitem{kondo2018lfp}
\bibinfo{author}{Kondo, H.}, \bibinfo{author}{Sasaki, T.},
  \bibinfo{author}{Barai, P.} \& \bibinfo{author}{Srinivasan, V.}
\newblock \bibinfo{title}{Comprehensive study of the polarization behavior of
  {LiFePO$_4$} electrodes based on a many-particle model}.
\newblock \emph{\bibinfo{journal}{J. Electrochem. Soc.}}
  \textbf{\bibinfo{volume}{165}}, \bibinfo{pages}{A2047--A2057}
  (\bibinfo{year}{2018}).

\bibitem{aravindan2013lmp}
\bibinfo{author}{Aravindan, V.}, \bibinfo{author}{Gnanaraj, J.},
  \bibinfo{author}{Lee, Y.} \& \bibinfo{author}{Madhavi, S.}
\newblock \bibinfo{title}{{LiMnPO$_4$} -- a next generation cathode material
  for lithium-ion batteries}.
\newblock \emph{\bibinfo{journal}{J. Mater. Chem. A}}
  \textbf{\bibinfo{volume}{1}}, \bibinfo{pages}{3518--3539}
  (\bibinfo{year}{2013}).

\bibitem{ohzuku1990lmo}
\bibinfo{author}{Ohzuku, T.}, \bibinfo{author}{Kitagawa, M.} \&
  \bibinfo{author}{Hirai, T.}
\newblock \bibinfo{title}{Electrochemistry of manganese dioxide in lithium
  nonaqueous cell. {III}. {X}-ray diffractional study on the reduction of
  spinel-related manganese dioxide}.
\newblock \emph{\bibinfo{journal}{J. Electrochem. Soc.}}
  \textbf{\bibinfo{volume}{137}}, \bibinfo{pages}{769--775}
  (\bibinfo{year}{1990}).

\bibitem{yabuuchi2004nmc}
\bibinfo{author}{Yabuuchi, N.} \& \bibinfo{author}{Ohzuku, T.}
\newblock \bibinfo{title}{Solid-state redox reactions of
  {LiCo$_{1/3}$Ni$_{1/3}$Mn$_{1/3}$O$_2$} for advanced lithium-ion batteries}.
\newblock \bibinfo{howpublished}{206th Meeting of The Electrochemical Society
  (2004 Joint International Meeting, Honolulu), Abstract 324,
  \url{https://www.electrochem.org/dl/ma/206/pdfs/0324.pdf}}
  (\bibinfo{year}{2004}).

\bibitem{ohzuku1994lco}
\bibinfo{author}{Ohzuku, T.} \& \bibinfo{author}{Ueda, A.}
\newblock \bibinfo{title}{Solid-state redox reactions of {LiCoO$_2$}
  ({R$\bar{3}$m}) for 4 volt secondary lithium cells}.
\newblock \emph{\bibinfo{journal}{J. Electrochem. Soc.}}
  \textbf{\bibinfo{volume}{141}}, \bibinfo{pages}{2972--2977}
  (\bibinfo{year}{1994}).

\bibitem{yamada2001lfp}
\bibinfo{author}{Yamada, A.}, \bibinfo{author}{Chung, S.~C.} \&
  \bibinfo{author}{Hinokuma, K.}
\newblock \bibinfo{title}{Optimized {LiFePO$_4$} for lithium battery cathodes}.
\newblock \emph{\bibinfo{journal}{J. Electrochem. Soc.}}
  \textbf{\bibinfo{volume}{148}}, \bibinfo{pages}{A224--A229}
  (\bibinfo{year}{2001}).

\bibitem{mp_hubbard_docs}
\bibinfo{author}{{Materials Project}}.
\newblock \bibinfo{title}{Hubbard {U} values}.
\newblock
  \bibinfo{howpublished}{\url{https://docs.materialsproject.org/methodology/materials-methodology/calculation-details/gga+u-calculations/hubbard-u-values}}
  (\bibinfo{year}{2026}).
\newblock \bibinfo{note}{Accessed 21 October 2026}.

\bibitem{wang2006ggau}
\bibinfo{author}{Wang, L.}, \bibinfo{author}{Maxisch, T.} \&
  \bibinfo{author}{Ceder, G.}
\newblock \bibinfo{title}{Oxidation energies of transition metal oxides within
  the {GGA}+{$U$} framework}.
\newblock \emph{\bibinfo{journal}{Phys. Rev. B}} \textbf{\bibinfo{volume}{73}},
  \bibinfo{pages}{195107} (\bibinfo{year}{2006}).

\bibitem{reimers1992lco}
\bibinfo{author}{Reimers, J.~N.} \& \bibinfo{author}{Dahn, J.~R.}
\newblock \bibinfo{title}{Electrochemical and \textit{in situ} {X}-ray
  diffraction studies of lithium intercalation in {Li$_x$CoO$_2$}}.
\newblock \emph{\bibinfo{journal}{J. Electrochem. Soc.}}
  \textbf{\bibinfo{volume}{139}}, \bibinfo{pages}{2091--2097}
  (\bibinfo{year}{1992}).

\bibitem{tan2021lco}
\bibinfo{author}{Tan, J.} \emph{et~al.}
\newblock \bibinfo{title}{Electrochemically driven phase transition in
  {LiCoO$_2$} cathode}.
\newblock \emph{\bibinfo{journal}{Materials}} \textbf{\bibinfo{volume}{14}},
  \bibinfo{pages}{242} (\bibinfo{year}{2021}).

\bibitem{zhou2004olivine}
\bibinfo{author}{Zhou, F.}, \bibinfo{author}{Cococcioni, M.},
  \bibinfo{author}{Kang, K.} \& \bibinfo{author}{Ceder, G.}
\newblock \bibinfo{title}{The {Li} intercalation potential of {LiMPO$_4$} and
  {LiMSiO$_4$} olivines with {M} = {Fe}, {Mn}, {Co}, {Ni}}.
\newblock \emph{\bibinfo{journal}{Electrochem. Commun.}}
  \textbf{\bibinfo{volume}{6}}, \bibinfo{pages}{1144--1148}
  (\bibinfo{year}{2004}).

\bibitem{furness2020r2scan}
\bibinfo{author}{Furness, J.~W.}, \bibinfo{author}{Kaplan, A.~D.},
  \bibinfo{author}{Ning, J.}, \bibinfo{author}{Perdew, J.~P.} \&
  \bibinfo{author}{Sun, J.}
\newblock \bibinfo{title}{Accurate and numerically efficient r$^2${SCAN}
  meta-generalized gradient approximation}.
\newblock \emph{\bibinfo{journal}{J. Phys. Chem. Lett.}}
  \textbf{\bibinfo{volume}{11}}, \bibinfo{pages}{8208--8215}
  (\bibinfo{year}{2020}).

\bibitem{manthiram2017outlook}
\bibinfo{author}{Manthiram, A.}
\newblock \bibinfo{title}{An outlook on lithium ion battery technology}.
\newblock \emph{\bibinfo{journal}{ACS Cent. Sci.}}
  \textbf{\bibinfo{volume}{3}}, \bibinfo{pages}{1063--1069}
  (\bibinfo{year}{2017}).

\bibitem{lei2014nacoo2}
\bibinfo{author}{Lei, Y.}, \bibinfo{author}{Li, X.}, \bibinfo{author}{Liu, L.}
  \& \bibinfo{author}{Ceder, G.}
\newblock \bibinfo{title}{Synthesis and stoichiometry of different layered
  sodium cobalt oxides}.
\newblock \emph{\bibinfo{journal}{Chem. Mater.}} \textbf{\bibinfo{volume}{26}},
  \bibinfo{pages}{5288--5296} (\bibinfo{year}{2014}).

\bibitem{delmas1981nacoo2}
\bibinfo{author}{Delmas, C.}, \bibinfo{author}{Braconnier, J.~J.},
  \bibinfo{author}{Fouassier, C.} \& \bibinfo{author}{Hagenmuller, P.}
\newblock \bibinfo{title}{Electrochemical intercalation of sodium in
  {Na$_x$CoO$_2$} bronzes}.
\newblock \emph{\bibinfo{journal}{Solid State Ionics}}
  \textbf{\bibinfo{volume}{3--4}}, \bibinfo{pages}{165--169}
  (\bibinfo{year}{1981}).

\bibitem{okamoto2015mgspinel}
\bibinfo{author}{Okamoto, S.} \emph{et~al.}
\newblock \bibinfo{title}{Intercalation and push-out process with
  spinel-to-rocksalt transition on {Mg} insertion into spinel oxides in
  magnesium batteries}.
\newblock \emph{\bibinfo{journal}{Adv. Sci.}} \textbf{\bibinfo{volume}{2}},
  \bibinfo{pages}{1500072} (\bibinfo{year}{2015}).

\bibitem{hatakeyama2019mgmn2o4}
\bibinfo{author}{Hatakeyama, T.} \emph{et~al.}
\newblock \bibinfo{title}{Electrochemical phase transformation accompanied with
  {Mg} extraction and insertion in a spinel {MgMn$_2$O$_4$} cathode material}.
\newblock \emph{\bibinfo{journal}{Phys. Chem. Chem. Phys.}}
  \textbf{\bibinfo{volume}{21}}, \bibinfo{pages}{23749--23757}
  (\bibinfo{year}{2019}).

\bibitem{choudhary2024atomgpt}
\bibinfo{author}{Choudhary, K.}
\newblock \bibinfo{title}{{AtomGPT}: Atomistic generative pretrained
  transformer for forward and inverse materials design}.
\newblock \emph{\bibinfo{journal}{J. Phys. Chem. Lett.}}
  \textbf{\bibinfo{volume}{15}}, \bibinfo{pages}{6909--6917}
  (\bibinfo{year}{2024}).

\bibitem{lee2026agapi}
\bibinfo{author}{Lee, J.} \emph{et~al.}
\newblock \bibinfo{title}{{AGAPI-Agents}: An open-access agentic {AI} platform
  for accelerated materials design on {AtomGPT.org}}.
\newblock \emph{\bibinfo{journal}{J. Phys. Chem. Lett.}}
  \textbf{\bibinfo{volume}{17}}, \bibinfo{pages}{7221--7231}
  (\bibinfo{year}{2026}).

\bibitem{choudhary2025slakonet}
\bibinfo{author}{Choudhary, K.}
\newblock \bibinfo{title}{{SlaKoNet}: A unified {Slater}--{Koster}
  tight-binding framework using neural network infrastructure for the periodic
  table}.
\newblock \emph{\bibinfo{journal}{J. Phys. Chem. Lett.}}
  \textbf{\bibinfo{volume}{16}}, \bibinfo{pages}{11109--11119}
  (\bibinfo{year}{2025}).

\bibitem{kaplan2025matpes}
\bibinfo{author}{Kaplan, A.~D.} \emph{et~al.}
\newblock \bibinfo{title}{A foundational potential energy surface dataset for
  materials}.
\newblock \emph{\bibinfo{journal}{arXiv preprint arXiv:2503.04070}}
  (\bibinfo{year}{2025}).

\bibitem{sun2016metastability}
\bibinfo{author}{Sun, W.} \emph{et~al.}
\newblock \bibinfo{title}{The thermodynamic scale of inorganic crystalline
  metastability}.
\newblock \emph{\bibinfo{journal}{Sci. Adv.}} \textbf{\bibinfo{volume}{2}},
  \bibinfo{pages}{e1600225} (\bibinfo{year}{2016}).

\bibitem{perdew1996pbe}
\bibinfo{author}{Perdew, J.~P.}, \bibinfo{author}{Burke, K.} \&
  \bibinfo{author}{Ernzerhof, M.}
\newblock \bibinfo{title}{Generalized gradient approximation made simple}.
\newblock \emph{\bibinfo{journal}{Phys. Rev. Lett.}}
  \textbf{\bibinfo{volume}{77}}, \bibinfo{pages}{3865--3868}
  (\bibinfo{year}{1996}).

\bibitem{dudarev1998ldau}
\bibinfo{author}{Dudarev, S.~L.}, \bibinfo{author}{Botton, G.~A.},
  \bibinfo{author}{Savrasov, S.~Y.}, \bibinfo{author}{Humphreys, C.~J.} \&
  \bibinfo{author}{Sutton, A.~P.}
\newblock \bibinfo{title}{Electron-energy-loss spectra and the structural
  stability of nickel oxide: An {LSDA+U} study}.
\newblock \emph{\bibinfo{journal}{Phys. Rev. B}} \textbf{\bibinfo{volume}{57}},
  \bibinfo{pages}{1505--1509} (\bibinfo{year}{1998}).

\end{thebibliography}


\begin{thebibliography}{10}
\expandafter\ifx\csname url\endcsname\relax
  \def\url#1{\texttt{#1}}\fi
\expandafter\ifx\csname urlprefix\endcsname\relax\def\urlprefix{URL }\fi
\providecommand{\bibinfo}[2]{#2}
\providecommand{\eprint}[2][]{\url{#2}}

\bibitem{si-choudhary2021alignn}
\bibinfo{author}{Choudhary, K.} \& \bibinfo{author}{DeCost, B.}
\newblock \bibinfo{title}{Atomistic line graph neural network for improved
  materials property predictions}.
\newblock \emph{\bibinfo{journal}{npj Comput. Mater.}}
  \textbf{\bibinfo{volume}{7}}, \bibinfo{pages}{185} (\bibinfo{year}{2021}).

\bibitem{si-choudhary2023alignnff}
\bibinfo{author}{Choudhary, K.} \emph{et~al.}
\newblock \bibinfo{title}{Unified graph neural network force-field for the
  periodic table: solid state applications}.
\newblock \emph{\bibinfo{journal}{Digit. Discov.}}
  \textbf{\bibinfo{volume}{2}}, \bibinfo{pages}{346--355}
  (\bibinfo{year}{2023}).

\bibitem{si-kresse1996vasp}
\bibinfo{author}{Kresse, G.} \& \bibinfo{author}{Furthm{\"u}ller, J.}
\newblock \bibinfo{title}{Efficient iterative schemes for \textit{ab initio}
  total-energy calculations using a plane-wave basis set}.
\newblock \emph{\bibinfo{journal}{Phys. Rev. B}} \textbf{\bibinfo{volume}{54}},
  \bibinfo{pages}{11169--11186} (\bibinfo{year}{1996}).

\bibitem{si-blochl1994paw}
\bibinfo{author}{Bl{\"o}chl, P.~E.}
\newblock \bibinfo{title}{Projector augmented-wave method}.
\newblock \emph{\bibinfo{journal}{Phys. Rev. B}} \textbf{\bibinfo{volume}{50}},
  \bibinfo{pages}{17953--17979} (\bibinfo{year}{1994}).

\bibitem{si-kresse1999paw}
\bibinfo{author}{Kresse, G.} \& \bibinfo{author}{Joubert, D.}
\newblock \bibinfo{title}{From ultrasoft pseudopotentials to the projector
  augmented-wave method}.
\newblock \emph{\bibinfo{journal}{Phys. Rev. B}} \textbf{\bibinfo{volume}{59}},
  \bibinfo{pages}{1758--1775} (\bibinfo{year}{1999}).

\bibitem{si-klimes2010optb88}
\bibinfo{author}{Klime{\v{s}}, J.}, \bibinfo{author}{Bowler, D.~R.} \&
  \bibinfo{author}{Michaelides, A.}
\newblock \bibinfo{title}{Chemical accuracy for the van der {W}aals density
  functional}.
\newblock \emph{\bibinfo{journal}{J. Phys.: Condens. Matter}}
  \textbf{\bibinfo{volume}{22}}, \bibinfo{pages}{022201}
  (\bibinfo{year}{2010}).

\bibitem{si-klimes2011optb88}
\bibinfo{author}{Klime{\v{s}}, J.}, \bibinfo{author}{Bowler, D.~R.} \&
  \bibinfo{author}{Michaelides, A.}
\newblock \bibinfo{title}{Van der {W}aals density functionals applied to
  solids}.
\newblock \emph{\bibinfo{journal}{Phys. Rev. B}} \textbf{\bibinfo{volume}{83}},
  \bibinfo{pages}{195131} (\bibinfo{year}{2011}).

\bibitem{si-mp_hubbard_docs}
\bibinfo{author}{{Materials Project}}.
\newblock \bibinfo{title}{Hubbard {U} values}.
\newblock
  \bibinfo{howpublished}{\url{https://docs.materialsproject.org/methodology/materials-methodology/calculation-details/gga+u-calculations/hubbard-u-values}}
  (\bibinfo{year}{2026}).
\newblock \bibinfo{note}{Accessed 21 October 2026}.

\bibitem{si-wang2006ggau}
\bibinfo{author}{Wang, L.}, \bibinfo{author}{Maxisch, T.} \&
  \bibinfo{author}{Ceder, G.}
\newblock \bibinfo{title}{Oxidation energies of transition metal oxides within
  the {GGA}+{$U$} framework}.
\newblock \emph{\bibinfo{journal}{Phys. Rev. B}} \textbf{\bibinfo{volume}{73}},
  \bibinfo{pages}{195107} (\bibinfo{year}{2006}).

\bibitem{si-ohzuku1990lmo}
\bibinfo{author}{Ohzuku, T.}, \bibinfo{author}{Kitagawa, M.} \&
  \bibinfo{author}{Hirai, T.}
\newblock \bibinfo{title}{Electrochemistry of manganese dioxide in lithium
  nonaqueous cell. {III}. {X}-ray diffractional study on the reduction of
  spinel-related manganese dioxide}.
\newblock \emph{\bibinfo{journal}{J. Electrochem. Soc.}}
  \textbf{\bibinfo{volume}{137}}, \bibinfo{pages}{769--775}
  (\bibinfo{year}{1990}).

\bibitem{si-yabuuchi2004nmc}
\bibinfo{author}{Yabuuchi, N.} \& \bibinfo{author}{Ohzuku, T.}
\newblock \bibinfo{title}{Solid-state redox reactions of
  {LiCo$_{1/3}$Ni$_{1/3}$Mn$_{1/3}$O$_2$} for advanced lithium-ion batteries}.
\newblock \bibinfo{howpublished}{206th Meeting of The Electrochemical Society
  (2004 Joint International Meeting, Honolulu), Abstract 324,
  \url{https://www.electrochem.org/dl/ma/206/pdfs/0324.pdf}}
  (\bibinfo{year}{2004}).

\bibitem{si-yoshizawa2003nmc}
\bibinfo{author}{Yoshizawa, H.} \& \bibinfo{author}{Ohzuku, T.}
\newblock \bibinfo{title}{Prismatic lithium-ion batteries of
  {LiCo$_{1/3}$Ni$_{1/3}$Mn$_{1/3}$O$_2$} with graphite}.
\newblock \emph{\bibinfo{journal}{Electrochemistry}}
  \textbf{\bibinfo{volume}{71}}, \bibinfo{pages}{1177--1181}
  (\bibinfo{year}{2003}).

\bibitem{si-ohzuku1994lco}
\bibinfo{author}{Ohzuku, T.} \& \bibinfo{author}{Ueda, A.}
\newblock \bibinfo{title}{Solid-state redox reactions of {LiCoO$_2$}
  ({R$\bar{3}$m}) for 4 volt secondary lithium cells}.
\newblock \emph{\bibinfo{journal}{J. Electrochem. Soc.}}
  \textbf{\bibinfo{volume}{141}}, \bibinfo{pages}{2972--2977}
  (\bibinfo{year}{1994}).

\bibitem{si-kondo2018lfp}
\bibinfo{author}{Kondo, H.}, \bibinfo{author}{Sasaki, T.},
  \bibinfo{author}{Barai, P.} \& \bibinfo{author}{Srinivasan, V.}
\newblock \bibinfo{title}{Comprehensive study of the polarization behavior of
  {LiFePO$_4$} electrodes based on a many-particle model}.
\newblock \emph{\bibinfo{journal}{J. Electrochem. Soc.}}
  \textbf{\bibinfo{volume}{165}}, \bibinfo{pages}{A2047--A2057}
  (\bibinfo{year}{2018}).

\bibitem{si-aravindan2013lmp}
\bibinfo{author}{Aravindan, V.}, \bibinfo{author}{Gnanaraj, J.},
  \bibinfo{author}{Lee, Y.} \& \bibinfo{author}{Madhavi, S.}
\newblock \bibinfo{title}{{LiMnPO$_4$} -- a next generation cathode material
  for lithium-ion batteries}.
\newblock \emph{\bibinfo{journal}{J. Mater. Chem. A}}
  \textbf{\bibinfo{volume}{1}}, \bibinfo{pages}{3518--3539}
  (\bibinfo{year}{2013}).

\bibitem{si-chen2022m3gnet}
\bibinfo{author}{Chen, C.} \& \bibinfo{author}{Ong, S.~P.}
\newblock \bibinfo{title}{A universal graph deep learning interatomic potential
  for the periodic table}.
\newblock \emph{\bibinfo{journal}{Nat. Comput. Sci.}}
  \textbf{\bibinfo{volume}{2}}, \bibinfo{pages}{718--728}
  (\bibinfo{year}{2022}).

\bibitem{si-deng2023chgnet}
\bibinfo{author}{Deng, B.} \emph{et~al.}
\newblock \bibinfo{title}{{CHG}{N}et as a pretrained universal neural network
  potential for charge-informed atomistic modelling}.
\newblock \emph{\bibinfo{journal}{Nat. Mach. Intell.}}
  \textbf{\bibinfo{volume}{5}}, \bibinfo{pages}{1031--1041}
  (\bibinfo{year}{2023}).

\bibitem{si-batatia2022mace}
\bibinfo{author}{Batatia, I.}, \bibinfo{author}{Kov{\'a}cs, D.~P.},
  \bibinfo{author}{Simm, G. N.~C.}, \bibinfo{author}{Ortner, C.} \&
  \bibinfo{author}{Cs{\'a}nyi, G.}
\newblock \bibinfo{title}{{MACE}: Higher order equivariant message passing
  neural networks for fast and accurate force fields}.
\newblock In \emph{\bibinfo{booktitle}{Advances in Neural Information
  Processing Systems}}, vol.~\bibinfo{volume}{35},
  \bibinfo{pages}{11423--11436} (\bibinfo{year}{2022}).

\bibitem{si-merchant2023gnome}
\bibinfo{author}{Merchant, A.} \emph{et~al.}
\newblock \bibinfo{title}{Scaling deep learning for materials discovery}.
\newblock \emph{\bibinfo{journal}{Nature}} \textbf{\bibinfo{volume}{624}},
  \bibinfo{pages}{80--85} (\bibinfo{year}{2023}).

\end{thebibliography}

\end{document}


\hbadness=1500
\setlength{\emergencystretch}{3em}
\setlength{\tabcolsep}{4pt}

\maketitle
\tableofcontents
\newpage

\section{ALIGNN-FF Architecture and Inference}
\label{si:alignnff}

The ALIGNN-FF surrogate used for the screening pass, the vacancy ranking of the DFT tier and the enumeration (model weights \texttt{alignnff\_wt01}) is the universal force-field variant of the Atomistic Line Graph Neural Network\cite{si-choudhary2021alignn,si-choudhary2023alignnff}. It maps a periodic atomic structure to a single scalar total energy in one forward pass; atomic forces and the stress tensor are obtained as autograd derivatives of that energy with respect to atomic positions and lattice strain. No relaxation, electronic-structure iteration, or explicit ion-removal energy is computed by the network itself: every voltage in the screening stage is a difference of these total energies combined with a lithium-metal reference via the formulae of Supplementary Note~\ref{si:voltage} (an effective $-0.957$~eV/atom in the screening pass, inferred from its outputs and close to the tabulated JARVIS-DFT value of $-0.925$~eV/atom; the force field's own energy of the JARVIS-DFT lithium entry JVASP-25117 for the training labels of the regressor, which were generated with the default checkpoint of the ALIGNN-FF package rather than \texttt{alignnff\_wt01}; see Methods), and every vacancy ranking in the DFT pipeline is a comparison of these total energies. Voltages at the DFT tier use DFT energies with the in-house references of Supplementary Note~\ref{si:libcc}.

\subsection{Graph construction}

A POSCAR (lattice vectors, species, fractional coordinates) is converted into a coupled pair of graphs. The atom graph $G$ has one node per atom; edges are constructed via a periodic-image-aware $k$-nearest-neighbor search with a radial cutoff of \SI{8}{\angstrom} and a maximum of 12 neighbors per atom (\texttt{cutoff = 8.0}, \texttt{max\_neighbors = 12} in the released \texttt{config.json}). Each edge carries the bond vector $\mathbf{r}_{ij}$ and its norm. The line graph $L(G)$ has one node per bond of $G$; two bond-nodes are connected when they share an atom, so each edge of $L(G)$ corresponds to a triplet $(i, j, k)$ and carries the bond angle $\theta_{ijk} = \angle(\mathbf{r}_{ji}, \mathbf{r}_{jk})$. The line graph is what distinguishes ALIGNN from purely pairwise GNN force fields and gives the network direct access to three-body angular information.

\subsection{Initial featurization}

Atom nodes are initialized with the 92-dimensional CGCNN element fingerprint (period, group, electronegativity, covalent radius, valence, etc.), then embedded into a 256-dimensional hidden space. Bond edges of $G$ are initialized by expanding $\lVert\mathbf{r}_{ij}\rVert$ over 80 Gaussian radial basis centers and passing the result through a two-layer MLP (\texttt{edge\_input\_features = 80}, \texttt{embedding\_features = 64}, \texttt{hidden\_features = 256}). Triplet edges of $L(G)$ are initialized by expanding $\cos\theta_{ijk}$ over 40 Gaussian basis centers on $[-1,1]$ and applying the same MLP head, also producing 256-dim features.

\subsection{Message passing}

The model alternates two layer types, all in the 256-dim hidden space and all using residual updates with LayerNorm and SiLU nonlinearities:
\begin{itemize}
    \item \textbf{Four ALIGNN layers ($\sim$2.64M parameters total).} Each layer performs two edge-gated graph convolutions: (i) on $G$, atom features are updated with messages gated by the bond features, which are updated in turn; (ii) on $L(G)$, whose nodes are the bonds, the bond features are updated again with messages gated by the triplet features, which are themselves updated. Angular information thus enters the bond features at the end of each layer and reaches the atoms in the next one.
    \item \textbf{Four edge-gated GCN layers ($\sim$1.32M parameters total).} Plain edge-gated graph convolution on $G$ only. These refine atom and bond features without using triplets and add depth at lower compute cost.
\end{itemize}

\subsection{Energy readout and derivatives}

After message passing, the per-atom embeddings are averaged over all atoms to give a 256-dim graph-level feature, which is mapped to a single scalar by a final \texttt{Linear(256, 1)} head (the $\texttt{fc}$ submodule, 257 parameters) and multiplied by the number of atoms (\texttt{energy\_mult\_natoms}), so that the predicted total energy $E$ in eV is extensive. The voltage regressor of the main text uses the same mean-pooled readout without the multiplication, so its output is a per-structure value. Atomic forces are obtained as $\mathbf{F}_i = -\partial E / \partial \mathbf{r}_i$ via PyTorch autograd, and the stress as $\boldsymbol{\sigma} = -V^{-1}\,\partial E / \partial \boldsymbol{\varepsilon}$ with respect to the lattice strain $\boldsymbol{\varepsilon}$. The \texttt{AlignnAtomwiseCalculator} ASE wrapper exposes these three quantities as a drop-in interatomic-potential calculator.

\subsection{Model size and training data}

The complete model stores 4{,}026{,}873 parameters (4{,}026{,}753 trainable plus 120 fixed radial-basis centers) and a 16.2 MB serialized checkpoint. The breakdown by submodule is:

\begin{table}[ht]
\centering
\caption{Parameter breakdown of the \texttt{alignnff\_wt01} model used in this work.}
\label{tab:alignn_params}
\small
\begin{tabular}{lrr}
\toprule
Submodule & Parameters & Size (fp32) \\
\midrule
ALIGNN layers (4)            & 2{,}639{,}872 & 10.07 MiB \\
Edge-gated GCN layers (4)    & 1{,}319{,}936 &  5.04 MiB \\
Atom embedding (CGCNN $\to$ 256)        &     24{,}320 &  0.09 MiB \\
Bond edge embedding (RBF + MLP)         &     22{,}544 &  0.09 MiB \\
Triplet edge embedding (RBF + MLP)      &     19{,}944 &  0.08 MiB \\
Energy readout (\texttt{Linear(256, 1)})    &           257 & $<$0.01 MiB \\
\midrule
\textbf{Total}                & \textbf{4{,}026{,}873} & \textbf{15.36 MiB} \\
\bottomrule
\end{tabular}
\end{table}

The pretrained weights were trained on the universal-force-field dataset distributed with the \texttt{alignn} package (energies and forces computed at the optB88-vdW level on JARVIS-DFT 3D structures; the stress weight in its training configuration is zero). The ALIGNN regressor reported in the main text (Fig.~2, MAE = \SI{0.17}{\V}) uses the same \texttt{alignn\_atomwise} architecture with force and stress outputs disabled, trained for 250 epochs on ALIGNN-FF average delithiation voltages of $7{,}610$ lithium-containing JARVIS-DFT structures (the database itself contains no voltage subset); both models share the same atom/\allowbreak line-graph backbone described above.

\subsection{Inference cost}

Inference for the 32- and 56-atom enumeration cells takes about 2--4~s per single point on one CPU thread; a complete delithiation profile (16--17 single-point evaluations per material, no relaxation) takes tens of seconds to about a minute. This is what makes the per-step ALIGNN-FF vacancy ranking in Supplementary Note~\ref{si:delithiation} tractable: every remaining Li site is evaluated as an independent forward pass, and the full sweep across $\sim$10--16 candidate vacancies fits inside the time budget of a single VASP SCF cycle.

\section{VASP Calculation Setup}
\label{si:vasp}

All first-principles calculations were performed with the Vienna \emph{ab initio} simulation package (VASP)\cite{si-kresse1996vasp} using projector augmented-wave (PAW) pseudopotentials\cite{si-blochl1994paw,si-kresse1999paw} and the spin-polarized (ISPIN = 2) formalism. Input files for all calculations in this work, including the twelve prospective materials, are provided in the public repository at \url{https://github.com/atomgptlab/batterymat}; each INCAR there is the one used for the run that produced the recorded energy (for the fully lithiated LMP endpoint, the recompute in \texttt{dft\_inputs/extra\_2026-10-26/LMP-Li16-altmag}).

\subsection{Plane-wave basis, precision, and electronic SCF}

The plane-wave cutoff is fixed at \SI{520}{\eV} (ENCUT = 520), the PREC tag is set to \texttt{Accurate}, and smearing is Gaussian (ISMEAR = 0) with a width of \SI{0.05}{\eV} (SIGMA = 0.05). Two input templates were used (per-chain counts are given in Supplementary Note~\ref{si:convergence}). The earlier one uses EDIFF = $10^{-4}$~eV (except the first steps of the LiMnPO$_4$ chain and of the superseded optPBE-vdW LiCoO$_2$ run, which used $10^{-6}$~eV), conjugate-gradient ionic relaxation (IBRION = 2), an iteration cap of NELM = 500, cold starts and the default mixing; the current one uses EDIFF = $10^{-4}$~eV with quasi-Newton relaxation (IBRION = 1), NELM = 100, a restart from the previous step's wave function and charge density (ISTART = 1, ICHARG = 1) and the conservative mixing described below. Nonspherical gradient contributions are included (LASPH = .TRUE.) and LMAXMIX = 4 is used for accurate projection of the $d$-electron density onto the onsite PAW spheres. Symmetry is disabled (ISYM = 0) to allow the electronic solution to relax freely under block-structured antiferromagnetic initialization (Supplementary Note~\ref{si:magmom}). The current template applies conservatively tuned charge and magnetization mixing (AMIX = 0.2, BMIX = $10^{-4}$, AMIX\_MAG = 0.4, BMIX\_MAG = $10^{-4}$) together with NELMIN = 4 to stabilize DFT$+U$ SCF on the highly oxidized deep-delithiation steps. It was used for $14$ of $17$ LFP steps, $14$ of $17$ NMC steps, $3$ of $17$ LMO steps, $3$ of the $17$ LMP steps, every step of the reported LiCoO$_2$ chain, and every other cathode calculation in this work except the original optPBE-vdW LiCoO$_2$ run, whose first seven steps used the earlier template (Supplementary Note~\ref{si:lco}); the earlier template was used for the remaining benchmark steps.

\subsection{Ionic relaxation}

Ionic relaxation uses the conjugate-gradient algorithm (IBRION = 2) in the earlier template and a quasi-Newton algorithm (IBRION = 1) in the current one. The ISIF tag is selected per step based on the dimensionality of the host framework:
\begin{itemize}
    \item \textbf{Step 0 (lithiated parent):} ISIF = 3 (full cell + ionic relaxation) for all materials.
    \item \textbf{Steps 1$+$ (delithiated):} ISIF = 2 (ionic relaxation at fixed cell shape and volume) for 3D-bonded frameworks (phosphates, spinel), and ISIF = 3 for layered materials whose interlayer spacing must relax on Li removal.
\end{itemize}
ISIF = 2 is used for hosts outside the layered whitelist, so that the cell is fixed at its fully lithiated shape. ISIF = 3 is retained for whitelisted layered cathodes (LiCoO$_2$; the NMC cell is stored with P1 symmetry in JARVIS-DFT and is relaxed at ISIF = 2 after step 0, like the non-layered hosts) so that the interlayer spacing can respond to delithiation; the NMC comparison of Supplementary Note~\ref{si:nmc} shows that fixing the cell of a layered host raises its full-range average by 0.22 to \SI{0.31}{\V} (0.21 to \SI{0.36}{\V} including the two half ranges). The ionic relaxation is converged on the default energy criterion (ten times EDIFF) in both templates, except for the first steps of the LMP chain and of the superseded optPBE-vdW LiCoO$_2$ run, most of which used EDIFF $=10^{-6}$~eV with the maximum-force criterion EDIFFG $= -0.03$~eV/\AA; NSW is set per material according to
\begin{equation}
\mathrm{NSW} = \min(300,\,\max(200,\,2 N_{\mathrm{atoms}})),
\end{equation}
which gives 200--300 ionic steps per calculation.

\subsection{Exchange-correlation functional selection}

BatteryMat auto-selects the exchange-correlation functional from the parent structure's spacegroup. Materials whose spacegroup number lies in the layered whitelist \{12, 15, 166, 194\} (C2/m, C2/c, R$\overline{3}$m, P6$_3$/mmc) are flagged as layered and routed to the optB88-vdW functional\cite{si-klimes2010optb88,si-klimes2011optb88}, which restores the $c$-axis interlayer binding that pure PBE underestimates. Non-layered materials keep the PBE functional. Both functionals add DFT$+U$ corrections on transition-metal $d$ electrons (Supplementary Note~\ref{si:dftu}).

For the optB88-vdW calculations the following tags are appended to the PBE base: \texttt{GGA = BO}, \texttt{PARAM1 = 0.1833333333}, \texttt{PARAM2 = 0.22}, \texttt{LUSE\_VDW = .TRUE.}, \texttt{AGGAC = 0.0}. \textit{Correction to the preprint version of this work (arXiv:2607.06645v1).} The earlier version of this note, and of the input generator, specified \texttt{GGA = OR} with \texttt{LUSE\_VDW = .TRUE.} and \texttt{AGGAC = 0.0} and described it as the optB88 setting. That combination selects the optPBE-vdW functional; optB88-vdW requires \texttt{GGA = BO} with the two \texttt{PARAM} values above. The generator has been corrected. Two sets of runs were affected: the LCO benchmark of the preprint, which has been rerun with the optB88-vdW tags (the original run is reported as an optPBE-vdW comparison in Supplementary Note~\ref{si:lco}), and the first pass of the four prospective materials routed to the vdW functional (\ce{LiV2F7}, \ce{LiCa2Ag} and the two \ce{Rb2LiFeF6} entries, which are the same structure stored twice in JARVIS-DFT), which were rerun with the optB88-vdW tags before being reported: $6.13$~V for both \ce{Rb2LiFeF6} entries ($6.03$~V in the first pass), $4.92$~V for \ce{LiV2F7} ($4.84$~V) and $0.45$~V for \ce{LiCa2Ag} ($0.41$~V). The first-pass values were internally consistent, because the matching optPBE-vdW lithium reference was used, and the differences of about $0.04$ to $0.1$~V change none of the conclusions of the prospective test. The roughly \SI{1}{\eV\per atom} lithium-reference discrepancy mentioned in the preprint arises from combining the tabulated JARVIS-DFT reference, an optB88-vdW value, with PBE cathode energies (Supplementary Note~\ref{si:libcc}), not from the \texttt{GGA} tag.

\subsection{\texorpdfstring{DFT$+U$}{DFT+U} parameters (Dudarev scheme)}
\label{si:dftu}

DFT$+U$ corrections are applied in the Dudarev scheme (LDAUTYPE = 2), with LDAU = .TRUE. and LDAUL = 2 (d-orbital) for every element in Table~\ref{tab:hubbardu} and LDAUL = $-1$ for all other elements. LDAUJ is set to zero for every element (Dudarev: only the effective $U_{\mathrm{eff}} = U - J$ matters). The per-element $U$ values are the Materials Project values\cite{si-mp_hubbard_docs}, fitted to binary-oxide formation enthalpies following Wang et al.\cite{si-wang2006ggau}.

\begin{table}[ht]
\centering
\caption{Per-element DFT$+U$ values (Dudarev scheme, LDAUTYPE = 2) used throughout this work.}
\label{tab:hubbardu}
\begin{tabular}{lc}
\toprule
Element & $U_{\mathrm{eff}}$ (eV) \\
\midrule
Mn & 3.90 \\
Fe & 5.30 \\
Co & 3.32 \\
Ni & 6.20 \\
V  & 3.25 \\
Cr & 3.70 \\
Mo & 4.38 \\
W  & 6.20 \\
\bottomrule
\end{tabular}
\end{table}

\subsection{Parallelization and performance tags}

The four 112-atom benchmark chains ran on 64 MPI ranks; the revision chains, including the reported LiCoO$_2$ chain, reduce the rank count to the number of atoms on small cells and adjust NCORE so that it divides the ranks, which does not affect the energies. Performance-critical tags are \texttt{LREAL = Auto} (real-space PAW projection, safe for supercells $>20$ atoms, with an accuracy impact of roughly \SI{0.001}{\eV/atom}; the metal references use \texttt{LREAL = .FALSE.}), \texttt{NCORE = 8} (band parallelization, $\sqrt{64}$), \texttt{KPAR = 2} (k-point parallelization, must divide both the number of k-points and MPI ranks), and \texttt{NSIM = 8} (number of bands optimized simultaneously in the Davidson/CG loop; default is 4). \texttt{LORBIT = 11} produces site- and orbital-projected moments and DOS.

\subsection{\texorpdfstring{$k$}{k}-point sampling}

A $\Gamma$-centered Monkhorst--Pack mesh is generated automatically. The base mesh for the primitive unit cell is $3\times 3\times 3$, and the mesh density scales inversely with each supercell multiplier: for a supercell of dimensions $(n_x, n_y, n_z)$ the generated mesh is $(\max(1, \lfloor 3/n_x \rceil), \max(1, \lfloor 3/n_y \rceil), \max(1, \lfloor 3/n_z \rceil))$. For the $2\times 2\times 2$ LCO and LMO supercells this gives a $2\times 2\times 2$ mesh; for the $2\times 2\times 1$ supercells of LFP, LMP, and NMC it gives $2\times 2\times 3$.

\section{Supercell Construction}
\label{si:supercell}

Delithiation is performed on diagonal supercells. The per-axis multiplier is chosen so that every lattice-vector norm exceeds a target minimum (\SI{7}{\angstrom} in this work):
\begin{equation}
n_i = \left\lceil \frac{L_{\min}}{\lVert\mathbf{a}_i\rVert}\right\rceil,
\quad L_{\min} = \SI{7}{\angstrom}.
\end{equation}
This rule uses the vector norm, not the perpendicular inter-plane distance, so non-orthogonal cells such as rhombohedral LiCoO$_2$ may have smaller actual inter-plane separations than \SI{7}{\angstrom}. A default cap of 300 atoms per supercell is enforced; if the cap is exceeded, the largest dimension is iteratively reduced and a warning is emitted. Only diagonal supercells (diagonal transformation matrices) are supported by the JARVIS \texttt{make\_supercell} routine used here.

The supercells of the five materials are listed in Table~\ref{tab:supercells}.

\begin{table}[ht]
\centering
\caption{Per-material supercell parameters, total atom counts, and number of Li atoms available for delithiation. The LFP and LMP entries are a \textit{Pnma} polymorph with tetrahedral lithium, not olivine; the olivine endpoints use $2\times2\times1$ supercells of the same size (Supplementary Note~\ref{si:olivine}).}
\label{tab:supercells}
\small
\setlength{\tabcolsep}{4pt}
\begin{tabular}{llccccl}
\toprule
JID & Material & Dim. & $N_{\mathrm{atoms}}$ & $N_{\mathrm{Li,total}}$ & Functional & Spacegroup (primitive) \\
\midrule
JVASP-42723   & LiFePO$_4$       & $2\times 2\times 1$ & 112 & 16 & PBE$+U$        & Pnma  \\
JVASP-116897  & LiMnPO$_4$       & $2\times 2\times 1$ & 112 & 16 & PBE$+U$        & Pnma  \\
JVASP-141792  & LiMn$_2$O$_4$    & $2\times 2\times 2$ & 112 & 16 & PBE$+U$        & Fd$\overline{3}$m \\
JVASP-144791  & Li$_4$Mn$_3$Co$_2$Ni$_3$O$_{16}$& $2\times 2\times 1$ & 112 & 16 & PBE$+U$        & P1 (layered)  \\
JVASP-2017    & LiCoO$_2$        & $2\times 2\times 2$ & 32  & 8  & optB88-vdW$+U$ & R$\overline{3}$m  \\
\bottomrule
\end{tabular}
\end{table}

\paragraph{Note on NMC.} The NMC cell (JVASP-144791) is layered in structure but is stored in JARVIS-DFT with triclinic P1 symmetry, so the spacegroup rule routes it to PBE$+U$ and relaxes steps 1 onward at ISIF = 2. An optB88-vdW$+U$ rerun of the same cell and cell-relaxed endpoints for both functionals (Supplementary Note~\ref{si:nmc}) show that the dispersion correction raises the full-range average by 0.27 to \SI{0.36}{\V}, that the fixed-cell treatment raises it by 0.22 to \SI{0.31}{\V}, and that the excess rise between the two halves of the lithium range is present in every variant.

\section{Magnetic Initialization}
\label{si:magmom}

We initialize the transition-metal moments in block-structured antiferromagnetic (AFM) MAGMOM patterns; whether this is the lowest-energy magnetic order was tested only for LiMnPO$_4$, where the per-atom alternating order is \SI{0.16}{\eV} per 112-atom cell lower than the block order at full lithiation (Supplementary Note~\ref{si:lmp}). Signs alternating atom by atom in the supercell (e.g.\ +5, $-$5, +5, \dots ordered by supercell index) caused DFT$+U$ SCF divergence in early trials on some hosts; block-structured patterns (e.g.\ \texttt{4*5 4*-5}, which encodes four Fe atoms with $+5~\mu_B$ followed by four with $-5~\mu_B$) converged reliably.

\subsection{Default per-element moments}

The per-element magnitudes (in $\mu_B$) used on the first step are Mn = 4.0, Fe = 5.0, Ni = 2.0, Co = 0.6, V = 3.0, Cr = 3.0, Mo = 3.0, W = 2.0. Non-magnetic atoms (Li, O, P) are given a moment of 0.

\subsection{Supercell mapping}

For a supercell the per-atom MAGMOM is built by mapping each atom back to its primitive-cell equivalent (modulo the supercell index) and taking the primitive moment. Consecutive magnetic atoms of the primitive cell alternate in sign, so each primitive site maps to a block of equal moments in the supercell, producing runs such as \texttt{4*5 4*-5}; in LiCoO$_2$, whose primitive cell holds a single Co atom, every Co starts at $+0.6~\mu_B$.

\subsection{MAGMOM inheritance on delithiation}

On steps~1 and onward the MAGMOM is not regenerated from scratch. It is read from the previous step's INCAR, the removed Li's entry is dropped, and the remaining list is written out verbatim. This preserves the block-structured AFM sublattice as the cell is delithiated and prevents SCF divergence. The LiMnPO$_4$ chain predates this inheritance: its steps~1--15 (and the step-16 attempts) were initialized with a per-atom alternating pattern (\texttt{4 -4 4 -4 \dots}), which converged at every step but to a different antiferromagnetic arrangement of the Mn sublattice than step~0. The fully lithiated endpoint was therefore recomputed in the alternating order, which is \SI{0.16}{\eV} per supercell lower, and the reported chain uses that energy (Supplementary Note~\ref{si:lmp}). The superseded optPBE-vdW LiCoO$_2$ run likewise used alternating $\pm0.6~\mu_B$ Co moments after step~0.

\subsection{As-used supercell MAGMOM}

Table~\ref{tab:magmom} shows the block-structured MAGMOM strings actually written to the step~0 INCAR for each material, in VASP's count-prefix notation.

\begin{table}[ht]
\centering
\caption{Step-0 (fully lithiated) MAGMOM strings. The count-prefix notation \texttt{n*m} denotes \texttt{n} consecutive atoms with moment \texttt{m} in units of $\mu_B$. Trailing zeros correspond to the anion atoms. For LMP the reported step-0 energy comes from a rerun initialized with the per-atom alternating pattern of steps 1--16 (Supplementary Note~\ref{si:lmp}).}
\label{tab:magmom}
\small
\begin{tabular}{ll}
\toprule
JID & MAGMOM \\
\midrule
JVASP-116897-LMP & \texttt{16*0 4*4 4*-4 4*4 4*-4 80*0} \\
JVASP-42723-LFP  & \texttt{16*0 4*5 4*-5 4*5 4*-5 80*0} \\
JVASP-141792-LMO & \texttt{16*0 8*4 8*-4 8*4 8*-4 64*0} \\
JVASP-144791-NMC & \texttt{16*0 4*4 4*-4 4*4 4*-0.6 4*0.6 4*-2 4*2 4*-2 64*0} \\
JVASP-2017-LCO-B88 & \texttt{8*0 8*0.6 16*0} \\
\bottomrule
\end{tabular}
\end{table}

\section{Sequential Supercell Delithiation Workflow}
\label{si:delithiation}

All delithiation happens on the supercell of Supplementary Note~\ref{si:supercell}. The end-to-end loop is \texttt{init} $\rightarrow$ (run DFT $\rightarrow$ \texttt{record} $\rightarrow$ \texttt{next}) $\times\,N$ $\rightarrow$ \texttt{voltage}, implemented in \texttt{dft\_prep.py}.

\begin{description}
\item[init.] Look up the material in the JARVIS-DFT database, build the supercell, auto-detect the functional (layered $\to$ optB88-vdW, else PBE), and emit VASP inputs for step~0 (fully lithiated parent). This is the only step that touches the JARVIS database.
\item[run DFT externally.] Inputs are copied to the HPC cluster; POTCAR is generated from the POTCAR\_spec (a list of PAW labels: Li\_sv, Mn\_pv, etc.) loaded from JARVIS \texttt{default\_potcars.json}; VASP runs produce a CONTCAR (relaxed structure) and an OUTCAR (containing the total energy).
\item[record.] Parse the OUTCAR and log the converged DFT total energy into \texttt{energies.json} for that step. Can auto-read from \texttt{results/OUTCAR} if the energy argument is omitted.
\item[next.] Read the CONTCAR from the previous step. Use the ALIGNN-FF force field to rank every Li vacancy individually; every Li site is evaluated, with no symmetry deduplication applied, because the supercell symmetry is broken after the first vacancy. Remove the Li whose vacancy has the lowest ALIGNN-FF total energy, inherit MAGMOM from the previous INCAR (with the removed Li's entry dropped), choose ISIF per Supplementary Note~\ref{si:vasp}, and write VASP inputs for the next step. Loop back to \textbf{run DFT externally}.
\item[voltage.] Once at least two energies are recorded, compute the step voltages and lower convex hull, and emit per-material voltage-curve plots (\texttt{voltage\_curve\_JVASP-XXXXX.png} for the line graph, \texttt{discharge\_curve\_JVASP-XXXXX.png} for the staircase discharge plot).
\end{description}

The \texttt{energies.json} schema is:
\begin{verbatim}
{"jid": "JVASP-XXXXX", "dim": [2,2,2], "n_li_total": 16,
 "functional": "pbe", "layered": false, "e_li_metal": -1.9031,
 "steps": [{"step": 0, "n_li": 16, "energy": -770.558,
            "removed_li_index": null}, ...]}
\end{verbatim}

\section{Lithium Reference Energy}
\label{si:libcc}

The metallic-lithium chemical potential $\mu_{\mathrm{Li}}$ enters every voltage equation, so it must be computed with the same functional, PAW pseudopotential and plane-wave settings as the cathode calculations. A mismatched reference shifts the entire voltage curve: the tabulated JARVIS-DFT value of $-0.925$~eV/atom is an optB88-vdW energy, and combining it with PBE$+U$ cathode energies shifts every voltage by $\sim\SI{1}{\V}$.

The in-house Li reference is BCC Li in the one-atom primitive cell with the following settings: Li\_sv PAW, ENCUT = 520, EDIFF = $10^{-6}$, ISMEAR = 1 (Methfessel--Paxton), SIGMA = 0.1, $k$-mesh = $17\times 17\times 17$ $\Gamma$-centered, ISPIN = 1 (the PBE and optB88-vdW values are single-point calculations on the optB88-vdW-relaxed BCC cell, the latter with EDIFF $= 10^{-7}$; the optPBE-vdW value and the optB88-vdW check run are full cell relaxations). The converged total energies per atom are:
\begin{align}
\mu_{\mathrm{Li}}^{\mathrm{PBE}} &= -1.9031~\mathrm{eV/atom}, \\
\mu_{\mathrm{Li}}^{\mathrm{optB88\text{-}vdW}} &= -0.9778~\mathrm{eV/atom}, \\
\mu_{\mathrm{Li}}^{\mathrm{optPBE\text{-}vdW}} &= -0.9646~\mathrm{eV/atom}.
\end{align}
The optB88-vdW calculation uses \texttt{GGA = BO}, \texttt{PARAM1 = 0.1833333333}, \texttt{PARAM2 = 0.22}, \texttt{LUSE\_VDW = .TRUE.}, \texttt{AGGAC = 0.0} (an independent relaxation gives $-0.97791$~eV/atom, within $0.2$~meV); the optPBE-vdW value uses \texttt{GGA = OR} with the same vdW tags and was used for the original LCO run and for the first pass of the four vdW-routed prospective materials, all since rerun with optB88-vdW. The preprint labeled the $-0.9646$~eV/atom value as optB88-vdW; see the correction in Supplementary Note~\ref{si:vasp}. Input files are in \texttt{dft\_inputs/\allowbreak JVASP-913-Li/\allowbreak Li\_sv\_PBE/} (PBE), \texttt{Li\_sv/\allowbreak static/} (optB88-vdW single point, on the cell of \texttt{Li\_sv/\allowbreak relax/}), \texttt{Li\_sv\_optB88vdW/} (optB88-vdW check) and \texttt{Li\_sv\_optPBEvdW/}. Every DFT voltage reported in the manuscript uses the functional-matched $\mu_{\mathrm{Li}}$.

\section{Voltage Curve Construction}
\label{si:voltage}

\subsection{Step voltage}

For the delithiation of a cathode from a more-lithiated state ``hi'' to a less-lithiated state ``lo'',
\begin{equation}
\mathrm{Li}_{\mathrm{hi}} \longrightarrow \mathrm{Li}_{\mathrm{lo}} + \Delta n\,\mathrm{Li}(\mathrm{metal}),
\end{equation}
where $\Delta n = n_{\mathrm{Li,hi}} - n_{\mathrm{Li,lo}}$. Conservation of energy gives $\Delta E = E_{\mathrm{lo}} + \Delta n\,\mu_{\mathrm{Li}} - E_{\mathrm{hi}}$, and the voltage in the zero-temperature, zero-pressure approximation is the energy required per extracted lithium, $V = \Delta E / (\Delta n\,e)$:
\begin{equation}
V = \frac{E_{\mathrm{lo}} - E_{\mathrm{hi}}}{\Delta n} + \mu_{\mathrm{Li}},
\label{eq:Vstep}
\end{equation}
which is the form used in \texttt{dft\_prep.py}. Here $E$ is the VASP total energy in eV and $\mu_{\mathrm{Li}}$ is the Li-metal reference from Supplementary Note~\ref{si:libcc}. All step voltages in this SI use $\Delta n = 1$.

\subsection{Formation energies and lower convex hull}

Let $x = n_{\mathrm{Li}} / n_{\mathrm{Li,total}}$ denote the Li fraction, and $E(x)$ the supercell total energy at that fraction. The formation energy relative to the $x = 0$ and $x = 1$ endpoints is
\begin{equation}
\Delta E(x) = E(x) - x\,E(1) - (1-x)\,E(0).
\end{equation}
The lower convex hull of the set $\{(x, \Delta E(x))\}$ identifies the thermodynamically stable compositions and their connecting plateaus. For each plateau bounded by stable compositions $x_i$ and $x_{i+1}$ (with $x_i > x_{i+1}$), the equilibrium voltage is
\begin{equation}
V_{\mathrm{plateau}} = \frac{E(x_{i+1}) - E(x_i)}{(x_i - x_{i+1})\,n_{\mathrm{Li,total}}} + \mu_{\mathrm{Li}}.
\end{equation}
This hull-based voltage is reported as the equilibrium discharge curve and is compared against experiment in the main manuscript.

\subsection{Average voltage}

The average voltage over the full delithiation window $x\in[0,1]$ is the mean of the per-step voltages (Eq.~\ref{eq:Vstep}) weighted by $\Delta n$, which for single-Li steps reduces to the arithmetic mean of the step voltages.

\section{Per-Material DFT Data}

This note tabulates the raw DFT data underlying the benchmark voltages of the main text, together with the functional and cell-relaxation comparisons for LiCoO$_2$ and the NMC variant; the energies of the prospective, spot-check, sodium and magnesium runs are in the code release (\texttt{energies.json} for each chain and \texttt{analysis/spotcheck\_results.csv}). For each material we list, per step: the step index, number of Li atoms in the supercell ($n_{\mathrm{Li}}$), Li fraction $x = n_{\mathrm{Li}} / n_{\mathrm{Li,total}}$, VASP total energy (eV), the Li index removed to reach this step (0-indexed into the previous step's Li list), and the step voltage from Eq.~\ref{eq:Vstep} (in V). All energies are converged SCF total energies from the production run.

\subsection{\texorpdfstring{LiFePO$_4$ polymorph (JVASP-42723, tetrahedral Li, PBE$+U$)}{LiFePO4 polymorph (JVASP-42723, tetrahedral Li, PBE+U)}}
\label{si:lfp}

Functional PBE$+U$ with $U_{\mathrm{eff}}(\mathrm{Fe}) = \SI{5.30}{\eV}$; supercell $2\times 2\times 1$, 112 atoms, $N_{\mathrm{Li,total}} = 16$; $\mu_{\mathrm{Li}} = \SI{-1.9031}{\eV/atom}$. All 17 steps complete. Mean step voltage \SI{3.60}{\V}.

\begin{table}[ht]
\centering
\caption{LiFePO$_4$ delithiation data.}
\label{tab:lfp_data}
\small
\begin{tabular}{rrrrrr}
\toprule
Step & $n_{\mathrm{Li}}$ & $x$ & $E_{\mathrm{tot}}$ (eV) & Removed Li idx & $V$ (V) \\
\midrule
 0 & 16 & 1.000 & $-760.94778$ & --  &  --  \\
 1 & 15 & 0.938 & $-755.54232$ & 8    & 3.502 \\
 2 & 14 & 0.875 & $-750.13009$ & 9    & 3.509 \\
 3 & 13 & 0.813 & $-744.74485$ & 4    & 3.482 \\
 4 & 12 & 0.750 & $-739.37268$ & 5    & 3.469 \\
 5 & 11 & 0.688 & $-734.03671$ & 3    & 3.433 \\
 6 & 10 & 0.625 & $-728.51410$ & 1    & 3.620 \\
 7 &  9 & 0.563 & $-723.07443$ & 9    & 3.537 \\
 8 &  8 & 0.500 & $-717.30743$ & 7    & 3.864 \\
 9 &  7 & 0.438 & $-711.82332$ & 2    & 3.581 \\
10 &  6 & 0.375 & $-706.42717$ & 2    & 3.493 \\
11 &  5 & 0.313 & $-700.75762$ & 4    & 3.766 \\
12 &  4 & 0.250 & $-695.13568$ & 4    & 3.719 \\
13 &  3 & 0.188 & $-689.24146$ & 3    & 3.991 \\
14 &  2 & 0.125 & $-683.72940$ & 2    & 3.609 \\
15 &  1 & 0.063 & $-678.27469$ & 1    & 3.552 \\
16 &  0 & 0.000 & $-672.96607$ & 0    & 3.406 \\
\bottomrule
\end{tabular}
\end{table}

The profile is flat at $\sim$\SI{3.5}{\V} across the middle of the window, with a modest rise at deep delithiation ($x \lesssim 0.25$). The entry is not olivine but a polymorph with tetrahedral lithium (Supplementary Note~\ref{si:olivine}); its average lies \SI{0.04}{\V} above the olivine value computed with the same protocol, and the finite supercell cannot support a true two-phase coexistence in either structure.

\subsection{\texorpdfstring{LiMnPO$_4$ polymorph (JVASP-116897, tetrahedral Li, PBE$+U$)}{LiMnPO4 polymorph (JVASP-116897, tetrahedral Li, PBE+U)}}
\label{si:lmp}

Functional PBE$+U$ with $U_{\mathrm{eff}}(\mathrm{Mn}) = \SI{3.90}{\eV}$; supercell $2\times 2\times 1$, 112 atoms, $N_{\mathrm{Li,total}} = 16$; $\mu_{\mathrm{Li}} = \SI{-1.9031}{\eV/atom}$. All seventeen steps converged (step~16 after a rerun, see below). The average over full delithiation is \SI{3.93}{\V} (olivine, with the same protocol: \SI{3.92}{\V}; Supplementary Note~\ref{si:olivine}).

\paragraph{Fully delithiated endpoint and magnetic order.} The first runs of step~16 (fully delithiated MnPO$_4$, in which all manganese is Mn$^{3+}$, high-spin $d^4$) were restarted repeatedly, the last time with a cold start, halved charge and magnetic mixing, and the ALGO = All solver. The run whose output survives was set up with a POTCAR set that does not match the structure (Li\_sv, Mn\_pv and P for a Mn, P, O cell; NELECT $= 576$ instead of $672$), which explains its energy of about \SI{-511.8}{\eV}, roughly \SI{190}{\eV} above the value expected from step~15. The endpoint was rerun from the step-15 relaxed structure with the correct POTCAR (Mn\_pv, P, O), the step-15 INCAR with a cold start, and the alternating MAGMOM of steps 1--15; it converged in 46 ionic steps at \SI{-704.89376}{\eV}, with Mn moments of $\pm3.97~\mu_B$ and at most $0.07~\mu_B$ on oxygen. Step~0 converged to the block order (\texttt{4*4 4*-4 4*4 4*-4}), whereas steps 1--16 are in the per-atom alternating order. Recomputed at the step-0 geometry and cell in the alternating order (ISIF = 2, five ionic steps), the fully lithiated endpoint is \SI{0.161}{\eV} per supercell lower ($-798.22301$ against $-798.06163$~eV). Table~\ref{tab:lmp_data} and the main text use the alternating-order value, so that every state of the chain has the same magnetic order; with the block-order step~0 the full-range average would be \SI{3.92}{\V} instead of \SI{3.93}{\V}, and the first step voltage \SI{3.50}{\V} instead of \SI{3.66}{\V}. Both runs are in \texttt{dft\_inputs/extra\_2026-10-26/} of the repository.

\begin{table}[ht]
\centering
\caption{LiMnPO$_4$ delithiation data. Step~0 is the fully lithiated endpoint recomputed in the alternating magnetic order of the later steps (the block-order step-0 run gives $-798.06163$~eV and a first step voltage of 3.495~V; see text).}
\label{tab:lmp_data}
\small
\begin{tabular}{rrrrrr}
\toprule
Step & $n_{\mathrm{Li}}$ & $x$ & $E_{\mathrm{tot}}$ (eV) & Removed Li idx & $V$ (V) \\
\midrule
 0 & 16 & 1.000 & $-798.22301$ & --  &  --  \\
 1 & 15 & 0.938 & $-792.66377$ & 0    & 3.656 \\
 2 & 14 & 0.875 & $-786.87415$ & 0    & 3.887 \\
 3 & 13 & 0.813 & $-781.24507$ & 0    & 3.726 \\
 4 & 12 & 0.750 & $-776.11056$ & 2    & 3.231 \\
 5 & 11 & 0.688 & $-770.55513$ & 4    & 3.652 \\
 6 & 10 & 0.625 & $-764.78390$ & 3    & 3.868 \\
 7 &  9 & 0.563 & $-759.17500$ & 7    & 3.706 \\
 8 &  8 & 0.500 & $-753.41416$ & 8    & 3.858 \\
 9 &  7 & 0.438 & $-747.53558$ & 4    & 3.975 \\
10 &  6 & 0.375 & $-741.48813$ & 0    & 4.144 \\
11 &  5 & 0.313 & $-735.44201$ & 0    & 4.143 \\
12 &  4 & 0.250 & $-729.42942$ & 0    & 4.109 \\
13 &  3 & 0.188 & $-723.31686$ & 0    & 4.209 \\
14 &  2 & 0.125 & $-717.09139$ & 1    & 4.322 \\
15 &  1 & 0.063 & $-710.84636$ & 0    & 4.342 \\
16 &  0 & 0.000 & $-704.89376$ & 0    & 4.050 \\
\bottomrule
\end{tabular}
\end{table}

\subsection{\texorpdfstring{LiMn$_2$O$_4$ (JVASP-141792, spinel, PBE$+U$)}{LiMn2O4 (JVASP-141792, spinel, PBE+U)}}
\label{si:lmo}

Functional PBE$+U$ with $U_{\mathrm{eff}}(\mathrm{Mn}) = \SI{3.90}{\eV}$; supercell $2\times 2\times 2$, 112 atoms, $N_{\mathrm{Li,total}} = 16$; $\mu_{\mathrm{Li}} = \SI{-1.9031}{\eV/atom}$. All 17 steps complete. Mean step voltage \SI{4.08}{\V}.

\begin{table}[ht]
\centering
\caption{LiMn$_2$O$_4$ delithiation data.}
\label{tab:lmo_data}
\small
\begin{tabular}{rrrrrr}
\toprule
Step & $n_{\mathrm{Li}}$ & $x$ & $E_{\mathrm{tot}}$ (eV) & Removed Li idx & $V$ (V) \\
\midrule
 0 & 16 & 1.000 & $-770.55839$ & --  &  --  \\
 1 & 15 & 0.938 & $-764.55243$ & 1    & 4.103 \\
 2 & 14 & 0.875 & $-758.50927$ & 7    & 4.140 \\
 3 & 13 & 0.813 & $-752.60997$ & 12   & 3.996 \\
 4 & 12 & 0.750 & $-746.65687$ & 6    & 4.050 \\
 5 & 11 & 0.688 & $-740.68039$ & 2    & 4.073 \\
 6 & 10 & 0.625 & $-734.70155$ & 3    & 4.076 \\
 7 &  9 & 0.563 & $-729.00933$ & 1    & 3.789 \\
 8 &  8 & 0.500 & $-723.21011$ & 1    & 3.896 \\
 9 &  7 & 0.438 & $-717.41584$ & 0    & 3.891 \\
10 &  6 & 0.375 & $-711.32434$ & 0    & 4.188 \\
11 &  5 & 0.313 & $-705.13426$ & 2    & 4.287 \\
12 &  4 & 0.250 & $-699.09585$ & 2    & 4.135 \\
13 &  3 & 0.188 & $-692.93691$ & 0    & 4.256 \\
14 &  2 & 0.125 & $-686.95928$ & 2    & 4.075 \\
15 &  1 & 0.063 & $-680.65063$ & 1    & 4.406 \\
16 &  0 & 0.000 & $-674.90141$ & 0    & 3.846 \\
\bottomrule
\end{tabular}
\end{table}

The hull shows the characteristic two-plateau spinel response: a lower plateau near \SI{4.00}{\V} and an upper plateau near \SI{4.17}{\V}, mirroring the experimental two-step profile, whose open-circuit voltage is \SI{4.11}{\V} where two cubic phases coexist ($0.27<x<0.60$) and centered at \SI{3.94}{\V} in the one-phase region ($0.60<x<1$)\cite{si-ohzuku1990lmo}; the computed step lies at $x=0.44$.

\subsection{\texorpdfstring{Li$_4$Mn$_3$Co$_2$Ni$_3$O$_{16}$ (JVASP-144791, layered, P1 in JARVIS-DFT, PBE$+U$)}{Li4Mn3Co2Ni3O16 (JVASP-144791, layered, P1 in JARVIS-DFT, PBE+U)}}
\label{si:nmc}

Nominal composition Li$_4$Mn$_3$Co$_2$Ni$_3$O$_{16}$ in the 112-atom supercell; functional PBE$+U$ with $U_{\mathrm{eff}} = \SI{3.90}{\eV}$ (Mn), \SI{3.32}{\eV} (Co), \SI{6.20}{\eV} (Ni); supercell $2\times 2\times 1$, 112 atoms, $N_{\mathrm{Li,total}} = 16$; $\mu_{\mathrm{Li}} = \SI{-1.9031}{\eV/atom}$. All 17 steps complete. Mean step voltage \SI{4.40}{\V}.

\begin{table}[ht]
\centering
\caption{Li$_4$Mn$_3$Co$_2$Ni$_3$O$_{16}$ delithiation data.}
\label{tab:nmc_data}
\small
\begin{tabular}{rrrrrr}
\toprule
Step & $n_{\mathrm{Li}}$ & $x$ & $E_{\mathrm{tot}}$ (eV) & Removed Li idx & $V$ (V) \\
\midrule
 0 & 16 & 1.000 & $-647.76742$ & --  &  --  \\
 1 & 15 & 0.938 & $-642.20577$ & 7    & 3.659 \\
 2 & 14 & 0.875 & $-636.93058$ & 5    & 3.372 \\
 3 & 13 & 0.813 & $-631.05676$ & 5    & 3.971 \\
 4 & 12 & 0.750 & $-625.02124$ & 8    & 4.132 \\
 5 & 11 & 0.688 & $-619.02464$ & 4    & 4.093 \\
 6 & 10 & 0.625 & $-613.08149$ & 5    & 4.040 \\
 7 &  9 & 0.563 & $-607.11961$ & 5    & 4.059 \\
 8 &  8 & 0.500 & $-601.03279$ & 4    & 4.184 \\
 9 &  7 & 0.438 & $-594.13237$ & 1    & 4.997 \\
10 &  6 & 0.375 & $-587.23771$ & 0    & 4.992 \\
11 &  5 & 0.313 & $-580.41171$ & 1    & 4.923 \\
12 &  4 & 0.250 & $-573.54717$ & 0    & 4.961 \\
13 &  3 & 0.188 & $-566.61139$ & 3    & 5.033 \\
14 &  2 & 0.125 & $-559.70774$ & 0    & 5.001 \\
15 &  1 & 0.063 & $-553.56873$ & 1    & 4.236 \\
16 &  0 & 0.000 & $-546.91740$ & 0    & 4.748 \\
\bottomrule
\end{tabular}
\end{table}

\paragraph{Comparison with experiment.} Two regimes are visible: $\sim$\SI{4.0}{\V} for $x > 0.5$ and $\sim$\SI{5.0}{\V} for $x < 0.5$. Per transition metal the cell spans Li$_{0.5}$TMO$_2$ to TMO$_2$, the upper half of the lithium range of NMC-111. Over that half the open-circuit voltage of NMC-111 measured by Yabuuchi and Ohzuku\cite{si-yabuuchi2004nmc} (intermittent charge, \SI{1}{\hour} on and \SI{5}{\hour} off, to \SI{4.71}{\V}, which they reach at Li$_{0.03}$ by charge count) averages \SI{4.19}{\V} from Li$_{0.5}$ to Li$_{0.25}$ ($x = 1\to0.5$ here), \SI{4.58}{\V} from Li$_{0.25}$ to Li$_{0.03}$ and \SI{4.37}{\V} overall; the values were read from the data of their Fig.~3. PBE$+U$ gives \SI{3.94}{\V}, \SI{4.86}{\V} and \SI{4.40}{\V}. The overall agreement ($+$\SI{0.03}{\V}) therefore combines a first half \SI{0.25}{\V} low with a second half \SI{0.28}{\V} high: the computed rise between the halves, \SI{0.92}{\V}, is more than twice the measured \SI{0.39}{\V}. The optB88-vdW$+U$ comparison and the cell-relaxed endpoints below show that the excess rise persists with the dispersion-corrected functional and with the cell relaxed; a hybrid-functional or meta-GGA calculation would show whether it is a GGA$+U$ artifact of the most oxidized states. Over the range in which NMC-111 is normally cycled, from full lithiation, the measured average is \SI{3.85}{\V} (3.00--\SI{4.45}{\V} at C/10)\cite{si-yoshizawa2003nmc}; the cell does not cover that range.

\paragraph{optB88-vdW$+U$ comparison.} The same supercell, MAGMOM and $U$ values were rerun with optB88-vdW$+U$ ($\mu_{\mathrm{Li}} = \SI{-0.9778}{\eV/atom}$, ISIF = 3 at step 0 and ISIF = 2 thereafter, as for the PBE$+U$ chain). All 17 steps converged without intervention (Table~\ref{tab:nmc_vdw}). The average voltage is \SI{4.76}{\V} over the full range against \SI{4.40}{\V} with PBE$+U$; over $x = 1\to0.5$ it is \SI{4.28}{\V} (PBE$+U$ \SI{3.94}{\V}) and over $x = 0.5\to0$ \SI{5.24}{\V} (PBE$+U$ \SI{4.86}{\V}). The two regimes are shifted by nearly the same amount, so the jump at $x = 0.5$ does not originate in the dispersion treatment. Hull plateaus: 4.10, 4.20, 4.34, 4.46~V for $x > 0.5$ and 5.18 to 5.34~V for $x < 0.5$.

\paragraph{Cell-relaxed endpoints.} Because both chains relax the cell only at step 0, the $x = 0.5$ and $x = 0$ structures of each chain were relaxed again with ISIF = 3 (same functional, $U$, MAGMOM and $k$-mesh, NSW = 300; all four converged). The averages over a composition range depend only on its two endpoints, so these four relaxations give the cell-relaxed averages directly (Table~\ref{tab:nmc_variants}). Relaxing the cell lowers every average in the table, including those over the two halves, by 0.21 to \SI{0.36}{\V} (0.22 to \SI{0.31}{\V} over the full range); at equal cell treatment optB88-vdW$+U$ lies 0.25 to \SI{0.38}{\V} above PBE$+U$ (0.27 to \SI{0.36}{\V} over the full range); and every variant rises by about \SI{0.9}{\V} between the two halves of the range.

\begin{table}[ht]
\centering
\caption{NMC-variant average voltages (V) for the two functionals with the cell fixed after step 0 (the automated protocol for this P1 entry) and with the $x = 0.5$ and $x = 0$ endpoints relaxed with ISIF = 3. Relaxing the cell contracts it, relative to the relaxed $x = 1$ cell, by 4.9\% and 9.1\% (PBE$+U$) and by 5.1\% and 9.8\% (optB88-vdW$+U$) at $x = 0.5$ and $x = 0$. Experiment: open-circuit voltage of NMC-111 over the same lithium range per transition metal (Li$_{0.5}\to$Li$_{0.25}$, Li$_{0.25}\to$Li$_{0.03}$ and Li$_{0.5}\to$Li$_{0.03}$)\cite{si-yabuuchi2004nmc}.}
\label{tab:nmc_variants}
\small
\setlength{\tabcolsep}{4pt}
\begin{tabular}{llccccc}
\toprule
Functional & Cell & $E(x{=}0.5)$ (eV) & $E(x{=}0)$ (eV) & $x = 1\to0.5$ & $x = 0.5\to0$ & $x = 1\to0$ \\
\midrule
PBE$+U$ & fixed & $-601.03279$ & $-546.91740$ & 3.94 & 4.86 & 4.40 \\
PBE$+U$ & relaxed & $-602.73942$ & $-550.48237$ & 3.73 & 4.63 & 4.18 \\
optB88-vdW$+U$ & fixed & $-389.35062$ & $-339.63082$ & 4.28 & 5.24 & 4.76 \\
optB88-vdW$+U$ & relaxed & $-391.41035$ & $-344.55144$ & 4.02 & 4.88 & 4.45 \\
\midrule
Experiment & -- & -- & -- & 4.19 & 4.58 & 4.37 \\
\bottomrule
\end{tabular}
\end{table}

\begin{table}[ht]
\centering
\caption{NMC-variant delithiation data with optB88-vdW$+U$. $V$ is the step voltage.}
\label{tab:nmc_vdw}
\small
\begin{tabular}{rrrrr}
\toprule
Step & $n_{\mathrm{Li}}$ & $x$ & $E_{\mathrm{tot}}$ (eV) & $V$ (V) \\
\midrule
 0 & 16 & 1.000 & $-431.40042$ & -- \\
 1 & 15 & 0.938 & $-426.26474$ & 4.158 \\
 2 & 14 & 0.875 & $-421.23843$ & 4.049 \\
 3 & 13 & 0.813 & $-416.04379$ & 4.217 \\
 4 & 12 & 0.750 & $-410.87728$ & 4.189 \\
 5 & 11 & 0.688 & $-405.42199$ & 4.477 \\
 6 & 10 & 0.625 & $-400.23609$ & 4.208 \\
 7 &  9 & 0.563 & $-394.76026$ & 4.498 \\
 8 &  8 & 0.500 & $-389.35062$ & 4.432 \\
 9 &  7 & 0.438 & $-383.17599$ & 5.197 \\
10 &  6 & 0.375 & $-376.95016$ & 5.248 \\
11 &  5 & 0.313 & $-370.87568$ & 5.097 \\
12 &  4 & 0.250 & $-364.69676$ & 5.201 \\
13 &  3 & 0.188 & $-358.47937$ & 5.240 \\
14 &  2 & 0.125 & $-352.25512$ & 5.246 \\
15 &  1 & 0.063 & $-345.95238$ & 5.325 \\
16 &  0 & 0.000 & $-339.63082$ & 5.344 \\
\bottomrule
\end{tabular}
\end{table}

\FloatBarrier
\subsection{\texorpdfstring{LiCoO$_2$ (JVASP-2017, layered R$\overline{3}$m, optB88-vdW$+U$)}{LiCoO2 (JVASP-2017, layered R-3m, optB88-vdW+U)}}
\label{si:lco}

Functional optB88-vdW$+U$ (\texttt{GGA = BO}, \texttt{PARAM1 = 0.1833333333}, \texttt{PARAM2 = 0.22}) with $U_{\mathrm{eff}}(\mathrm{Co}) = \SI{3.32}{\eV}$; supercell $2\times 2\times 2$, 32 atoms, $N_{\mathrm{Li,total}} = 8$; $\mu_{\mathrm{Li}} = \SI{-0.9778}{\eV/atom}$ (optB88-vdW Li\_sv reference). All 9 steps converged without intervention (ISIF~=~3 at every step, as for all whitelisted layered hosts). Final cell volumes: 258.0, 257.0, 257.8, 257.4, 258.1, 256.2, 251.1, 247.8, 241.6 \AA$^3$ from $x=1$ to $x=0$, i.e.\ within 1\% of the lithiated cell down to $x=3/8$ and contracting gradually to $-6.4$\% at the empty endpoint.

\begin{table}[ht]
\centering
\caption{LiCoO$_2$ delithiation data, optB88-vdW$+U$ (reported benchmark).}
\label{tab:lco_data}
\small
\begin{tabular}{rrrrrr}
\toprule
Step & $n_{\mathrm{Li}}$ & $x$ & $E_{\mathrm{tot}}$ (eV) & Removed Li idx & $V$ (V) \\
\midrule
0 & 8 & 1.000 & $-134.27144$ & -- & -- \\
1 & 7 & 0.875 & $-128.92973$ & 3 & 4.364 \\
2 & 6 & 0.750 & $-123.94777$ & 2 & 4.004 \\
3 & 5 & 0.625 & $-118.68284$ & 4 & 4.287 \\
4 & 4 & 0.500 & $-113.52493$ & 4 & 4.180 \\
5 & 3 & 0.375 & $-107.70730$ & 3 & 4.840 \\
6 & 2 & 0.250 & $-102.32348$ & 0 & 4.406 \\
7 & 1 & 0.125 & $-96.13330$ & 1 & 5.212 \\
8 & 0 & 0.000 & $-90.81549$ & 0 & 4.340 \\
\bottomrule
\end{tabular}
\end{table}

The hull identifies plateaus at \SI{4.18}{\V}, \SI{4.23}{\V}, \SI{4.62}{\V} and \SI{4.78}{\V}; the full-range average is \SI{4.45}{\V} and the average over $x=1\to0.5$, the experimentally cycled range, is \SI{4.21}{\V}. Against the two-phase plateaus at which the measured open-circuit voltage is flat (\SI{3.92}{\V} for $3/4<x<1$ and \SI{4.50}{\V} for $0<x<1/4$\cite{si-ohzuku1994lco}), the DFT averages over the same ranges are \SI{4.18}{\V} and \SI{4.78}{\V}: the \SI{0.58}{\V} rise between the plateaus is reproduced to within \SI{0.02}{\V}, with a uniform offset of about $+$\SI{0.27}{\V}.

\paragraph{Functional comparison on the same supercell.} Table~\ref{tab:lco_functionals} compares the reported run with a PBE$+U$ series (per-step data in Table~\ref{tab:lco_pbe}) on the same supercell, $U$, $k$-mesh and input template, and with the run of the preprint, which used the optPBE-vdW setting (\texttt{GGA = OR}) while being described as optB88-vdW (Supplementary Note~\ref{si:vasp}). PBE$+U$ gives nearly equal averages over the two experimental plateau ranges and so misses the rise between them (its hull plateaus climb by only \SI{0.25}{\V} over the whole range); the cell expands by 3.5\% on full delithiation, relative to its relaxed fully lithiated volume, instead of contracting. The original optPBE-vdW run used mostly the earlier input template and fixed-cell relaxations for steps 1--3, and its fully delithiated endpoint was rerun with ISIF~=~2 after the ISIF~=~3 relaxation expanded the cell by $\sim$28\%; with the correct optB88-vdW input and the current template this expansion did not recur, so it is not a property of the protocol. Its values are given for completeness only and do not isolate the functional.

\begin{table}[ht]
\centering
\caption{LiCoO$_2$ average voltages (V) by functional, $2\times 2\times 2$ supercell. Experimental values from Ohzuku and Ueda\cite{si-ohzuku1994lco}; the $x=1\to0.5$ reference, 3.97~V, is the composition-weighted average of their open-circuit voltage over that range.}
\label{tab:lco_functionals}
\small
\begin{tabular}{lcccc}
\toprule
 & $3/4<x<1$ & $x=1\to0.5$ & $0<x<1/4$ & full range \\
\midrule
Experiment & 3.92 & 3.97 & 4.50 & -- \\
optB88-vdW$+U$ (reported) & 4.18 & 4.21 & 4.78 & 4.45 \\
PBE$+U$ & 3.82 & 3.81 & 3.81 & 3.84 \\
optPBE-vdW$+U$ (original run) & 4.18 & 4.01 & 4.47 & 4.18 \\
\bottomrule
\end{tabular}
\end{table}

\begin{table}[ht]
\centering
\caption{LiCoO$_2$ delithiation data, PBE$+U$ ($\mu_{\mathrm{Li}} = \SI{-1.9031}{\eV/atom}$).}
\label{tab:lco_pbe}
\small
\begin{tabular}{rrrrrr}
\toprule
Step & $n_{\mathrm{Li}}$ & $x$ & $E_{\mathrm{tot}}$ (eV) & Removed Li idx & $V$ (V) \\
\midrule
0 & 8 & 1.000 & $-183.30930$ & -- & -- \\
1 & 7 & 0.875 & $-177.63411$ & 7 & 3.772 \\
2 & 6 & 0.750 & $-171.86272$ & 6 & 3.868 \\
3 & 5 & 0.625 & $-166.42037$ & 5 & 3.539 \\
4 & 4 & 0.500 & $-160.43819$ & 4 & 4.079 \\
5 & 3 & 0.375 & $-155.01766$ & 0 & 3.517 \\
6 & 2 & 0.250 & $-148.80970$ & 2 & 4.305 \\
7 & 1 & 0.125 & $-143.09988$ & 1 & 3.807 \\
8 & 0 & 0.000 & $-137.38152$ & 0 & 3.815 \\
\bottomrule
\end{tabular}
\end{table}

\clearpage
\subsection{\texorpdfstring{Olivine endpoints and cell-relaxed endpoints of the PBE$+U$ benchmarks}{Olivine endpoints and cell-relaxed endpoints of the PBE+U benchmarks}}
\label{si:olivine}

\paragraph{The phosphate entries.} The JARVIS-DFT entries of the LFP and LMP chains, JVASP-42723 (mp-850917) and JVASP-116897 (mp-776296), have the \textit{Pnma} spacegroup of olivine but place lithium in tetrahedral sites (Li--O 1.96--2.14~\AA); their cells are $4.92\times6.09\times9.39$ and $4.92\times6.36\times9.39$~\AA, they lie 0.024 and 0.015~eV/atom above the JARVIS-DFT hull, and neither has an ICSD record (Table~\ref{tab:si_ids}). Olivine places lithium in octahedral sites; the olivine LiMnPO$_4$ entry JVASP-91261 (mp-18997, four ICSD records, 1.5~meV/atom above the hull) has a $4.74\times6.11\times10.44$~\AA\ cell. Olivine LiFePO$_4$ is not in the JARVIS-DFT snapshot used here. The chains of Supplementary Notes~\ref{si:lfp} and~\ref{si:lmp} are therefore calculations on this polymorph; the main text uses them for the protocol and the force-field comparison and compares the olivine endpoints below with experiment.

\paragraph{Olivine endpoints.} Olivine LiMnPO$_4$ was taken from JVASP-91261 and olivine LiFePO$_4$ was built from the same cell with Fe on the Mn sites, each as a $2\times2\times1$ supercell (Li$_{16}M_{16}$P$_{16}$O$_{64}$, 112 atoms) with the settings of the chains: PBE$+U$ with $U_\mathrm{eff}(\mathrm{Fe})=5.30$ and $U_\mathrm{eff}(\mathrm{Mn})=3.90$~eV, the current input template, a $2\times2\times3$ $k$-mesh, a cold start and up to 300 ionic steps. The initial magnetic order is antiferromagnetic with zero net moment, with every nearest-neighbor pair of transition metals in a layer (3.92~\AA) and every nearest pair in adjacent layers (5.51~\AA) antiparallel. The fully lithiated cell was relaxed with ISIF = 3; the fully delithiated endpoint, with every lithium removed, was relaxed twice, at the relaxed $x=1$ cell with ISIF = 2 (the protocol of the chains) and with ISIF = 3. All six relaxations converged (15 to 54 ionic steps), kept \textit{Pnma} symmetry, octahedral lithium at $x=1$ (Li--O 2.10--2.25~\AA) and the antiferromagnetic order (Fe$^{2+}$ 3.76, Fe$^{3+}$ 4.33, Mn$^{2+}$ 4.65 and Mn$^{3+}$ 3.88--3.91~$\mu_B$), and formed no O--O bond. The fully lithiated olivine cells lie 29 (Fe) and 37 (Mn) meV/atom below the step-0 cells of the polymorph chains. Table~\ref{tab:olivine} gives the energies and averages: with the protocol of the chains the olivines lie within \SI{0.05}{\V} of the polymorph chains (3.60 and \SI{3.93}{\V}), and relaxing the $x=0$ cell lowers them by 0.10 and \SI{0.11}{\V}.

\begin{table}[ht]
\centering
\caption{Olivine endpoints ($2\times2\times1$ supercells, PBE$+U$, $\mu_{\mathrm{Li}} = \SI{-1.9031}{\eV/atom}$). Energies in eV, voltages in V. Fixed: the $x=0$ endpoint at the relaxed $x=1$ cell (ISIF = 2, the protocol of the chains); relaxed: the $x=0$ cell relaxed as well (ISIF = 3), with its volume change relative to $x=1$.}
\label{tab:olivine}
\small
\setlength{\tabcolsep}{4pt}
\begin{tabular}{lccccccc}
\toprule
Material & $E(x{=}1)$ & $E_\mathrm{fixed}(x{=}0)$ & $V_\mathrm{fixed}$ & $E_\mathrm{relaxed}(x{=}0)$ & $V_\mathrm{relaxed}$ & Volume & $V^{\mathrm{exp}}$ \\
\midrule
LiFePO$_4$ & $-764.16916$ & $-676.88849$ & 3.55 & $-678.53622$ & 3.45 & $-4.8$\% & 3.43 \\
LiMnPO$_4$ & $-802.41966$ & $-709.29623$ & 3.92 & $-711.01380$ & 3.81 & $-5.8$\% & 4.10 \\
\bottomrule
\end{tabular}
\end{table}

\paragraph{Cell-relaxed endpoints.} The LFP, LMP and LMO chains, like the NMC chain, relax the cell only at step 0. As for NMC (Table~\ref{tab:nmc_variants}), their $x=0.5$ and $x=0$ structures were relaxed again with ISIF = 3 (the INCAR of the chain step with NSW = 300 and a cold start; same functional, $U$, MAGMOM and $k$-mesh), which together with step 0 gives the cell-relaxed averages directly (Table~\ref{tab:cellrelax}). The conjugate-gradient relaxations of LMP and LMO at $x=0.5$ stalled and were rerun with the quasi-Newton optimizer and the current template from an intermediate geometry. The cold starts did not always reproduce the electronic state of the chain at the same geometry (LFP at $x=0.5$ started \SI{0.41}{\eV} above the chain step, LMO \SI{0.10}{\eV} below); these offsets move the half-range averages by at most \SI{0.05}{\V} and do not enter the full-range values. Relaxing the cell lowers the full-range average by \SI{0.14}{\V} on LMO and by 0.30 and \SI{0.44}{\V} on the two phosphate polymorphs, against 0.10 and \SI{0.11}{\V} on the olivines and \SI{0.22}{\V} on NMC; with relaxed cells the five benchmarks lie within \SI{0.3}{\V} of experiment (Table~\ref{tab:exp_benchmark}).

\begin{table}[ht]
\centering
\caption{Average voltages (V) of the LFP and LMP polymorph chains and the LMO chain with the cell fixed after step 0 (the automated protocol) and with the $x = 0.5$ and $x = 0$ endpoints relaxed with ISIF = 3 ($\mu_{\mathrm{Li}} = \SI{-1.9031}{\eV/atom}$). Relative to the relaxed $x = 1$ cell the relaxed cells change volume by $-1.6$\% and $-2.6$\% (LFP), $-2.5$\% and $-6.9$\% (LMP) and $-2.3$\% and $-6.0$\% (LMO) at $x = 0.5$ and $x = 0$.}
\label{tab:cellrelax}
\small
\setlength{\tabcolsep}{4pt}
\begin{tabular}{llccccc}
\toprule
Chain & Cell & $E(x{=}0.5)$ (eV) & $E(x{=}0)$ (eV) & $x = 1\to0.5$ & $x = 0.5\to0$ & $x = 1\to0$ \\
\midrule
LFP polymorph & fixed & $-717.30743$ & $-672.96607$ & 3.55 & 3.64 & 3.60 \\
LFP polymorph & relaxed & $-719.00914$ & $-677.75144$ & 3.34 & 3.25 & 3.30 \\
LMP polymorph & fixed & $-753.41416$ & $-704.89376$ & 3.70 & 4.16 & 3.93 \\
LMP polymorph & relaxed & $-755.27873$ & $-711.99261$ & 3.46 & 3.51 & 3.49 \\
LiMn$_2$O$_4$ & fixed & $-723.21011$ & $-674.90141$ & 4.02 & 4.14 & 4.08 \\
LiMn$_2$O$_4$ & relaxed & $-723.73915$ & $-677.10992$ & 3.95 & 3.93 & 3.94 \\
\bottomrule
\end{tabular}
\end{table}

\section{Summary of Experimental Benchmarks}

Table~\ref{tab:exp_benchmark} summarizes the DFT-predicted average voltages and experimental reference values used in the main-text benchmarking figures.

\begin{table}[ht]
\centering
\caption{Summary of DFT average voltages, experimental average voltages, and residuals, with the cell fixed after full lithiation (the automated protocol, as in the benchmark table of the main text) and with the cell also relaxed at the compositions that fix each average. Experimental values are literature average or plateau voltages, as detailed in the footnotes and in the benchmark table of the main text.}
\label{tab:exp_benchmark}
\small
\setlength{\tabcolsep}{3pt}
\begin{tabular}{llccccc>{\raggedright\arraybackslash}p{3.0cm}}
\toprule
Material & Functional & $V_{\mathrm{fixed}}^{\mathrm{DFT}}$ & $\Delta V$ & $V_{\mathrm{relaxed}}^{\mathrm{DFT}}$ & $\Delta V$ & $V_{\mathrm{avg}}^{\mathrm{exp}}$ & Experimental reference \\
\midrule
LiFePO$_4$\textsuperscript{a} & PBE$+U$ & 3.55 & $+0.12$ & 3.45 & $+0.02$ & 3.43 & Kondo 2018\cite{si-kondo2018lfp} \\
LiMnPO$_4$\textsuperscript{a} & PBE$+U$ & 3.92 & $-0.18$ & 3.81 & $-0.29$ & $\sim$4.10 & Aravindan 2013\cite{si-aravindan2013lmp} \\
LiMn$_2$O$_4$\textsuperscript{b} & PBE$+U$ & 4.08 & $+0.04$ & 3.94 & $-0.10$ & 4.04 & Ohzuku 1990\cite{si-ohzuku1990lmo} \\
Li$_4$Mn$_3$Co$_2$Ni$_3$O$_{16}$\textsuperscript{c} & PBE$+U$ & 4.40 & $+0.03$ & 4.18 & $-0.19$ & 4.37 & Yabuuchi \& Ohzuku 2004\cite{si-yabuuchi2004nmc} \\
LiCoO$_2$\textsuperscript{d} & optB88-vdW$+U$ & 4.21 & $+0.24$ & 4.21 & $+0.24$ & 3.97 & Ohzuku \& Ueda 1994\cite{si-ohzuku1994lco} \\
\bottomrule
\end{tabular}

\vspace{2pt}\begin{minipage}{\linewidth}\footnotesize \textsuperscript{a}Olivine phase from its two endpoints (Supplementary Note~\ref{si:olivine}); the chains of the polymorph entries JVASP-42723 and JVASP-116897 give 3.60 and 3.93~V with the cell fixed and 3.30 and 3.49~V with it relaxed. \textsuperscript{b}Composition-weighted open-circuit voltage (\SI{5}{\hour} rest): \SI{4.11}{\V} for $0.27<x<0.60$, taken to extend to $x=0$, and a one-phase region centered at \SI{3.94}{\V} for $0.60<x<1$. \textsuperscript{c}Per transition metal the cell spans Li$_{0.5}$TMO$_2$ to TMO$_2$; the reference is the open-circuit voltage of NMC-111 over the same range (to Li$_{0.03}$, the end of the measured charge at \SI{4.71}{\V}). Over the conventional cycling range from full lithiation the measured average is \SI{3.85}{\V}\cite{si-yoshizawa2003nmc}; see Supplementary Note~\ref{si:nmc}. \textsuperscript{d}Compared over $x=1\to0.5$, the experimentally cycled range; the experimental value is the composition-weighted average of the measured open-circuit voltage (\SI{10}{\hour} rest) over that range, obtained by digitizing the published curve (uncertainty about \SI{0.02}{\V}). Full-range DFT average 4.45~V; see Supplementary Note~\ref{si:lco}. The LCO chain relaxes the cell at every step, so both columns are the same.\end{minipage}
\end{table}

With the cell fixed after full lithiation all residuals are at most \SI{0.24}{\V}; with the cell also relaxed they are at most \SI{0.29}{\V} (LiMnPO$_4$). The LCO residual is close to the offset of $+$0.26 to $+$\SI{0.28}{\V} on its two-phase plateaus (Supplementary Note~\ref{si:lco}). The small NMC residual combines a first half \SI{0.25}{\V} low and a second half \SI{0.28}{\V} high; an optB88-vdW$+U$ rerun raises the average by \SI{0.36}{\V}, relaxing the cell at $x = 0.5$ and $x = 0$ lowers the PBE$+U$ average by \SI{0.22}{\V}, and neither removes the excess rise between the halves (Supplementary Note~\ref{si:nmc}).

\clearpage
\section{Top-50 candidates of the JARVIS-DFT screen}
\label{si:top50}

Table~\ref{tab:si_top50} lists the 50 highest-scoring candidates of the lithium-containing JARVIS-DFT screen described in the main text (filters: $3.0\le V_\mathrm{avg}\le 4.5$~V, $Q_\mathrm{grav}>100$~mAh/g per formula unit, $E_\mathrm{hull}\le 0.05$~eV/atom, $V_\mathrm{max}\le 5.5$~V, redox-active transition metal present; $682$ survivors of $7{,}193$). $V_\mathrm{avg}$ and $V_\mathrm{max}$ are ALIGNN-FF surrogate values; $Q_\mathrm{grav}=n_\mathrm{Li}F/(3.6\,M)$ is the theoretical capacity per formula unit for full lithium extraction; $E_\mathrm{hull}$ is the JARVIS-DFT value. The complete ranked list (\texttt{cathode\_candidates\_ranked.csv}) and the superseded per-cell-normalized ranking of the preprint (\texttt{cathode\_candidates\_ranked\_v1\_cellnorm.csv}, 71 entries) are both included in the code release; the prospective DFT campaign of the main text was selected on the latter.

{\small
\setlength{\tabcolsep}{4pt}
\begin{longtable}{rllcccccc}
\caption{Top-50 candidates of the JARVIS-DFT lithium cathode screen.}
\label{tab:si_top50}\\
\toprule
Rank & JARVIS ID & Composition & $V_\mathrm{avg}$ (V) & $V_\mathrm{max}$ (V) & $Q_\mathrm{grav}$ (mAh/g) & $E_\mathrm{hull}$ (eV/atom) & Score \\
\midrule
\endfirsthead
\toprule
Rank & JARVIS ID & Composition & $V_\mathrm{avg}$ (V) & $V_\mathrm{max}$ (V) & $Q_\mathrm{grav}$ (mAh/g) & $E_\mathrm{hull}$ (eV/atom) & Score \\
\midrule
\endhead
\bottomrule
\endfoot
1 & JVASP-117295 & \ce{Li3FeO3} & 4.39 & 5.47 & 645 & 0.001 & 0.639 \\
2 & JVASP-11735 & \ce{Li2MnO3} & 4.45 & 5.48 & 459 & 0.000 & 0.542 \\
3 & JVASP-122347 & \ce{KLi4FeO5} & 4.35 & 5.46 & 529 & 0.011 & 0.492 \\
4 & JVASP-122531 & \ce{Li3CrClO4} & 4.08 & 5.12 & 467 & 0.000 & 0.465 \\
5 & JVASP-31181 & \ce{Li3RuO4} & 4.06 & 5.00 & 433 & 0.000 & 0.439 \\
6 & JVASP-42740 & \ce{Li3CrO4} & 3.61 & 5.49 & 588 & 0.000 & 0.433 \\
7 & JVASP-148139 & \ce{Li5Mn3O8} & 4.09 & 5.34 & 409 & 0.002 & 0.419 \\
8 & JVASP-112917 & \ce{Li5Mn3O8} & 4.03 & 5.05 & 409 & 0.000 & 0.418 \\
9 & JVASP-118606 & \ce{Li2NiO2} & 3.73 & 5.13 & 513 & 0.000 & 0.414 \\
10 & JVASP-49957 & \ce{Li2RhO3} & 4.24 & 5.23 & 325 & 0.000 & 0.412 \\
11 & JVASP-147246 & \ce{Li5Mn3O8} & 4.06 & 5.30 & 409 & 0.004 & 0.400 \\
12 & JVASP-44367 & \ce{Li2VBPO7} & 4.38 & 5.05 & 245 & 0.000 & 0.396 \\
13 & JVASP-112912 & \ce{Li4Co3O7} & 4.20 & 5.11 & 339 & 0.003 & 0.391 \\
14 & JVASP-48209 & \ce{Li2VP2O8} & 4.44 & 5.17 & 210 & 0.000 & 0.386 \\
15 & JVASP-133673 & \ce{Li2CrP2O7} & 4.35 & 5.20 & 224 & 0.000 & 0.375 \\
16 & JVASP-144773 & \ce{Li4Mn3TeO12} & 4.36 & 5.19 & 209 & 0.000 & 0.369 \\
17 & JVASP-44908 & \ce{LiCrC2O6} & 4.49 & 4.88 & 150 & 0.000 & 0.361 \\
18 & JVASP-119348 & \ce{LiCoNiO4} & 4.49 & 5.08 & 142 & 0.000 & 0.358 \\
19 & JVASP-117435 & \ce{LiVOF3} & 4.30 & 4.99 & 205 & 0.000 & 0.352 \\
20 & JVASP-9699 & \ce{LiCuO} & 4.01 & 5.23 & 310 & 0.000 & 0.352 \\
21 & JVASP-45906 & \ce{LiMnCoO4} & 4.46 & 5.05 & 145 & 0.000 & 0.351 \\
22 & JVASP-137746 & \ce{LiVF4} & 4.30 & 5.44 & 200 & 0.000 & 0.349 \\
23 & JVASP-119318 & \ce{Li2CoNi3O8} & 4.45 & 5.00 & 142 & 0.000 & 0.348 \\
24 & JVASP-137727 & \ce{LiFe2O4} & 4.43 & 4.90 & 147 & 0.000 & 0.345 \\
25 & JVASP-112763 & \ce{Li3MnCoO5} & 4.29 & 5.50 & 375 & 0.017 & 0.343 \\
26 & JVASP-112435 & \ce{LiMn2O3F} & 4.41 & 5.30 & 146 & 0.000 & 0.342 \\
27 & JVASP-140204 & \ce{LiNi2O4} & 4.42 & 4.67 & 142 & 0.000 & 0.340 \\
28 & JVASP-137586 & \ce{Li4MnCo2O7} & 4.13 & 5.30 & 343 & 0.009 & 0.337 \\
29 & JVASP-145787 & \ce{Li7Mn2Co3O12} & 4.22 & 5.37 & 356 & 0.014 & 0.334 \\
30 & JVASP-43658 & \ce{Li2CrPCO7} & 4.10 & 5.24 & 243 & 0.000 & 0.332 \\
31 & JVASP-147136 & \ce{Li4CuNi3P4O16} & 4.38 & 5.04 & 166 & 0.002 & 0.331 \\
32 & JVASP-46250 & \ce{Li2CrO4} & 3.90 & 5.15 & 413 & 0.009 & 0.330 \\
33 & JVASP-144539 & \ce{Li4MnCo2O7} & 4.17 & 5.14 & 343 & 0.012 & 0.329 \\
34 & JVASP-142144 & \ce{Li4MnCo2O7} & 4.17 & 5.32 & 343 & 0.012 & 0.328 \\
35 & JVASP-143574 & \ce{Li4FeCu3P4O16} & 4.36 & 5.25 & 164 & 0.002 & 0.326 \\
36 & JVASP-46758 & \ce{Li6VCoP2C2O14} & 3.78 & 5.22 & 348 & 0.000 & 0.325 \\
37 & JVASP-142155 & \ce{Li7Mn2Co3O12} & 4.17 & 5.28 & 356 & 0.014 & 0.322 \\
38 & JVASP-141222 & \ce{Li4MnCo2O7} & 4.17 & 5.47 & 343 & 0.013 & 0.322 \\
39 & JVASP-122361 & \ce{Li4Cu3NiP4O16} & 4.42 & 5.14 & 163 & 0.005 & 0.321 \\
40 & JVASP-146723 & \ce{LiCrPO4} & 4.23 & 5.26 & 174 & 0.000 & 0.318 \\
41 & JVASP-117258 & \ce{Li2TiCr3O8} & 4.28 & 4.54 & 155 & 0.000 & 0.318 \\
42 & JVASP-113155 & \ce{Li2TiCr3O8} & 4.28 & 4.61 & 155 & 0.000 & 0.317 \\
43 & JVASP-46772 & \ce{Li3VPCO7} & 3.72 & 4.92 & 355 & 0.000 & 0.315 \\
44 & JVASP-143293 & \ce{Li4CoNi3O8} & 3.93 & 4.98 & 274 & 0.000 & 0.314 \\
45 & JVASP-141578 & \ce{Li2CoNiO4} & 3.93 & 5.07 & 274 & 0.000 & 0.313 \\
46 & JVASP-96563 & \ce{Li2MgMn3O8} & 4.23 & 4.39 & 162 & 0.000 & 0.311 \\
47 & JVASP-146620 & \ce{Li2FeNi3O8} & 4.28 & 4.75 & 143 & 0.000 & 0.310 \\
48 & JVASP-43821 & \ce{LiCrP2O7} & 4.35 & 5.03 & 115 & 0.000 & 0.310 \\
49 & JVASP-46694 & \ce{Li3Mn5OF11} & 4.24 & 5.38 & 154 & 0.000 & 0.309 \\
50 & JVASP-50374 & \ce{Li2TiTeO6} & 4.14 & 4.83 & 188 & 0.000 & 0.307 \\
\end{longtable}}

\section{Regressor residuals, training curves and tier-1-front ranking}
\label{si:regressor}

\begin{figure}[H]
\centering
\includegraphics[width=\textwidth]{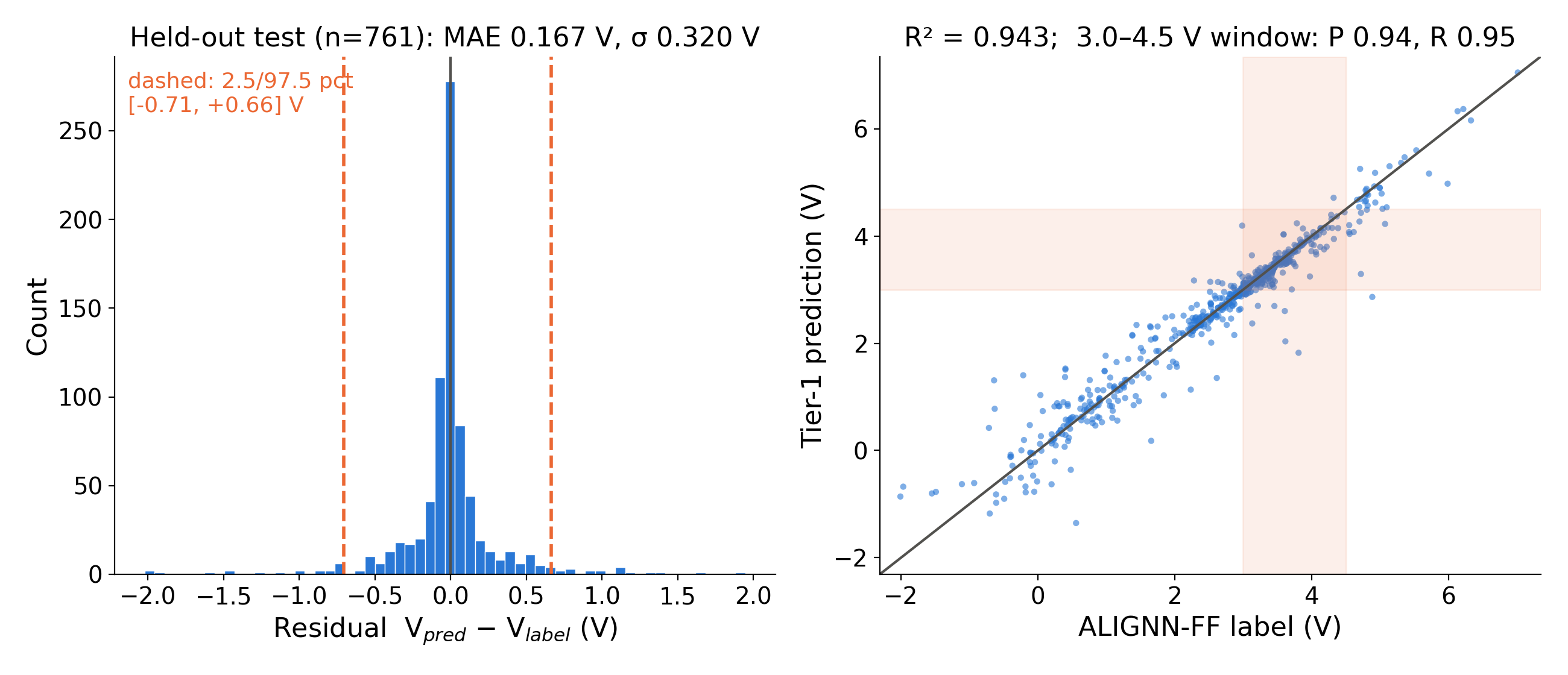}
\caption{\textbf{Held-out residuals of the ALIGNN regressor.} Left: distribution of $V_\mathrm{pred}-V_\mathrm{label}$ on the $761$ test structures (MAE $0.167$~V, standard deviation $0.320$~V; dashed lines mark the central 95\% interval, $[-0.71,+0.66]$~V). Right: parity with the $3.0$--$4.5$~V screening window shaded; precision $0.94$ and recall $0.95$ for the in-window decision.}
\label{fig:si-residuals}
\end{figure}

\begin{figure}[H]
\centering
\includegraphics[width=\textwidth]{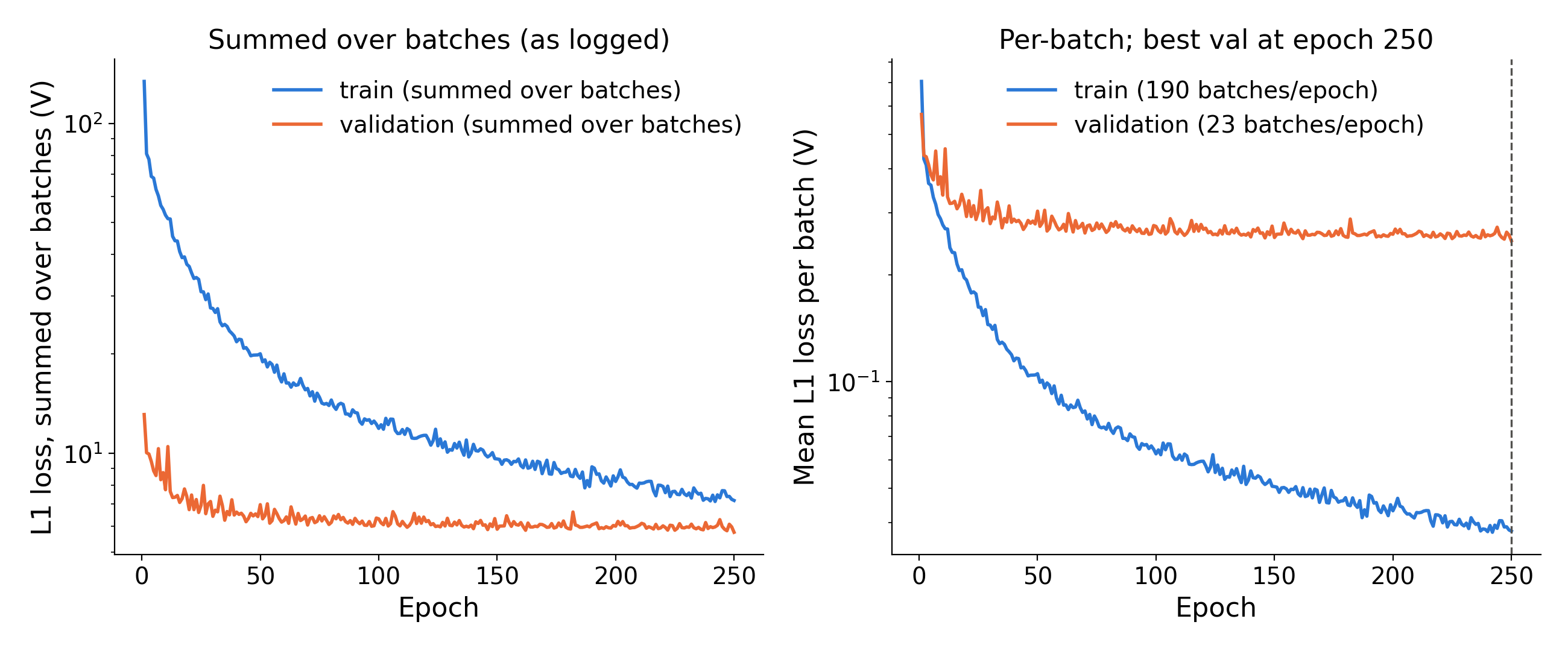}
\caption{\textbf{Training and validation loss of the ALIGNN regressor.} L1 (mean-absolute-error) loss summed over batches (left) and per batch (right) for the $250$ training epochs; the best validation epoch is the last one. The final training MAE is $0.04$~V against $0.25$~V on validation and $0.17$~V on the held-out test set.}
\label{fig:si-loss}
\end{figure}

\begin{figure}[H]
\centering
\includegraphics[width=\textwidth]{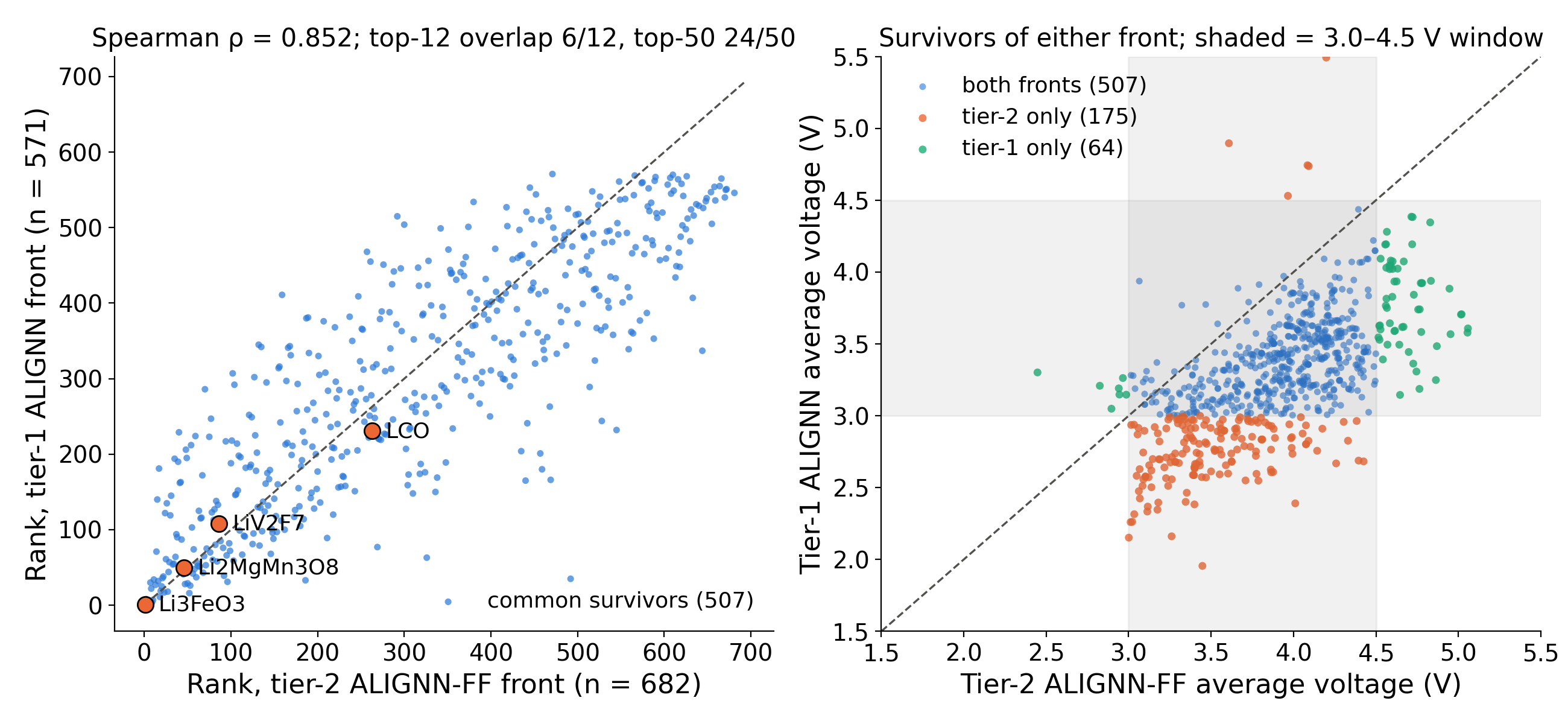}
\caption{\textbf{The regressor as the screening front.} Left: rank under the tier-1 (ALIGNN regressor) front versus rank under the tier-2 (ALIGNN-FF) front for the $507$ materials that survive the identical filters under both (the $5.5$~V maximum-voltage cap is taken from the force-field profile in both) (Spearman $0.85$); the benchmark LCO and three prospective candidates are labeled. Right: tier-1 versus tier-2 average voltage for the survivors of either front; the $175$ tier-2-only survivors span the lower part of the window on the tier-2 scale (median $3.45$~V) and $168$ of them fall below $3.0$~V on the tier-1 scale, while $57$ of the $64$ tier-1-only survivors lie above $4.5$~V on the tier-2 scale (median $4.61$~V), reflecting the $0.63$~V offset between the two force-field label sets. Two tier-2-only survivors with tier-1 voltages above $6.4$~V lie outside the plotted range.}
\label{fig:si-front}
\end{figure}

Figure~\ref{fig:si-residuals} shows the residual distribution of the ALIGNN average-voltage regressor on the $761$ held-out test structures and the parity against the ALIGNN-FF labels with the $3.0$--$4.5$~V screening window marked. Figure~\ref{fig:si-loss} shows the training and validation losses over the $250$ epochs. Figure~\ref{fig:si-front} compares the ranking obtained with the regressor as the screening front against the force-field ranking of the main text.

\subsection{Regressor architecture and training details}

The atom graph $G$ of the ALIGNN regressor has nodes for every atom in the cell and edges between every pair of atoms within a cutoff radius of $8$~\AA, capped at $12$ neighbors per atom; bond features are radial basis functions over $80$ distance centers. The line graph $L(G)$ has nodes for every bond in $G$ and edges between bond pairs that share an atom; triplet features are radial basis functions over $40$ angle centers. Atom features are seeded from the 92-dimensional CGCNN element fingerprint. The forward pass is four ALIGNN layers (each of which first updates atom and bond features on $G$ and then bond and triplet features on $L(G)$, so that angular information enters the bond features at the end of each layer) followed by four edge-gated GCN layers on $G$ alone, with $256$ hidden features. Atom features are averaged (mean-pooled) to a graph vector and projected to a single scalar by a final linear layer; the scalar is the predicted average voltage for the structure. The model has $\sim 4$~M parameters and $16$~MB of weights at fp32. The one-cycle schedule and $2{,}000$ warm-up steps named in the training configuration file have no effect for this model class, because the ALIGNN training script attaches the scheduler to an optimizer instance that it then replaces, so the learning rate stays at $10^{-3}$ throughout.

The grouped-split retraining of the main-text section ``The regressor as the screening front'' ran on GPU nodes without the DGL graph library and therefore used the pure-PyTorch implementation in the ALIGNN code base (\texttt{alignn\_atomwise\_pure}), with the same layer counts, hidden size, $8$~\AA\ cutoff and $12$-neighbor cap ($4{,}026{,}753$ parameters against $4{,}026{,}873$), a radius-based rather than $k$-nearest neighbor search, and the angular (line-graph) cutoff set equal to the $8$~\AA\ cutoff; the five random folds, trained identically, are the baseline against which the grouped folds are compared.

\section{Benchmark crystal structures}
\label{si:structures}

\begin{figure}[H]
\centering
\includegraphics[width=\textwidth]{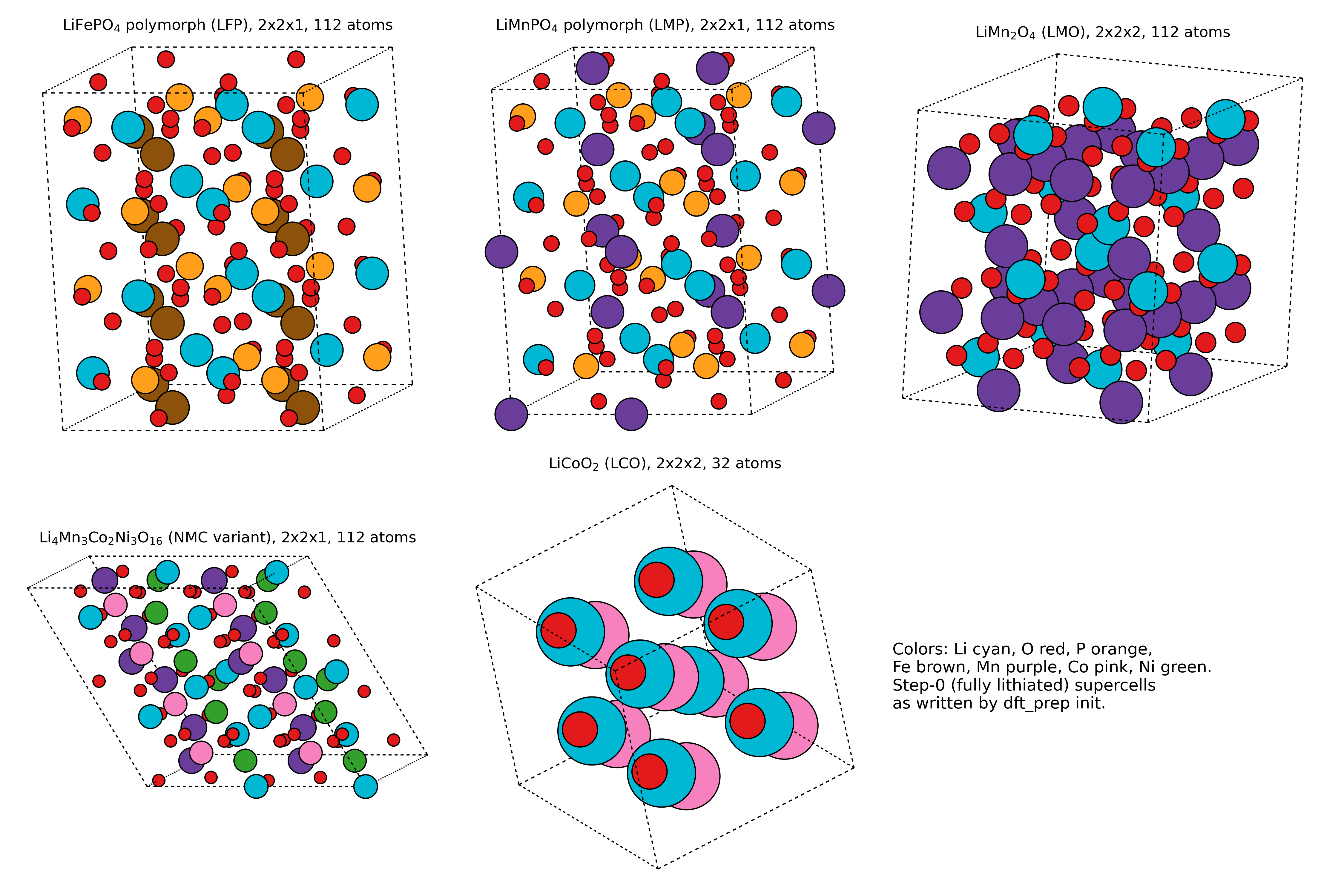}
\caption{\textbf{Step-0 supercells of the five benchmark cathodes.} Ball projections of the input structures for the DFT campaigns (colors: Li cyan, O red, P orange, Fe brown, Mn purple, Co pink, Ni green). The NMC entry (JVASP-144791) has composition \ce{Li4Mn3Co2Ni3O16} (Li:TM $=1{:}2$). Database identifiers and ICSD codes are listed in Table~\ref{tab:si_ids}. The LFP and LMP cells are the JARVIS-DFT polymorph entries used for the chains, not olivine (Supplementary Note~\ref{si:olivine}).}
\label{fig:si-structures}
\end{figure}

Figure~\ref{fig:si-structures} shows the fully lithiated step-0 supercells of the five benchmark cathodes exactly as written by \texttt{dft\_prep init}. The JVASP-144791 cell has composition \ce{Li4Mn3Co2Ni3O16} (Li:TM $=1{:}2$) and is a partially delithiated layered variant rather than NMC-111; see the main text for the consequences.

\section{Database identifiers for the materials discussed}
\label{si:ids}

Table~\ref{tab:si_ids} lists, for the benchmark, olivine, top-12 and prospective-campaign materials of the main text, the JARVIS-DFT identifier, composition, spacegroup, hull energy and the ICSD collection codes recorded in the JARVIS-DFT metadata. The MP source column gives the entry from which JARVIS-DFT took the starting structure (its \texttt{reference} field); the JARVIS entry is an optB88-vdW re-relaxation of that structure, so the identifier marks its source rather than a structure match, and JVASP-118606 (JQE-568816) and JVASP-154749 (auid-07879d18a5b5ea96) were not taken from the Materials Project; where JARVIS records no ICSD code the entry is a JARVIS-generated or database-derived structure and is marked as such.

{\small
\setlength{\tabcolsep}{4pt}
{\footnotesize\setlength{\tabcolsep}{3.5pt}
\begin{longtable}{llllll>{\raggedright\arraybackslash}p{0.12\textwidth}}
\caption{Identifiers for the benchmark, olivine, top-12 and prospective-campaign materials.}
\label{tab:si_ids}\\
\toprule
Set & JARVIS ID & Composition & Spacegroup & $E_\mathrm{hull}$ (eV/at.) & MP source & ICSD codes \\
\midrule
\endfirsthead
\toprule
Set & JARVIS ID & Composition & Spacegroup & $E_\mathrm{hull}$ (eV/at.) & MP source & ICSD codes \\
\midrule
\endhead
\bottomrule
\endfoot
Benchmark & JVASP-42723 & \ce{LiFePO4} & Pnma & 0.024 & mp-850917 & none \\
Benchmark & JVASP-116897 & \ce{LiMnPO4} & Pnma & 0.015 & mp-776296 & none \\
Benchmark & JVASP-141792 & \ce{LiMn2O4} & Fd$\bar{3}$m & 0.000 & mp-1272804 & none \\
Benchmark & JVASP-144791 & \ce{Li4Mn3Co2Ni3O16} & P1 & 0.079 & mp-867541 & none \\
Benchmark & JVASP-2017 & \ce{LiCoO2} & R$\bar{3}$m & 0.023 & mp-24850 & 246411, 246412, 246413 (+20 more) \\
Olivine & JVASP-91261 & \ce{LiMnPO4} & Pnma & 0.002 & mp-18997 & 38208, 25834, 97763, 99858 \\
Olivine & JVASP-121734 & \ce{LiMnPO4} & Pnma & 0.002 & mp-755784 & none \\
Top-12, prosp. & JVASP-117295 & \ce{Li3FeO3} & P4$_{2}$/mnm & 0.001 & mp-771894 & none \\
Top-12 & JVASP-11735 & \ce{Li2MnO3} & C2/m & 0.000 & mp-18988 & 202639 \\
Top-12 & JVASP-122347 & \ce{KLi4FeO5} & P$\bar{1}$ & 0.011 & mp-756561 & none \\
Top-12 & JVASP-122531 & \ce{Li3CrClO4} & P4/nmm & 0.000 & mp-756743 & none \\
Top-12 & JVASP-31181 & \ce{Li3RuO4} & P2/c & 0.000 & mp-37692 & none \\
Top-12 & JVASP-42740 & \ce{Li3CrO4} & I$\bar{4}$2m & 0.000 & mp-850964 & none \\
Top-12 & JVASP-148139 & \ce{Li5Mn3O8} & P2/m & 0.002 & mp-1174646 & none \\
Top-12 & JVASP-112917 & \ce{Li5Mn3O8} & P$\bar{1}$ & 0.000 & mp-1174741 & none \\
Top-12 & JVASP-118606 & \ce{Li2NiO2} & Immm & 0.000 & \texttt{JQE-568816} & none \\
Top-12 & JVASP-49957 & \ce{Li2RhO3} & C2/m & 0.000 & mp-754870 & none \\
Top-12 & JVASP-147246 & \ce{Li5Mn3O8} & C2/m & 0.004 & mp-1174519 & none \\
Top-12 & JVASP-44367 & \ce{Li2VBPO7} & P2$_{1}$/m & 0.000 & mp-772254 & none \\
Prospective & JVASP-117419 & \ce{LiCr2P2O8} & Pc & 0.000 & mp-753999 & none \\
Prospective & JVASP-154749 & \ce{Rb2LiFeF6} & R$\bar{3}$m & 0.010 & AFLOW (see text) & none \\
Prospective & JVASP-96563 & \ce{Li2MgMn3O8} & P4$_{3}$32 & 0.000 & mp-585492 & 90648 \\
Prospective & JVASP-116849 & \ce{LiV2F7} & C2/c & 0.000 & mp-756227 & none \\
Prospective & JVASP-95533 & \ce{LiCoH24C8N8O12} & P$\bar{3}$m1 & 0.000 & mp-600348 & 110737, 110738 \\
Prospective & JVASP-141543 & \ce{Rb2LiFeF6} & R$\bar{3}$m & 0.000 & mp-22663 & none \\
Prospective & JVASP-97428 & \ce{LiH4SNO4} & Pna2$_{1}$ & 0.000 & mp-557560 & 86232, 160683, 73989 \\
Prospective & JVASP-81829 & \ce{LiMnIr2} & Immm & 0.712 & mp-1097473 & none \\
Prospective & JVASP-85076 & \ce{LiCaTl2} & Fm$\bar{3}$m & 0.021 & mp-1097074 & none \\
Prospective & JVASP-77457 & \ce{LiCa2Ag} & C2/m & 0.121 & mp-1097380 & none \\
Prospective & JVASP-80802 & \ce{LiY2Ga} & Immm & 0.313 & mp-1095811 & none \\
\end{longtable}
}}

\section{JARVIS-DFT screening distributions}
\label{si:screening-dist}

\begin{figure}[H]
\centering
\includegraphics[width=0.85\textwidth]{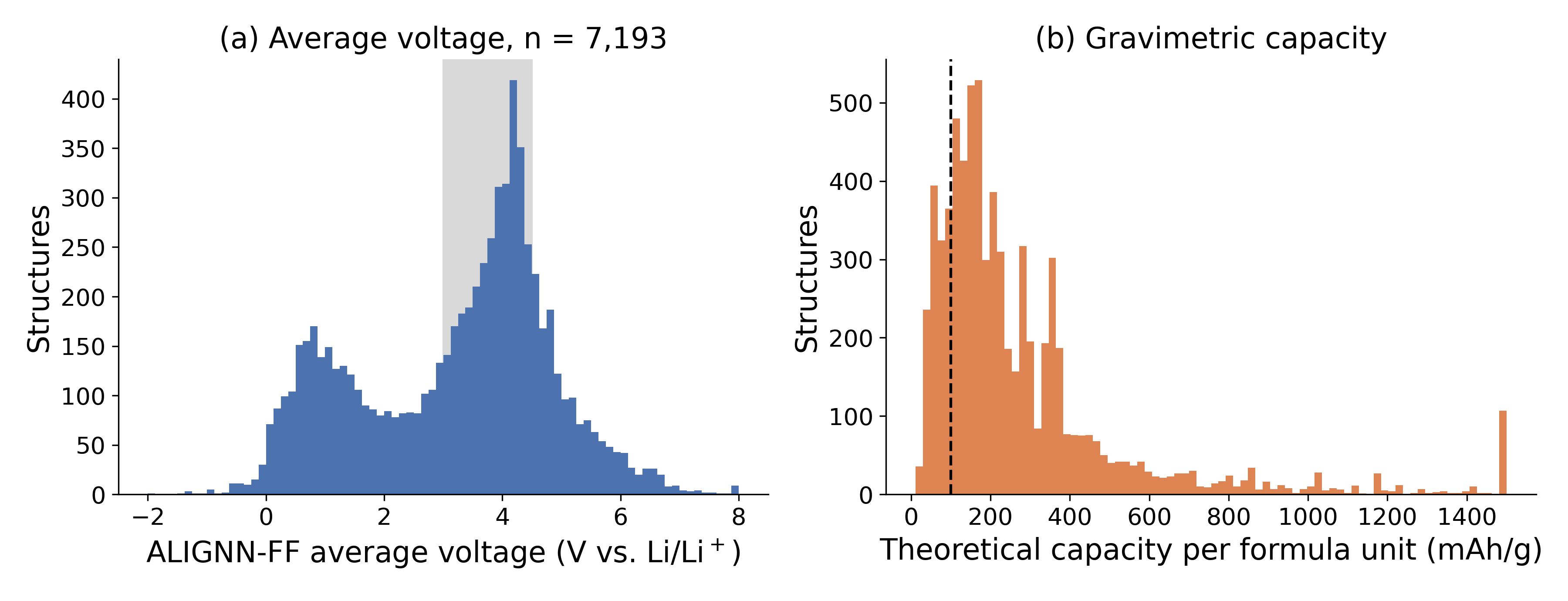}
\caption{\textbf{Screening distributions across the lithium-containing JARVIS-DFT pool.} (a) ALIGNN-FF-predicted average intercalation voltage over the $7{,}193$-entry lithium pool; the shaded band is the $3.0$--$4.5$~V ranking window. (b) Theoretical gravimetric capacity per formula unit over the same pool; the dashed line marks $100$~mAh/g. Values outside the plotted ranges are accumulated in the end bins ($8$ voltages; $105$ capacities above $1{,}500$~mAh/g).}
\label{fig:distribution}
\end{figure}

The high-throughput ALIGNN-FF screen of the lithium-containing JARVIS-DFT pool (described in the main text) yields the average-voltage and gravimetric-capacity distributions in Figure~\ref{fig:distribution}. The voltage distribution is bimodal, with a minor mode near $1$~V and its main peak near $4.2$~V; the $3.0$--$4.5$~V ranking band captures the main peak together with the polyanion operating window (${\sim}3.5$~V). The capacity distribution (theoretical, per formula unit, for extraction of all lithium) is right-skewed because $Q_\mathrm{grav}$ scales as $n_\mathrm{Li}/M$, and lithium-rich light-element hosts reach compositional ceilings far above the $144$--$274$~mAh/g of the five benchmark chemistries; the dashed line marks the $100$~mAh/g screening threshold.

\section{Exhaustive lithium-vacancy enumeration}
\label{si:enumeration}

\begin{figure}[H]
\centering
\includegraphics[width=\textwidth]{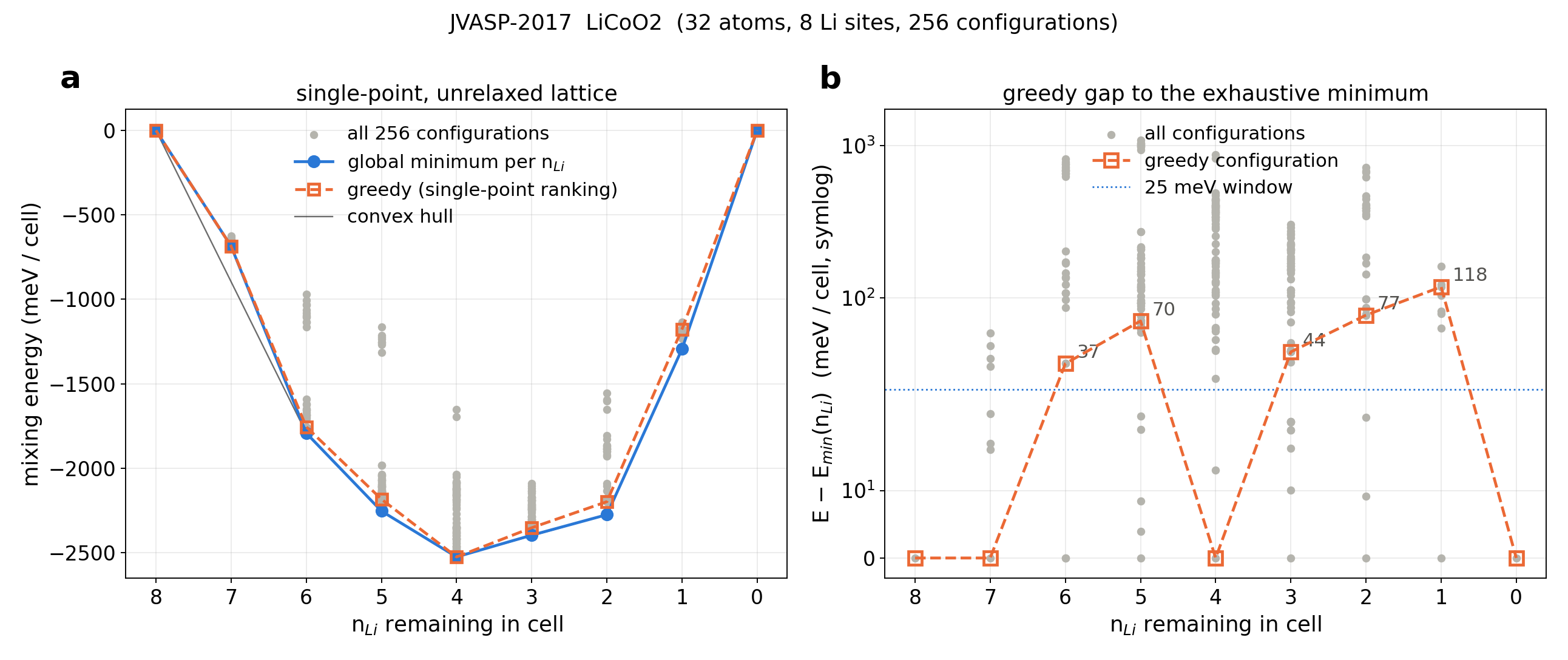}\\[0.6em]
\includegraphics[width=\textwidth]{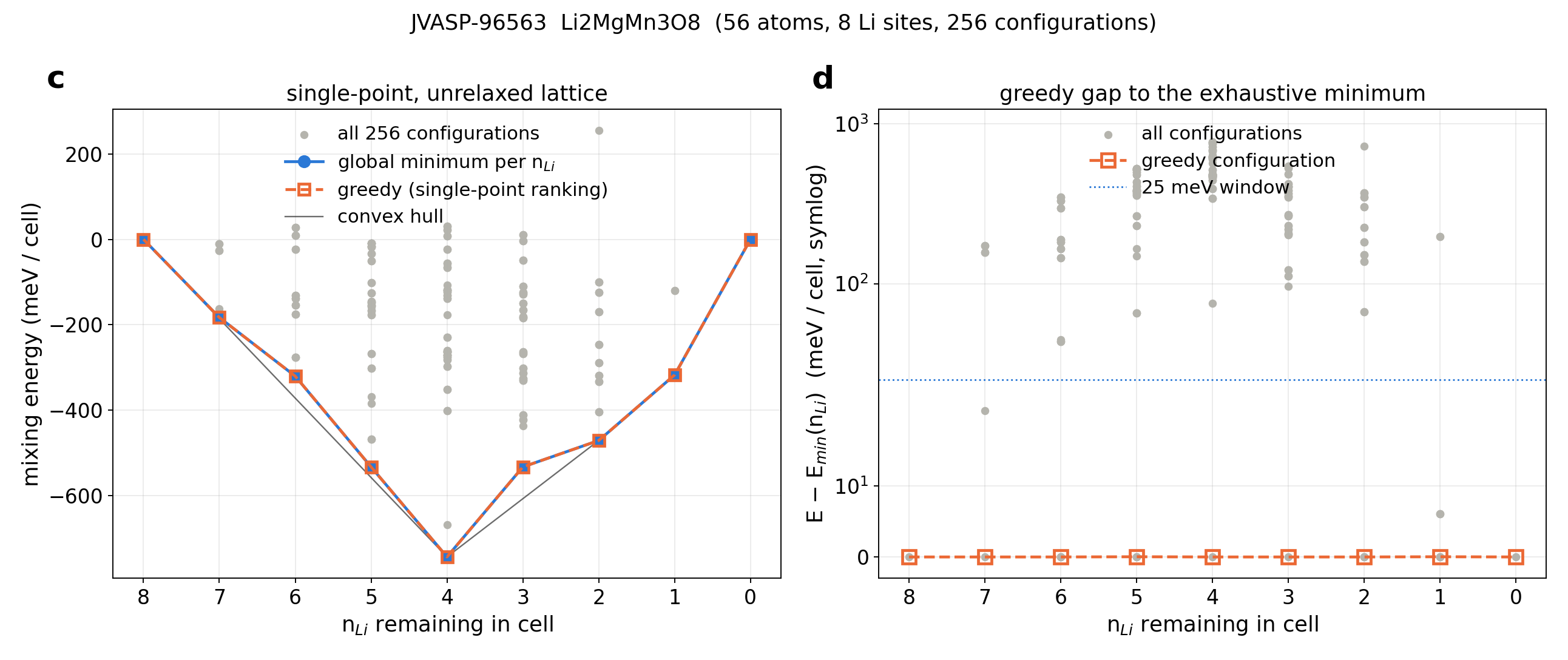}
\caption{\textbf{Greedy vacancy rule versus exhaustive enumeration.} (a, b) \ce{LiCoO2}, $2{\times}2{\times}2$ supercell, $8$ lithium sites. (c, d) \ce{Li2MgMn3O8}, $8$ lithium sites. (a, c) Mixing energy of every configuration versus lithium content, with the greedy path and the per-composition minimum. (b, d) Energy of the greedy configuration above the exhaustive minimum at each composition.}
\label{fig:si-enum}
\end{figure}

Figure~\ref{fig:si-enum} shows the ALIGNN-FF single-point energies of all $256$ lithium-vacancy configurations of the $2{\times}2{\times}2$ \ce{LiCoO2} supercell and of the \ce{Li2MgMn3O8} cell as a function of lithium content, with the greedy path of the DFT-tier vacancy rule and the per-composition exhaustive minimum overlaid; the numerical tables for \ce{LiCoO2} are in \texttt{analysis/enumeration/enumeration\_summary.md} and the per-configuration energies of both cells in \texttt{analysis/enumeration/results\_JVASP-*.json} in the code release.

Where the greedy path missed the exhaustive minimum of the \ce{LiCoO2} cell (five of the seven intermediate compositions), the greedy configuration lay $37$--$118$~meV per cell ($5$--$15$~meV per formula unit) above the minimum and ranked between fourth and twelfth of the $8$ to $56$ configurations at its composition. The hull plateaus of the greedy and exhaustive paths differ by at most $0.12$~V (on the highest plateau) and by $0.07$~V or less elsewhere. Two further observations bound the interpretation. First, symmetry-equivalent configurations are not degenerate in ALIGNN-FF: the eight equivalent single vacancies of the \ce{LiCoO2} cell spread over $59$~meV per cell and those of \ce{Li2MgMn3O8} over $173$~meV per cell, a symmetry-breaking artifact of the fixed-neighbor-count graph construction, so two of the five \ce{LiCoO2} gaps are within the force field's own noise floor. Second, for \ce{LiCoO2} the exhaustive minima are not nested along any single removal sequence (the minimum at $n-1$ lithium is reached from the minimum at $n$ by removing one lithium only at the two ends of the range), so no sequential protocol, greedy or otherwise, can visit every minimum; a path-independent hull requires enumeration at each composition. The greedy rule is therefore an approximation whose error on these two cells is at most $15$~meV per formula unit in configuration energy and $0.12$~V in any single plateau, at the level of the single-point energies that the rule uses.

We also attempted the comparison with every configuration relaxed in ALIGNN-FF (FIRE on ions and cell, force tolerance $0.05$~eV/\AA, $300$-step cap). Only $3$ of the $512$ relaxations converged: after $300$ steps the median residual force was $0.97$~eV/\AA\ for \ce{LiCoO2} and $1.24$~eV/\AA\ for \ce{Li2MgMn3O8}, the energies had dropped by a median of $0.47$ and $0.86$~eV per cell below the single points, and some \ce{LiCoO2} cells had contracted by up to $12$\%. The force field already returns residual forces of $2.3$ and $2.0$~eV/\AA\ on the DFT-relaxed parent cells, so a force-converged relaxation of these cells with the \texttt{alignnff\_wt01} checkpoint is not attainable, and a greedy-versus-exhaustive comparison on the capped energies would measure force noise rather than ordering. This is the reason the DFT tier ranks vacancies by single-point energy and leaves relaxation to DFT; the single-point comparison is the relevant bound for the rule as used.

\section{Web-application force-field profiles}
\label{si:webapp}

\begin{figure}[H]
\centering
\includegraphics[width=0.85\textwidth]{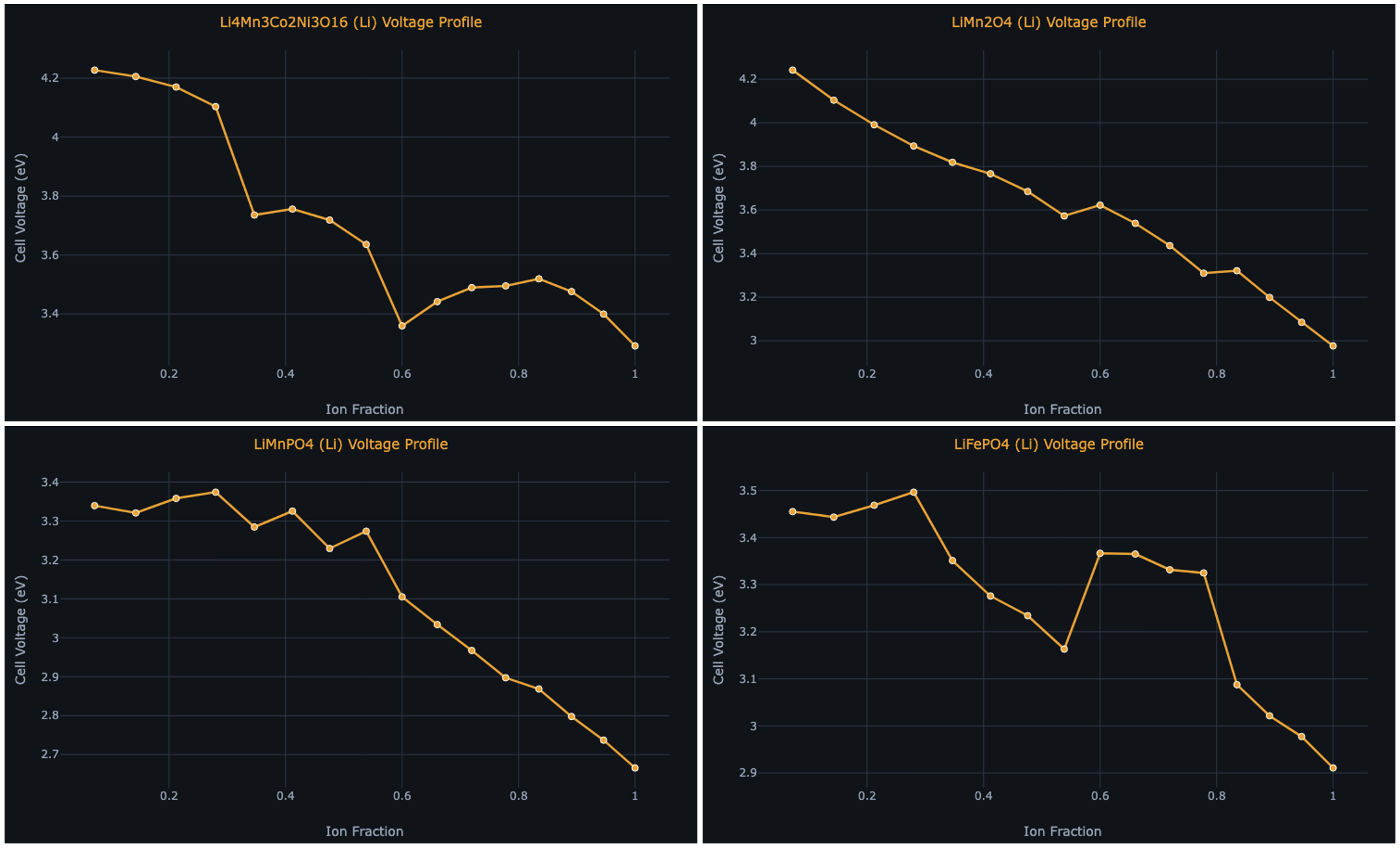}
\caption{\textbf{ALIGNN-FF voltage profiles returned by the web application for four of the five benchmark cathodes.} Panels show the predicted delithiation profile for \ce{Li4Mn3Co2Ni3O16} (NMC), \ce{LiMn2O4} (LMO), \ce{LiMnPO4} (LMP), and \ce{LiFePO4} (LFP); the vertical axes are in volts. These profiles follow the training-label protocol of the regressor (main-text section ``Average voltage prediction with ALIGNN''; their averages match the labels) and lie $0.07$ to $0.49$~V below the screening-pass averages of the same entries in main-text Fig.~5 (for LMP and LFP the JARVIS-DFT polymorph entries, not olivine); the two force-field passes differ in checkpoint and lithium reference (main-text section ``The regressor as the screening front'').}
\label{fig:si-webapp}
\end{figure}

Figure~\ref{fig:si-webapp} shows the ALIGNN-FF delithiation profiles returned by the AtomGPT Battery Explorer web application for four of the five benchmark cathodes.

\section{Convergence record of the benchmark chains}
\label{si:convergence}

This note gives the details behind the convergence summary in the Methods (``Endpoints, convergence and fallbacks'') of the main text. Apart from the fully delithiated LMP endpoint, which was rerun after a setup error in its pseudopotentials, ten benchmark relaxations ran in more than one job. One was a recovery: the first run of LFP step~1 reached the $500$-iteration self-consistency cap in two ionic steps and stopped, and the step was continued from its last structure with reduced charge mixing (\texttt{AMIX~=~0.1}, \texttt{BMIX~=~0.01}, \texttt{AMIX\_MAG~=~0.2}, \texttt{BMIX\_MAG~=~0.0001}). The other nine stopped before convergence when their jobs ended (job cancellations, where the job logs survive) and were continued from their last structure (LFP steps~0 and 2, LMP step~13, LMO steps~0 and 13, NMC steps~0 and 2) or restarted from the generated structure (LMP step~2, LMO step~16); two steps of the superseded optPBE-vdW LCO run were continued in the same way.

The input template changed during the benchmark campaigns (Supplementary Note~\ref{si:vasp}): $14$ of $17$ LFP steps, $14$ of $17$ NMC steps, $3$ of $17$ LMO steps, $3$ of $17$ LMP steps (the rerun endpoint with a cold start among them; for step~0 the original relaxation is counted, while its reported energy comes from a five-step ISIF $=2$ recompute in the alternating magnetic order with the current template at the relaxed step-0 geometry) and all $9$ steps of the reported LCO chain use the current template, the remainder the earlier one. Four transitional steps that already used IBRION $=1$, NELM $=100$ and restarts, but not the conservative mixing, are counted with the earlier template; they are the four steps that were continued after the template change, LFP and NMC step~2 and LMO and LMP step~13. The first three steps of the LMP chain and of the superseded optPBE-vdW LCO run used \texttt{EDIFF} $= 10^{-6}$~eV, most of them with a force criterion \texttt{EDIFFG} $= -0.03$~eV/\AA.

The olivine endpoints of the main-text section ``DFT validation on five commercial cathodes'' were run with the current template and the cell-relaxed endpoints of the main-text section ``Benchmarking against experiment and error analysis'' with the INCAR of the corresponding chain step, all separately, with a cold start, up to $300$ ionic steps and ISIF $=3$, except for the olivine $x=0$ endpoint at the cell of $x=1$ (ISIF $=2$). All converged, two of the cell-relaxed ones (LMP and LMO at $x=0.5$, whose chain steps used the earlier conjugate-gradient template) only when rerun with the current template after the cell relaxation had stalled.

Convergence statistics extracted from the final run of each relaxation of the five benchmark chains (for LMP step~0, the original relaxation) are as follows: the median number of ionic steps per relaxation is $41$ (LFP), $41$ (LMP), $28$ (LMO), $17$ (NMC) and $22$ (LCO), the median number of self-consistency iterations per relaxation is $308$, $277$, $177$, $233$ and $420$, and the self-consistency cap was reached in $3$ (LFP), $2$ (NMC) and $2$ (LCO) individual ionic steps and in none for LMP and LMO. Counting the earlier runs of the split relaxations as well, the medians become $41$, $46$, $31$, $17$ and $22$ ionic steps and $340$, $283$, $191$, $235$ and $420$ self-consistency iterations, and the cap was also reached in two further ionic steps of LFP (the first run of step~1) and five of LMO (an interrupted run of step~16); the log of the earlier run of NMC step~2 was not kept, so that run is not counted.

\section{Cost per tier}
\label{si:cost}

\begin{table}[H]
\small
\centering
\caption{Order-of-magnitude cost per structure for each tier of the BatteryMat pipeline. The ALIGNN time is measured in this work (main-text section ``The regressor as the screening front''); the ALIGNN-FF time is a representative single-point time per delithiation step; DFT-tier numbers are characteristic of the supercell relaxations of the benchmark campaigns on CPU nodes ($64$ cores for the $112$-atom PBE$+U$ chains, $24$--$32$ for the $32$-atom LCO chain) and vary by an order of magnitude or more with delithiation depth (see text).}
\label{tab:speed}
\begin{tabular}{p{0.32\textwidth}ll}
\toprule
Tier & Cost & Hardware \\
\midrule
ALIGNN voltage scalar & ${\sim}10^{-1}$~s & CPU \\
ALIGNN-FF, per step & ${\sim}10^{0}$~s & CPU or GPU \\
optB88-vdW$+U$, 32~at.\ step & 7--290~core-h & 24--32-core CPU \\
PBE$+U$, 112~at.\ step & 30--1300~core-h & 64-core CPU \\
\bottomrule
\end{tabular}
\end{table}

Table~\ref{tab:speed} sets the relative cost of the three tiers explicitly. The two surrogate tiers are many orders of magnitude faster than a single PBE$+U$ supercell relaxation on a per-structure basis, which is what makes the database-scale force-field screen of the main-text section ``High-throughput screening across JARVIS-DFT'' feasible at all. The PBE$+U$ cost per step ranges from about $30$ to $1300$ core-hours in the present campaign; the most expensive step, the empty NMC endpoint (step~16, ${\sim}1300$ core-hours), needed $190$ ionic steps. In practice the framework spends almost all of its wall-clock budget on the third tier, which is reserved for the few candidates that survive the first two. The hierarchy is therefore not a constant-factor speedup over DFT; it is a strict sequential funnel that places the expensive tier last.

\section{Supplementary discussion}
\label{si:discussion}

\subsection{Cathode-level energy density}

The cathode-level metric that ultimately determines a cell's value is the gravimetric energy density $E_\mathrm{grav} \approx V_\mathrm{avg} \times Q_\mathrm{grav}$, modulated by the active-material loading fraction and the cathode's contribution to total cell mass. For the five DFT-validated chemistries, multiplying the DFT-predicted $V_\mathrm{avg}$ by the literature theoretical gravimetric capacity (full lithium extraction per formula unit) gives, in $\mathrm{Wh\,kg^{-1}}_\mathrm{cathode}$: LFP $603$, LMP $670$ (the olivine values of main-text Table~3), LMO $605$, LCO $1220$ (with the full-delithiation average of $4.45$~V, since the theoretical capacity assumes full extraction). For NMC the corresponding number depends on the assumed stoichiometry: with the experimental NMC-111 (Li:TM $= 1{:}1$, $278$~mAh/g) and the measured average voltage over its normal cycling window, $3.85$~V\cite{si-yoshizawa2003nmc}, the cathode-level energy density is ${\sim}1070$~Wh/kg, while the JVASP-144791 cell with Li:TM $= 1{:}2$ has roughly half the theoretical lithium content per formula mass. Across the four chemistries listed, the DFT residual propagates to an error of at most about $6\%$ in cathode-level energy density (LCO, $+0.24$~V over its cycled range), well below the residual contribution from packaging, electrolyte, and cycling losses at the device level. LFP and LMO, $3\%$ apart experimentally, are less than $1\%$ apart at the DFT tier and are not resolved.

\subsection{Using the framework on structures outside JARVIS-DFT}

A reader who wants to use this framework on a candidate not in JARVIS-DFT can do so as follows, supplying the structure to the DFT tier as a local JARVIS-format record that includes its spacegroup number (the \texttt{-{}-db-cache} option of the BatteryMat \texttt{dft\_prep} module): relax the proposed structure with DFT (or with a universal force field, subject to the relaxation caveat of the main-text section ``Greedy versus exhaustive lithium-vacancy ordering'' and Supplementary Note~\ref{si:enumeration}), pass the relaxed structure to the ALIGNN scalar regressor for an average-voltage estimate, then promote to ALIGNN-FF for a full delithiation profile and to PBE$+U$ or optB88-vdW$+U$ DFT for a quantitative voltage curve. The auto-functional logic, the in-house lithium reference, and the block-AFM MAGMOM construction are all exposed as command-line entry points in the BatteryMat code release. We expect the most useful applications to be (i) screening within a focused chemistry family, where the user starts from a few hundred candidate structures rather than the full JARVIS-DFT pool; (ii) checking proposed substitutions to commercial cathodes (e.g., Mg- or Al-doped NMC variants); and (iii) regenerating the screen as new JARVIS-DFT releases bring in additional structures.

\subsection{Choice of universal force field for the second tier}

ALIGNN-FF is one of several universal interatomic potentials now available; the choice of force field at the second tier of BatteryMat is not hard-wired, and the framework is engineered to accept any of them. The relevant comparison is between ALIGNN-FF (trained on JARVIS-DFT 3D energies and forces, with a 4-ALIGNN-layer + 4-GCN-layer backbone), M3GNet\cite{si-chen2022m3gnet} (trained on ten years of Materials Project relaxation trajectories, with three-body messages), CHGNet\cite{si-deng2023chgnet} (trained on the Materials Project trajectory dataset with explicit charge information through the magnetic-moment channel), and MACE\cite{si-batatia2022mace} (higher-order equivariant message passing with four-body messages). Each is available as a pretrained universal force field, with per-structure inference times in the tens of milliseconds to seconds range on modern GPUs. The GNoME graph networks\cite{si-merchant2023gnome}, scaled to discover roughly $381{,}000$ new stable materials on the updated convex hull, belong to the same class, but their weights are not public.

For BatteryMat specifically, the requirement on the second tier is the ability to (i) evaluate the energy of a partially delithiated supercell and (ii) rank lithium-vacancy positions by relative energy; relaxation is left to the DFT tier (main-text section ``Greedy versus exhaustive lithium-vacancy ordering''). ALIGNN-FF satisfies both through the \texttt{AlignnAtomwiseCalculator} interface. M3GNet and CHGNet expose energies through the pymatgen interface; MACE exposes them through the ASE calculator wrapper. Swapping the second tier between force fields therefore requires changing only the energy function of each module; we have not benchmarked the alternatives on the delithiation task and make no claim about their relative accuracy. The most useful direction for future work is not to choose one force field as canonical but to ensemble the surrogate predictions across the available models and propagate the spread into the screening confidence interval.

\bibliography{ref}